\documentclass[a4paper,12pt]{article}
\usepackage{graphicx,rotating,amsmath,multirow,xcolor,colortbl,xcolor}
\usepackage{hyperref,pdflscape,slashed,charter}
\pdfoutput=1
\hypersetup{colorlinks,bookmarksopen,bookmarksnumbered,citecolor=verdes,
linkcolor=blus,pdfstartview=FitH,urlcolor=rossos}

\allowdisplaybreaks

\newcommand{\fig}[1]{~\ref{fig:#1}}
\def\hhref#1{\href{http://arxiv.org/abs/#1}{arXiv:#1}} 

\usepackage{multicol}
\usepackage{color}
\definecolor{rosso}{cmyk}{0,1,1,0.4}
\definecolor{rossos}{cmyk}{0,1,1,0.55}
\definecolor{rossoc}{cmyk}{0,1,1,0.2}
\definecolor{blu}{cmyk}{1,1,0,0.3}
\definecolor{blus}{cmyk}{1,1,0,0.6}
\definecolor{bluc}{cmyk}{1,1,0,0.1}
\definecolor{verde}{cmyk}{0.92,0,0.59,0.25}
\definecolor{verdec}{cmyk}{0.92,0,0.59,0.15}
\definecolor{verdes}{cmyk}{0.92,0,0.59,0.4}
\definecolor{verdess}{cmyk}{0.92,0,0.59,0.8}

\def\eq#1{eq.~(\ref{#1})}

\newcommand{\br}{+ \nonumber \\ &&}

\definecolor{Gray}{gray}{0.95}
\newcommand{\bbox}[1]{\fcolorbox{gray}{Gray}{~$\displaystyle #1$~}}

\newcommand{\GeV}{\,{\rm GeV}}
\newcommand{\TeV}{\,{\rm TeV}}

\newcommand{\Tr}{\,{\rm Tr}}
\newcommand{\diag}{\,{\rm diag}}

\def\circa#1{\,\raise.3ex\hbox{$#1$\kern-.75em\lower1ex\hbox{$\sim$}}\,}

\newcommand{\beq}{\begin{equation}}
\newcommand{\eeq}{\end{equation}}

\newcommand{\bea}{\begin{eqnarray}}
\newcommand{\eea}{\end{eqnarray}}
\newcommand{\be}{\begin{equation}}
\newcommand{\ee}{\end{equation}}
\font\tenrsfs=rsfs10 at 12pt
\font\sevenrsfs=rsfs7
\font\fiversfs=rsfs5
\newfam\rsfsfam
\textfont\rsfsfam=\tenrsfs
\scriptfont\rsfsfam=\sevenrsfs
\scriptscriptfont\rsfsfam=\fiversfs
\def\mathscr#1{{\fam\rsfsfam\relax#1}}
\def\Lag{\mathscr{L}}

\newcommand{\ba}{\begin{array}}
\newcommand{\ea}{\end{array}}

\newcommand{\lsim}{\stackrel{<}{_\sim}}
\newcommand{\gsim}{\stackrel{>}{_\sim}}
\newcommand{\Vt}{\widetilde V}
\def\mpl{M_{\rm Pl}}
\def\identity{1\!{\rm l}}
\def\Re{{\rm Re}\,}

\renewcommand{\theequation}{\thesection.\arabic{equation}}

\newcommand{\mysection}[1]{\section{#1}\setcounter{equation}{0}}

\def\circa#1{\,\raise.3ex\hbox{$#1$\kern-.75em\lower1ex\hbox{$\sim$}}\,}
\makeatletter

\def\art{\@ifnextchar[{\eart}{\oart}}
\def\eart[#1]#2#3#4#5#6{{\rm #2}, {\em #3 \bf #4} {\rm (#6) #5} ({\em #1})}
\def\hepart[#1]#2{{\rm #2, \em#1}}
\newcommand{\oart}[5]{{\rm #1}, {\em #2 \bf #3} {\rm (#5) #4}}

\newcounter{alphaequation}[equation]
\def\thealphaequation{\theequation\hbox to
0.6em{\hfil\alph{alphaequation}\hfil}}
\def\eqnsystem#1{
\def\@eqnnum{{\rm (\thealphaequation)}}
\def\@@eqncr{\let\@tempa\relax \ifcase\@eqcnt \def\@tempa{& & &} \or
  \def\@tempa{& &}\or \def\@tempa{&}\fi\@tempa
  \if@eqnsw\@eqnnum\refstepcounter{alphaequation}\fi
\global\@eqnswtrue\global\@eqcnt=0\cr}
\refstepcounter{equation} \let\@currentlabel\theequation \def\@tempb{#1}
\ifx\@tempb\empty\else\label{#1}\fi
\refstepcounter{alphaequation}
\let\@currentlabel\thealphaequation
\global\@eqnswtrue\global\@eqcnt=0 \tabskip\@centering\let\\=\@eqncr
$$\halign to \displaywidth\bgroup \@eqnsel\hskip\@centering
$\displaystyle\tabskip\z@{##}$&\global\@eqcnt\@ne
\hskip2\arraycolsep\hfil${##}$\hfil& \global\@eqcnt\tw@\hskip2\arraycolsep
$\displaystyle\tabskip\z@{##}$\hfil
\tabskip\@centering&\llap{##}\tabskip\z@\cr}
\def\endeqnsystem{\@@eqncr\egroup$$\global\@ignoretrue} \makeatother

\newcommand{\SU}{\,{\rm SU}}
\newcommand{\U}{\,{\rm U}}

\begin{document}
\thispagestyle{empty}
\centerline{CERN-PH-TH-2014-247 \qquad ZU-TH-41/14\qquad    IFT-UAM/CSIC-14-127   \qquad IFUP-TH/2014}

\bigskip

\begin{center}
{\LARGE \bf 
\color{verdess}
Softened Gravity and the Extension of\\[3mm]
the Standard Model up to Infinite Energy}\\[1cm]

{\large\bf Gian F.\ Giudice$^{a}$, Gino Isidori$^b$, \\[2mm]
Alberto Salvio$^c$, Alessandro Strumia$^d$
}  
\\[7mm]
{\it $^a$ CERN, Theory Division, CH-1211 Geneva 23, Switzerland}\\[1mm]
{\it $^b$ Physik-Institut, Universit\"at Z\"urich, CH-8057, Z\"urich, Switzerland \\ and 
INFN, Laboratori Nazionali di Frascati, I-00044 Frascati, Italy}\\[1mm]
{\it $^c$ } {\em Departamento de F\'isica Te\'orica, Universidad Aut\'onoma de Madrid\\ and Instituto de F\'isica Te\'orica IFT-UAM/CSIC,  Cantoblanco, Madrid 28049, Spain}\\[3mm]
{\it $^d$ Dipartimento di Fisica dell'Universit{\`a} di Pisa and INFN, Italy\\ and 
National Institute of Chemical Physics and Biophysics, Estonia}\\[1mm]
\vspace{1cm}
{\large\bf
Abstract}

\end{center}
\begin{quote}
{\large\noindent
Attempts to solve naturalness by having the weak scale as the only breaking of classical scale invariance have to deal with two severe difficulties: gravity and the absence of Landau poles. We show that solutions to the first problem require premature modifications of gravity at scales no larger than $10^{11}$~GeV, while the second problem calls for many new particles at the weak scale. To build models that fulfill these properties,
we classify 4-dimensional Quantum Field Theories that satisfy Total Asymptotic Freedom (TAF):
the theory holds up to infinite energy, where all coupling constants flow to zero. We develop a technique to identify such theories and determine their low-energy predictions. Since the Standard Model turns out to be asymptotically free only under the unphysical conditions $g_1 = 0$,  $M_t = 186\GeV$, $M_\tau = 0$, $M_h = 163\GeV$, we explore some of its weak-scale extensions that satisfy the requirements for TAF. 
}

\end{quote}

\vfill
\vfill
\newpage

\tableofcontents

\setcounter{footnote}{0}

\newpage

\mysection{Introduction}

The naturalness problem of the Higgs mass is a central issue for the searches in the upcoming high-energy run of the LHC and, more generally, for defining the future strategy of particle physics. The conventional wisdom that new physics should intervene below the TeV and soften the quantum corrections to the Higgs mass has been seriously challenged by Run-1 LHC data. This has prompted a reconsideration of the standard approach and the study of scenarios in which the issue of naturalness is addressed without new dynamics at the weak scale affecting the quantum corrections to the Higgs mass. 

\smallskip

A severe obstacle to this logical path is the expectation that several open questions in high-energy physics (such as quantum gravity, gauge unification, inflation, neutrino masses, baryogenesis, strong CP problem, etc.) require the existence of new heavy particles. Such particles, if sufficiently coupled to the Higgs, introduce an unavoidable naturalness problem. 
A radical (but admittedly questionable) approach is to ignore many of these high-energy open questions or, at least, to believe that some of these questions can be resolved by introducing only new particles that either have masses below the weak scale or are sufficiently decoupled from the Higgs~\cite{pap}. 
If this is the case, the SM, or a mild modification of it, could be the final theory of particle physics~\cite{follie}. However, even such a radical approach cannot ignore one problem: gravity. The unknown dynamics of quantum gravity at the Planck mass $\mpl$ is expected to introduce an unavoidable naturalness problem.

Our ignorance about quantum gravity leaves the door open to unexpected solutions. After all, we do not know if the Newton constant $G_N$ simply describes a coupling constant or signals the presence of new degrees of freedom with mass $\mpl$. Moreover, gravity becomes strong only at high energies and thus a UV softening of gravity could bypass the naturalness problem. In particular, if the power-law running of the gravitational interaction shuts off at a scale $\Lambda_G$, the gravitational corrections to the Higgs mass would amount to $\delta M_h^2 \approx \ell \, G_N \Lambda_G^4$, where the coefficient $\ell$ includes couplings and loop factors. Since, in the limit in which we turn off SM interactions, the Higgs field appears in the energy-momentum tensor only with derivatives, minimally-coupled gravity respects a Higgs shift symmetry and cannot generate through quantum corrections terms in the Higgs potential at any order in perturbation theory. However, the shift symmetry is broken by SM interactions or by a direct Higgs coupling with the scalar curvature, and we expect that corrections to $M_h^2$ occur at least at two loops with $\ell \sim (4\pi)^{-4}$. Gravity would not pose a naturalness problem as long as $\delta M_h^2 \lsim M_h^2$, which holds only
if its conventional high-energy behaviour softens before reaching the scale $\Lambda_G \sim 4\pi (M_h \mpl )^{1/2} \sim 10^{11}$~GeV. 

In conclusion, any candidate for a theory of gravity that respects Higgs naturalness must satisfy the following properties: {\it (i)} premature UV softening of gravitational interactions below $\Lambda_G$; {\it (ii)} no heavy degrees of freedom with sizeable non-gravitational couplings to SM particles. We will refer to theories of this kind as {\it softened gravity}.

\smallskip

We do not know of any realistic working examples of softened gravity, but maybe such theories could be built out of some low-scale string theory~\cite{low} or fat-gravity~\cite{fat} models.
An attempt to construct a theory of softened gravity was presented in~\cite{agravity}. 
The modification of the high-energy gravitational behaviour required the presence of a spin-two ghost at the scale $\Lambda_G$ and thus the theory cannot be considered realistic. 
Nevertheless, this attempt teaches a useful lesson: for gravity not to introduce a hierarchy problem, general relativity must be modified at the scale $\Lambda_G$, {\it i.e.} well below $\mpl$. 

Even without specifying the model of softened gravity or knowing if any such theory exists, we can nonetheless reach a general conclusion about them. Since such theories must soften the ordinary power-law energy increase starting from a scale smaller or equal to $\Lambda_G$, the gravitational interactions felt by SM particles remain weak at all energies, with couplings at most $\sim \Lambda_G/M_{\rm Pl}\sim 10^{-7}$. This means that, in the context of the hypothetical theories with softened gravity, the SM sector (or any of its extensions) is not influenced by the gravitational sector at any scale. Thus, it is sensible to investigate the behaviour of the SM using the ordinary tools of Quantum Field Theory (QFT), even at energy scales larger than $\mpl$. This leads us to consider another important issue.

\smallskip

Once we accept that a theory of softened gravity exists, we must face another problem: the scale invariance is broken at the quantum level. This breaking manifests itself in the generation of dynamical scales, such as the confinement scale in asymptotically-free gauge theories, dimensional transmutation, or Landau poles in non-asymptotically free theories. In particular, 
in the SM the hypercharge gauge coupling $g_1$ --- and, consequently, all other coupling constants --- hit Landau poles. The appearance of Landau poles in the Renormalisation-Group (RG) evolution corresponds to a loss of perturbative control.\footnote{The exception found in~\cite{Sannino},
where the growth of the couplings leads to an  interacting fixed point rather than to a Landau pole,
is perturbatively calculable because the model has a small one-loop beta function. However this is not the case for  hypercharge,
which could reach an interacting fixed point only at a non-perturbative value, thereby leading to the formation
of condensates at an unnaturally large scale.
}
Thus, any interpretation of the dynamics lurking behind Landau poles lies completely beyond the perturbative regime and different non-perturbative physical situations can occur (see {\it e.g.}~\cite{Suslov,Seiberg}). However, it is expected that a Landau pole signals the presence of new dynamics and that the corresponding mass threshold reintroduces a Higgs naturalness problem. In a Quantum Field Theory (QFT), the change in the short-distance behaviour associated with the Landau pole is believed to affect the mass of scalars charged under the corresponding interaction, whether new particles at that scale exist or not~\cite{Skiba}.

Landau poles of SM coupling constants are usually ignored, since they occur at energies much larger than $\mpl$. However, as explained above, in softened gravity the RG evolution of SM couplings is unaffected by the gravitational sector at any energy scale. As a result, the problem of Landau poles cannot be ignored and must find a solution within the sector of the SM (or one of its extensions) at the weak scale. In softened gravity, the gravitational sector cannot be of any help in preventing the appearance of SM Landau poles because of its intrinsic weakness. 

So we are led to the conclusion that a theory of softened gravity satisfying Higgs naturalness must be made up of separate sectors, connected with each other only by very feeble interactions. 
The first is the {\it observable sector}, which contains a weak-scale extension of the SM, free from any Landau pole. Its only mass scale is of the order of the TeV.
The second is a {\it gravitational sector}, which must satisfy the  property of softening gravitational interactions at the scale $\Lambda_G$ or below. This sector could be weakly or strongly-coupled, but its interactions with the observable sector must be suppressed by the gravitational coupling, of size $\Lambda_G /\mpl \sim 10^{-7}$ or less. It has still to be proved that gravitational sectors with such properties exist.  {\it Additional sectors} containing new particles (such as the inflaton, the right-handed neutrino or the axion) may exist, as long as the coupling constants $c$, describing the interaction between the new heavy particles with mass $M$ and the observable sector, are sufficiently small.
If we write the contribution to the Higgs mass from heavy particles as $\delta M_h^2 \approx \ell \, c^2 M^2$, where $\ell$ counts SM couplings and loop factors, the naturalness principle (which states that $\delta M_h^2 \lsim M_h^2$)
 requires $c\lsim 10^{-7} (4\pi \sqrt{\ell})^{-1}(10^{10}\, {\rm GeV} /M)$. 
We assume that the couplings $c$ are too small to affect in any significant way the RG evolution of the coupling constants observable sector. 

The goal of this paper is the construction of viable models for the observable sector. This task, for the reasons explained above, is fairly independent of the details of the gravitational sector (and of other additional sectors), as long as the hypotheses of softened gravity are satisfied. To achieve this goal, we study the conditions under which a general 4-dimensional QFT can hold up to infinite energy, with
all dimensionless coupling constants (gauge couplings $g$, Yukawa couplings $y$, and scalar quartic couplings $\lambda$) 
remaining perturbative at any energy above a fixed scale $\mu_0$ and flowing to zero in the far UV. We will refer to this situation as {\it Total Asymptotic Freedom} (TAF).

In the first part of this paper we derive the conditions for TAF, by studying the
Renormalisation Group Equations (RGE) that describe how a QFT  behaves at largely different energies $\mu$. The issue had already been considered in the Seventies~\cite{Gross,Cheng:1973nv,K5,K224} (see also refs.~\cite{Oehme:1984yy,Zimmermann:2001pq} and references therein), but then abandoned because of the belief that the onset of quantum gravity makes any QFT prediction above $\mpl$ completely irrelevant. We critically revisit the problem, highlighting the importance, for the determination of the TAF conditions, of Yukawa couplings sitting at special RG trajectories with isolated UV behaviour. The relevance of these solutions is an aspect that has often been missed in the past. 

\medskip

In the second part of the paper we apply our results to phenomenologically relevant theories. In particular, we address the question of constructing viable observable sectors for theories with softened gravity. Such observable sectors must be extensions of the SM that satisfy the TAF conditions, while restricting all new particles to live near the weak scale. Since these models must be based on non-abelian gauge groups, they provide an immediate explanation for the observed charge quantisation. 
Although we find examples of models that satisfy our criteria, we view such examples only as proofs of existence
that illustrate how tough it is to build TAF theories at the weak scale.
First of all, putting together the constraints from asymptotically-free couplings and from realistic flavour structures requires rather elaborate constructions with special choices of the field quantum numbers and appropriate assumptions on alignment of different flavour-violating couplings.
Second, even in the most optimistic case, limits from precision and flavour-physics experiments place 
new particles  to be well above the TeV.   Such particles
give physical corrections to the Higgs mass,
dampening hopes for a fully natural theory. 

One redeeming aspect of this unsatisfactory situation is that we have identified the most important testable prediction of softened gravity. Since any scheme of softened gravity requires the problem of SM Landau poles to be solved within the observable sector at the weak scale, the general prediction of such theories is that new particles must exist in the TeV domain. The hypercharge Landau pole requires the enlargement of the SM gauge group with new vector particles; realistic flavour structures require new fermionic particles; the Higgs embedding in the extended gauge group and the need for a correct pattern of gauge symmetry breaking require new scalar particles. The general prediction of softened gravity is the existence of many new particles around the weak scale. 

This result is in open conflict with the claim, sometimes made in the literature, that the pure SM can be made natural without adding new particles at the weak scale. It also provides a way to distinguish these theories from solutions with an anthropic explanation of naturalness. Extra scalar particles at the weak scale, beyond a single Higgs boson, find no anthropic justification. The observation of ``odd and unexpected" particles at the weak scale, seemingly unrelated to dynamical explanations of naturalness, but with the appropriate quantum numbers to satisfy the TAF conditions, are indicators for non-anthropic and non-dynamical solutions of naturalness, belonging to the class of theories with softened gravity. We have reached the surprising conclusion that, in spite of its vagueness, softened gravity is experimentally testable. Although it is not guaranteed that the new particles predicted by softened gravity must be within the reach of the LHC (especially because of the stringent limits from flavour physics), a possible future hadron collider in the 100-TeV range can certain say the last word on the viability of modifications of gravity safe from the Higgs naturalness problem.

\bigskip

The paper is organised as follows.  In section~\ref{simple} we solve analytically the RGE in simple cases and derive the conditions for TAF. We also discuss how the exact solutions are related to the asymptotic behaviour valid for ultra-high energy. Building from these simple cases, in section~\ref{multi} we develop a general formalism to study the high-energy asymptotic behaviour of coupling constants and determine the conditions for a theory to be TAF. In section~\ref{SM} we consider, as a working example, the SM and find that it can satisfy TAF only under unphysical conditions.
Motivated by naturalness, in section~\ref{nonab} we consider TAF extensions of the SM at the weak scale, discuss their general features, and study some of their phenomenological constraints. In section~\ref{towards} we explore weak-scale TAF models based on the gauge group $\SU(2)_L\otimes\SU(2)_R\otimes\SU(4)_{\rm PS}$,
finding a proof of existence, and models based on  $\SU(3)_L\otimes\SU(3)_R\otimes\SU(3)_c$. Finally, we present our conclusions in section~\ref{concl}.

\mysection{Conditions for {TAF}: simple cases} \label{simple}
The RGE structure is such that generally, in the perturbative regime, no coupling can flow toward non-zero UV fixed points. A possible exception, which we ignore, occurs when the one-loop $\beta$-function is accidentally small and the interplay between one- and two-loop contributions can generate a non-trivial structure\footnote{This case was recently investigated in ref.~\cite{Litim:2015iea}.}.
Since the only UV fixed points relevant for our considerations correspond to vanishing couplings, we are justified to truncate the RGE at the one-loop order.

In this section we consider simple cases where full explicit solutions to the one-loop RGE can be obtained.
We define  $t = \ln(\mu^2/\mu_0^2)/{(4\pi)^2}$, so that
 $t= 0$ corresponds to the IR scale $\mu_0$ (which could be the weak scale or some higher reference scale) and $t\approx 0.5$ corresponds to the running from the weak to the Planck scale. Here we are interested in the behaviour for $t\to \infty$.

\subsection{One gauge coupling}
The one-loop RGE for gauge couplings is
\beq 
\frac{d}{dt} g^2 = -b g^4\ ,
\label{eqg0}
\eeq
where $b$ is the $\beta$-function coefficient. The solution of \eq{eqg0} is
\beq
\frac{1}{g^{2}}=\frac{1}{g_0^{2}} +bt,
\eeq
where $g_0$ is the gauge coupling at the IR scale $\mu_0$ ($t=0$). For $b<0$, we encounter a Landau pole at $t_*=-1/(g_0^2 b)$. For $b>0$, the gauge theory is asymptotically free and the asymptotic solution is
\beq
g^2 =\frac{1}{bt} ~~~~{\rm for}~t\to \infty \ .
\eeq
Thus the {TAF} condition is\footnote{Theories where $b=0$ can have a different asymptotic solution for gauge couplings, $g^2\propto 1/\sqrt{t}$, depending on the sign of the two-loop RGE coefficient for $g$.  However such solution cannot be extended to systems 
with Yukawa and quartic couplings, where non-vanishing RGE arise at one loop.}
\beq
\bbox{b>0  \qquad \hbox{({TAF} condition for the gauge coupling).}}
\eeq
{TAF} constrains the matter content of the theory and, in particular, excludes any Abelian gauge groups. So the SM does not satisfy {TAF}, because of hypercharge.

\subsection{One Yukawa coupling}
Next we consider Yukawa couplings, focusing on the case of a single (asymptotically free) gauge coupling $g$ and a single Yukawa $y$.
The RGE for  $y$ is
\beq 
\frac{d}{dt} y^2 = y^2(f_y  y^2 - f_g  g^2) \ ,
\label{eqy0}
\eeq
where $f_y$ and $f_g$ are non-negative constants in any QFT. It is convenient to express \eq{eqy0} in terms of new variables
\beq
Y\equiv \frac{g^2}{y^2} \ , ~~~~~x\equiv \ln \frac{g_0^2}{g^2} \ ,
\eeq
such that
\beq
\frac{dY}{dx} =A(Y-B)
\label{eqy1}
\eeq
where
\beq
A\equiv \frac{f_g}{b}-1\ , ~~~~~B\equiv \frac{f_y}{f_g -b} \ .
\eeq
$A$ and $B$ have always the same sign: positive if $f_g>b$ and negative otherwise.
The solution to \eq{eqy1} is
\beq
Y=(Y_0-B)\left( \frac{g_0^2}{g^2} \right)^A +B \ ,
\label{eqy3}
\eeq
where $Y_0=g_0^2/y_0^2$ and $y_0$ is the Yukawa coupling at the IR scale $\mu_0$ ($t=0$). 
Equation~(\ref{eqy3}) can be written more explicitly as
\beq
y^{-2}=\left[ y_0^{-2}- \frac{f_y}{(f_g -b) g_0^2} \right] \left( \frac{g_0^2}{g^2} \right)^{\frac{f_g}{b}} + \frac{f_y}{(f_g -b) g^2} \ .
\label{eqsoly}
\eeq

Landau poles in the Yukawa coupling exist if the equation $Y(t)=0$ admits solutions for some positive values of $t$. From \eq{eqy3}, we see that there are Landau poles when $A\le 0$ or when $A>0$ and $B>Y_0$. Therefore, in the case of the Yukawa coupling, the conditions for {TAF} are $A>0$ and $B\le Y_0$, which can be written explicitly as
\beq
\bbox{ f_g>b~~~{\rm and}~~~\frac{y_0^2}{g_0^2} \le \frac{f_g -b}{f_y }  \qquad \hbox{({TAF} conditions for the Yukawa coupling).}}
\label{TAFyuk}
\eeq

The one-loop coefficient $f_g$ is given by
\beq f_g =\frac32 (C_{2\psi_1}+C_{2\psi_2})\eeq 
where $C_{2\psi_{1,2}}$ are the quadratic Casimirs of the two fermions involved in the Yukawa coupling.
The Yukawa coupling is allowed by {TAF} provided that their $C_2$ Casimirs are large enough.
A gauge-neutral fermion (`right-handed neutrino') has $C_{2\psi_1}=0$: it can have Yukawa coupling 
compatibly with {TAF} provided that the other fermion has a large enough $C_{2\psi_2}$.


\begin{figure}[t]
\begin{center}
$$\includegraphics[width=0.45\textwidth]{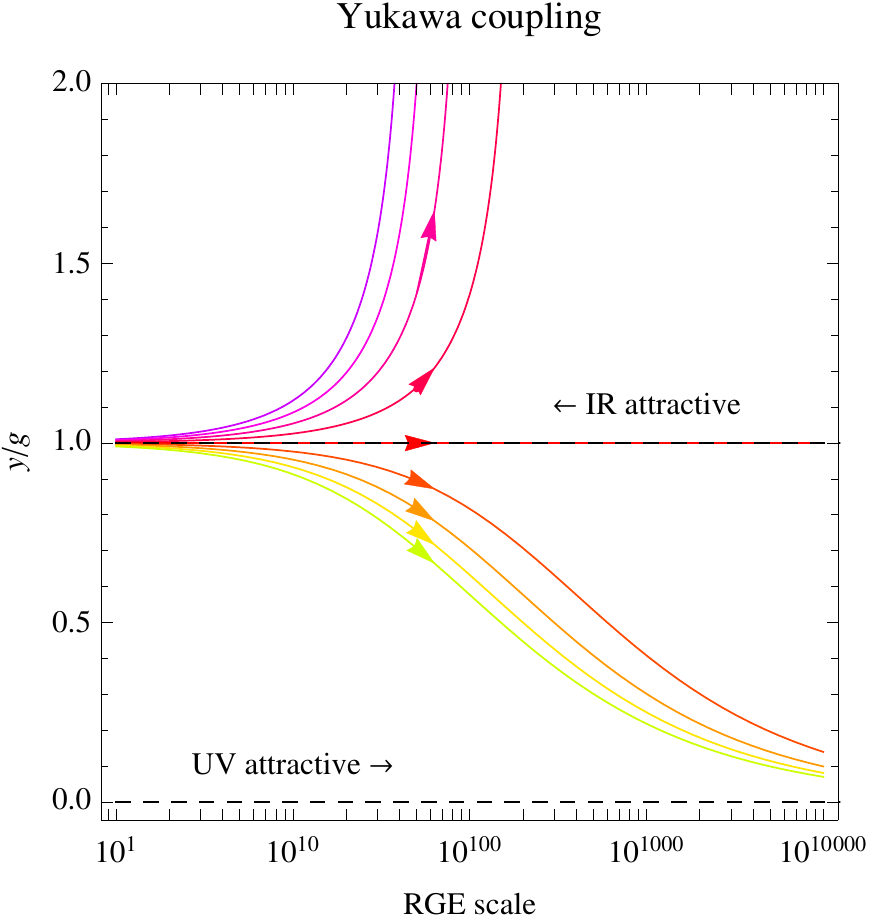}\qquad
\includegraphics[width=0.45\textwidth]{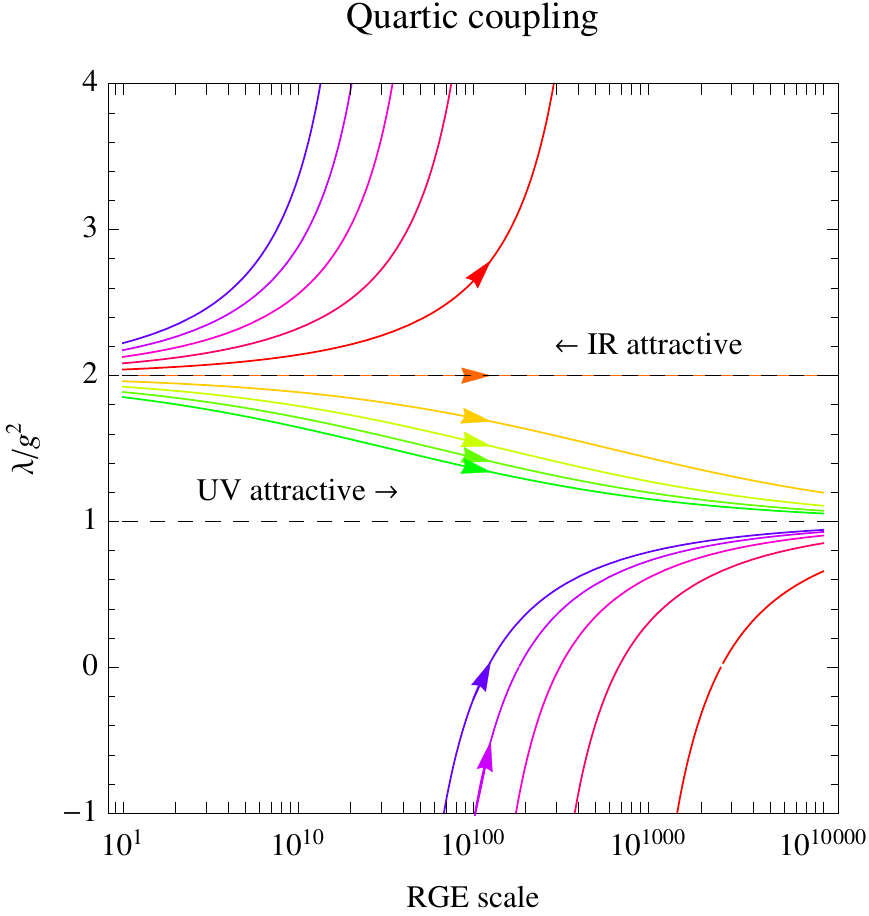}$$
\caption{\em Sample of the general behaviour of the RG running of the Yukawa coupling $y$ (left, for $A=B=b=1$) and scalar quartic $\lambda$ (right, for $C=b=1$, $D=3/4$, $E=1/16$) varying the initial conditions.
\label{fig:runs}}
\label{default1}
\end{center}
\end{figure}

Since $A>0$ for the {TAF} solution, the first term in \eq{eqy3} dominates in the UV ($t \to \infty$) over the constant term, as long as $B\ne Y_0$. Then, the asymptotic behaviour is
\beq
y^2 \propto \left( \frac 1t \right)^{\frac {f_g}{b}} ~~~~{\rm for~}t\to \infty
\label{asy0}
\eeq
and, at large $t$, the Yukawa coupling becomes negligible with respect to the gauge coupling, $g^2\propto 1/t$. The exception 
(missed in~\cite{Cheng:1973nv})
is for $B=Y_0$, when $g$ and $y$ scale in the same way with $t$. We will refer to this case as {\it fixed-flow}\footnote{We thank M.~Strassler for suggesting this name to us.}, because couplings run, but their ratio is RG invariant. At the fixed-flow, the Yukawa coupling is fixed at any energy in terms of the gauge coupling
\beq
y^2= \frac{f_g -b}{f_y} \ g^2 ~~~~{\rm for~any~}t \ .
\label{soly0}
\eeq
Equation~(\ref{soly0}) is an exact solution of the RGE, which corresponds to the Pendleton-Ross point~\cite{Pendleton:1980as} in the IR. We will see later that the existence of a fixed-flow solution for the Yukawa coupling is an important ingredient to satisfy the {TAF} requirement for scalar quartic couplings.

We illustrate the situation for $A>0$ in fig.~\ref{fig:runs}, where we plot the running of $y/g$ ($=1/\sqrt{Y}$) for $A=1$ and $B=1$, while varying the initial condition of the coupling constants. The fixed-flow in \eq{soly0} corresponds to the limiting case of a family of solutions ($Y_0\le B$) that suffer from Landau poles; in the special case $Y_0=B$, the Landau pole slides to $t=\infty$ and the {TAF} condition is satisfied. The fixed-flow is an IR attractor. For $Y_0>B$, the solutions have the asymptotic behaviour in \eq{asy0} and are attracted in the UV to the point $y/g=0$.  

\subsubsection*{Asymptotic solutions}
It is instructive to obtain the asymptotic behaviour of the solutions of the RGE without using their complete analytic expression. This procedure is redundant here, because we have solved exactly the RGE, but it will be useful later, whenever we are not able to solve analytically the full RGE. Making the ansatz $Y= ct^\alpha$ for the asymptotic behaviour of the Yukawa coupling, \eq{eqy1} turns into an algebraic equation
\beq
c\, \alpha \left( 1+\frac{1}{bg_0^2t}\right) =A(c-Bt^{-\alpha}) \qquad\hbox{for}\qquad t\to \infty \ .
\eeq
Asymptotic solutions can exist only for $\alpha \ge0$, and we find two possibilities. Either $\alpha =A$, which corresponds  
to \eq{asy0}; or $\alpha =0$ and $c=B$, which corresponds to \eq{soly0}. 

Moreover, the nature of the fixed-flow can be understood by analysing solutions of the form $Y=B+\Delta$, which represent small deformations of the fixed-flow solution. From \eq{eqy1}, we find that the perturbation $\Delta$ satisfies
\beq
\frac{d}{dx}|\Delta |=A|\Delta |.
\eeq
For $A>0$, the perturbation grows with $t$ and thus the fixed-flow is repulsive in the UV. 
So we can easily reproduce all the features of the asymptotic behaviour without solving analytically the RGE.

\subsection{One scalar quartic coupling}\label{secqua}
Next we turn to the scalar quartic coupling, considering the case of a single coupling of each kind ($g$, $y$, $\lambda$), where $g$ and $y$ are asymptotically free. The relevant RGE is
\beq 
\frac{d}{dt}\lambda = \lambda (s_\lambda  \lambda + s_{\lambda y}  y^2 - s_{\lambda g}  g^2) -s_y y^4 + s_g  g^4 \ ,
\label{eqlam0}
\eeq
where, for any QFT, all coefficients $s_i$ are positive. We first solve \eq{eqlam0} in special cases.

\subsubsection*{One quartic, without Yukawa or gauge couplings}
In this case, the solution to \eq{eqlam0} is
\beq
\lambda^{-1} =\lambda_0^{-1} -s_\lambda t \ ,
\eeq
where $\lambda_0$ is the quartic coupling at the IR scale $\mu_0$ ($t=0$). If $\lambda_0 >0$, a Landau pole is reached at the scale $t_*=1/(\lambda_0 s_\lambda )$. If $\lambda_0 <0$, there are no Landau poles, but the coupling $\lambda$ is negative at all scales and the scalar potential is unstable. This means that other interactions, beyond the scalar quartic, are needed to satisfy {TAF}. This agrees with the well-known result~\cite{Gross} that no renormalisable field theory without non-abelian gauge fields can be asymptotically free.

\subsubsection*{One quartic, with one gauge coupling and no Yukawas}
It is convenient to express the RGE \eq{eqlam0} in terms of new variables
\beq
\Lambda \equiv \frac{g^2}{\lambda} \ , ~~~~~x\equiv \ln \frac{g_0^2}{g^2} \ ,
\eeq
such that
\beq
\frac{d  \Lambda}{dx} =-C \left[ (\Lambda - D)^2-E \right]
\label{eqy1uffa}
\eeq
where
\beq
C\equiv \frac{s_g }{b }\ , ~~~~D\equiv \frac{s_{\lambda g}-b}{2\, s_g}
\ , ~~~~ E\equiv \frac{(s_{\lambda g}-b)^2-4s_\lambda s_g}{4\, s_g^2}
 \ .
 \label{defpar}
\eeq
Note that the definitions in \eq{defpar} imply $C>0$ and $D^2>E$.
Thereby, the two special  solutions where $\Lambda$ is RG invariant, $\Lambda(x)=\Lambda_\pm = D \pm \sqrt{E}$ are both positive if $D>0$,
and both negative otherwise.
Since the right-hand side of \eq{eqy1uffa} depends only on the variable $\Lambda$, we can easily integrate the RGE. 

For $E<0$, the solution is
\beq
\Lambda =\frac{\Lambda_0 \sqrt{-E}+[D(\Lambda_0 -D)+E] \tan (C\sqrt{-E}x)}{\sqrt{-E}+(\Lambda_0-D) \tan (C\sqrt{-E}x)}
~~~~{\rm for~} E<0 \ .
\eeq
As the equation $\Lambda (x)=0$ admits solutions for positive $x$, the coupling $\lambda$ hits Landau poles. 

For $E>0$, the solution is
\beq
\Lambda =\frac{ (D+\sqrt{E})(\Lambda_0-D+\sqrt{E}) \left( \frac{g_0^2}{g^2}\right)^{2C\sqrt{E}}-(D-\sqrt{E})(\Lambda_0-D-\sqrt{E})}
 { (\Lambda_0-D+\sqrt{E}) \left( \frac{g_0^2}{g^2}\right)^{2C\sqrt{E}}-(\Lambda_0-D-\sqrt{E})}
~~~~{\rm for~} E>0 \ .
\label{solsol}
\eeq 
In this case, Landau poles (solutions of $\Lambda =0$) are found for $D<-\sqrt{E}$ or $D>\Lambda_0+\sqrt{E}$. Thus, the {TAF} requirement of having no Landau poles at any $t$ is $E\ge 0$ and $\sqrt{E}\le D\le \Lambda_0+\sqrt{E}$ (recall that $|D|>\sqrt{E}$) or, more explicitly,
\beq
\bbox{s_{\lambda g} - b\ge 2\sqrt{s_\lambda s_g}  ~~~{\rm and}~~~\frac{\lambda_0}{g_0^2} \le \frac{(s_{\lambda g} -b)+  \sqrt{(s_{\lambda g}-b)^2-4 s_\lambda s_g}}{2 s_\lambda  } \qquad \parbox{0.26\textwidth}{({TAF} conditions for the scalar quartic coupling).}} 
\label{CAFlambdanoy}
\eeq

The solution in \eq{solsol} does not cross $\lambda =0$, whenever the {TAF} conditions in \eq{CAFlambdanoy} are satisfied. Therefore, there are no problems with potential stability at any $t$. 
The asymptotic behaviour of \eq{solsol} is $\Lambda = D+\sqrt{E}$, which can be written explicitly as
\beq 
\lambda = \frac{s_{\lambda g} -b-  \sqrt{(s_{\lambda g}-b)^2-4 s_\lambda s_g}}{2s_\lambda b \ t} ~~~~{\rm for}~t\to \infty \ .
\label{asinf}
\eeq

The asymptotic solution becomes an exact solution in the case $\Lambda_0=D+\sqrt{E}$. In this case we find a fixed-flow such that the ratio $\lambda /g^2$ is RG invariant. This corresponds to the condition
\beq 
\frac{\lambda}{g^2} = \frac{s_{\lambda g} -b-  \sqrt{(s_{\lambda g}-b)^2-4 s_\lambda s_g}}{2 s_\lambda  }~~~~{\rm for~any}~t \ .
\label{uvattra}
\eeq

As we approach $D=-\sqrt{E}$ or $\Lambda_0=D-\sqrt{E}$, the Landau pole slides to $t\to \infty$. The case $D=-\sqrt{E}$ is not interesting because it corresponds to $s_\lambda =0$, which is never verified in a QFT. Especially interesting is the case $D=\Lambda_0+\sqrt{E}$, in which the exact solution does not have the asymptotic behaviour of \eq{asinf}. This case gives another kind of fixed-flow of the ratio $\lambda /g^2$, given by
\beq 
\frac{\lambda}{g^2} = \frac{s_{\lambda g} -b+  \sqrt{(s_{\lambda g}-b)^2-4 s_\lambda s_g}}{2 s_\lambda  }~~~~{\rm for~any}~t \ .
\label{irattra}
\eeq
Note that, unlike the case of the Yukawa coupling, the quartic has only a single possible asymptotically-free behaviour, given by $\lambda \propto 1/t$. 
%

To visualise the situation for $E>0$, we show in fig.~\ref{fig:runs}b the ratio $\lambda /g^2$, obtained from \eq{solsol}, as a function of the RG parameter $t$ for different initial conditions $\Lambda_0$ (with the choice $C=b=1$, $D=3/4$, $E=1/16$, such that $\Lambda_-=1$ and $\Lambda_+=2$). The two fixed-flow solutions correspond to the two horizontal lines. The lower one, which is given by \eq{uvattra}, tracks the asymptotic behaviour of the other {TAF} solutions and therefore is a UV attractor. The upper one, given by \eq{irattra}, corresponds to the separating case between a family of solutions (with $\Lambda_0 > D-\sqrt{E}$) that blow up at finite $t$ and a family of asymptotically-free solutions (with $\Lambda_0 < D-\sqrt{E}$). In the separating case (with $\Lambda_0 = D-\sqrt{E}$), there is no Landau pole and the {TAF} condition can be satisfied. Therefore, the upper horizontal line, given by \eq{irattra}, is an isolated solution that cannot be continuously deformed in the asymptotic UV region without violating the {TAF} conditions. It corresponds to an IR attractor. 

\medskip

Can the condition \eq{CAFlambdanoy} for {TAF} be satisfied, in absence of Yukawa couplings?
Let us consider a generic scalar $\varphi $ in an irreducible representation $R$ with dimension $d_R$ and
real dimension $d_{RR}$ 
($d_{RR}=d_R$ for a real representation and $d_{RR} =2d_R$ for a complex representation)
with generators $T^a$ under a generic simple group  $G$.
We define the usual quadratic Casimirs as
$ (T^a T^a)_{ij} = C_R \delta_{ij}$ and $\Tr(T^a T^b) = \delta^{ab} T_R$, related by $d_R C_R = d_A T_R$,
where $A$ is the adjoint representation.
The RGE coefficient $s_{\lambda g}$ is given by
$  s_{\lambda g} = 6C_R$.
To compute the other RGE coefficients, we suppose that the group theory allows only  one quartic, given by the square of 
the quadratic, $V = \frac18 \lambda (\sum \varphi_i^2)^2 $, where $\varphi_i$ are the canonically normalised $d_{RR}$ real components of $\varphi$.
Then the other coefficients are
\beq s_\lambda = 4+\frac{d_{RR}}{2},\qquad s_g = \frac{3C_R \left[2C_R (2d_A+d_{RR})-d_A C_A\right] }{d_{A} (2+d_{RR})}.
\eeq
The TAF{} condition in \eq{CAFlambdanoy}  explicitly becomes
\beq \left(C_R - \frac{b}{6}\right)^2  > \frac{C_R}{6d_A} \frac{8+d_{RR}}{2+d_{RR}} \left[2C_R (d_{RR}+2d_A)-6 d_A C_A\right].\eeq
In the most favourable case where the gauge $\beta$-function coefficient $b$ is small and can be neglected, the {TAF} condition simplifies to 
\beq \frac{C_A}{C_R}> \frac{36}{8+d_{RR}}+2\frac{d_{RR}}{d_A}-2.\eeq
Such condition favours  representations smaller than the adjoint and large gauge groups.
For $G=\SU(n)$ one has $C_A=n$, $d_A=n^2-1$ and, for its fundamental $n$, $T_R=1/2$ and $d_R=d_{RR}/2=n$ such that $C_R = (n^2-1)/2n$.
For $G={\rm SO}(n)$ one has $C_A=(n-2)/2$, $d_A=n(n-1)/2$ and, for its fundamental $n$, $T_R=1/2$ and $d_R=d_{RR}=n$ such that $C_R=(n-1)/4$.
Therefore, in both cases the TAF{} condition is satisfied for   $n$ larger than a critical value that depends on $b$.

\medskip

The {TAF} condition for scalar quartics is more easily satisfied in presence of Yukawa couplings, as we now discuss.



\subsubsection*{One quartic, with one gauge and one Yukawa coupling} 
In terms of the variables introduced previously, \eq{eqlam0} for $g,y\ne 0$ becomes
\beq
\frac{d}{dx} \Lambda=-\frac{1}{b} \left[ (s_g -\frac{s_y}{Y^2}) \, \Lambda^2 -(s_{\lambda g} -b-\frac{s_{\lambda y}}{Y})\Lambda +s_\lambda \right] \ .
\label{eqyg1}
\eeq

Let us first consider the case in which the Yukawa is not on its fixed-flow. 
Although we cannot analytically integrate \eq{eqyg1}, we can consider asymptotic solutions for $t\to \infty$. 
In this regime, $y$ is negligible with respect to $g$ and therefore the asymptotic solutions to \eq{eqyg1} are $\Lambda = \Lambda_\pm$, {\it i.e.} the same as in the case without Yukawa, given by  
eqs.~(\ref{uvattra}) and (\ref{irattra}). The important difference is that, in this case, these solutions hold only in the $t\to\infty$ asymptotic region.
The nature of these two asymptotic behaviours can be studied by making small deformations, taking $\Lambda = \Lambda_{\pm}+\Delta_{\pm}$. From \eq{eqyg1} we find that the perturbation $\Delta_{\pm}$ satisfies
\beq
\frac{d}{dx}|\Delta_{\pm}|=\pm \frac{\sqrt{(s_{\lambda g}-b)^2-4 s_\lambda s_g}}{b}\,
 |\Delta_{\pm}| \ .
\label{delttt}
\eeq
Thus, $|\Delta_{-}|$ decreases at large $t$, while  $|\Delta_{+}|$ increases. This shows that $\Lambda_{-}$ is UV attractive and $\Lambda_{+}$ is UV repulsive. Also, $\Lambda_+$ defines the divider between solutions with Landau poles and asymptotically-free solutions. The divider is an isolated solution that behaves as $\Lambda_+$ for $t\to \infty$. All other {TAF} solutions converge to $\Lambda_-$ in the UV. 
These results are in agreement with what found in the previous section. In particular, the {TAF} conditions are still given by \eq{CAFlambdanoy}.

\bigskip

Next, let us consider the case in which the Yukawa coupling is on its fixed-flow, given by \eq{soly0}. Then \eq{eqyg1} becomes
formally identical to \eq{eqy1uffa}, after the replacement
\beq
s_g \to s_g -\frac{(f_g-b)^2}{f_y^2}\,s_y ~,~~~~s_{\lambda g}\to 
s_{\lambda g} -\frac{(f_g-b)}{f_y}\, s_{\lambda y} \ .
\eeq
Thus, the solutions to \eq{eqyg1} are analogous to those discussed in the previous section, once we replace the parameters $C$, $D$, $E$ defined in \eq{defpar} with
\beq
\hat{C} \equiv \frac 1b \left[ s_g -\frac{(f_g-b)^2}{f_y^2}\,s_y \right] ,~~~
\hat{D} \equiv \frac{1}{2b\hat{C}} \left[ s_{\lambda g} -\frac{(f_g-b)}{f_y}\, s_{\lambda y} -b \right] ,~~~
\hat{E} \equiv \hat{D}^2 - \frac{s_\lambda}{b\hat{C}}.
\label{defparhat}
\eeq
The important difference is that, unlike the case without Yukawa coupling, the parameter $\hat{C}$ can be either positive or negative. This gives rise to two classes of asymptotically-free solutions for the scalar quartic coupling.

The first class of solutions occurs for $\hat{C}>0$ (which implies $\hat{D}^2 > \hat{E}$). The conditions for the absence of Landau poles are 
\beq
\bbox{\hat{C}>0, ~~~\hat{E}>0  ~~~{\rm and}~~~\frac{\lambda_0}{g_0^2} \le \left(\hat{D}-\sqrt{\hat{E}}\right)^{-1} 
\qquad \parbox{0.34\textwidth}{({TAF} conditions for the scalar quartic coupling with Yukawa on fixed-flow).}} 
\label{CAFlambdaquasiy1}
\eeq
The exact solutions are given by \eq{solsol} (with the replacement $C,D,E \to \hat{C},\hat{D},\hat{E}$). From these solutions, we observe that the coupling $\lambda$ never crosses zero, so there is no instability issue. The asymptotic behaviour of these solutions for $t\to \infty$ is $\Lambda = \hat{D}+\sqrt{\hat{E}}$. In practical applications, the contribution from the Yukawa coupling on its fixed-flow can be very useful, because it is easier to satisfy the {TAF} conditions in \eq{CAFlambdaquasiy1}, rather than those in \eq{CAFlambdanoy}.

The second class of solutions occurs for $\hat{C}<0$ (which implies $\hat{D}^2 < \hat{E}$). The conditions for the absence of Landau poles are 
\beq
\bbox{\hat{C}<0 ~~~{\rm and}~~~\frac{\lambda_0}{g_0^2} \le \left(\hat{D}+\sqrt{\hat{E}}\right)^{-1} 
\qquad \parbox{0.45\textwidth}{({TAF} conditions for the scalar quartic coupling with Yukawa on fixed-flow).}}
\label{CAFlambdaquasiy2}
\eeq
The exact solutions are given again by \eq{solsol} (with the replacement $C,D,E \to \hat{C},\hat{D},\hat{E}$). In this case, if the coupling $\lambda$ starts positive in the IR, it will cross zero, becoming negative at high energy and raising a problem with the stability of the potential. The asymptotic behaviour of the solutions for $t\to \infty$ is $\Lambda = \hat{D}-\sqrt{\hat{E}}$.

\subsection{Multiple gauge couplings}\label{gauge}
The generalisation to semi-simple groups is trivial, since in the one-loop approximation each gauge coupling evolves independently
\beq 
\frac{d}{dt} g_i^2 = -b_i g_i^4\ .
\label{eqg0multi}
\eeq
Here the index $i$ scans over the couplings of the different simple gauge group factors. 
The full solutions are $g_i^2  = g_{0i}^2/(1+g_{0i}^2 b_i t)=1/b_i(t-t_i)$,
where $t_i = -1/b_i g_{0i}^2$ are the low-energy scales where the gauge coupling $g_i$ becomes non-perturbative. The asymptotic behaviour is
\beq\label{eq:gi}
g_i^2 \simeq \frac{1}{b_i t} ~~~~{\rm for}~t\to \infty \ .
\eeq
The {TAF} conditions are
\beq
\bbox{b_i>0  \qquad \hbox{({TAF} conditions for multiple gauge couplings).}}
\eeq

 \subsubsection*{One Yukawa, with multiple gauge couplings}
We  consider the case of a single Yukawa coupling with a semi-simple gauge group. Equation~(\ref{eqy0}) becomes
\beq \label{above}
\frac{d}{dt} y^2 = y^2\left( f_y  y^2 - \sum_i f_{g_i}  {g_i}^2 \right) \ ,
\eeq
where $f_y$ and $f_{g_i}$ are non-negative constants in any QFT.
The general solution of \eq{above},  analogous to \eq{eqsoly}, is
\beq
y^{-2}=\left( y_0^{-2}- f_y \ I \right) \prod_i
\left( \frac{g_{i0}^2}{g_i^2} \right)^{\frac{f_{g_i}}{b_i}} \ ,
\label{solymul}
\eeq
where
\beq
I(t) \equiv \int_0^t dt^\prime ~ \prod_i
\left( \frac{g_{i0}^2}{g_i^2} \right)^{-\frac{f_{g_i}}{b_i}}\ .
\label{deffy}
\eeq
When $\sum_i f_{g_i}/b_i \le1$, \eq{solymul} always encounters a Landau pole. For $\sum_i f_{g_i}/b_i >1$, we can define $I_\infty \equiv \lim_{t\to \infty} I(t)$ and
expand $I$ in the asymptotic region as
\beq
I(t)=I_\infty +\frac{\prod_i (g_{i0}^2b_i)^{-\frac{f_{g_i}}{b_i}}}{1-\sum_i \frac{f_{g_i}}{b_i}}~t^{1-\sum_i \frac{f_{g_i}}{b_i} }~~~~{\rm for}~t\to \infty \ .
\eeq
Then, the conditions for {TAF}, which generalise \eq{TAFyuk}, are
\beq
\bbox{\sum_i \frac{f_{g_i}}{b_i} > 1~~~{\rm and}~~~y_0^2 \le \frac{1}{f_y I_\infty }  \qquad\hbox{({TAF} condition for the Yukawa, with multiple $g_i$)}.}
\label{Cyuk}
\eeq
Whenever the second condition in \eq{Cyuk} is satisfied as a strict inequality, the asymptotic behaviour is
\beq
y^2 = \frac{t^{-\sum_i \frac{f_{g_i}}{b_i} }}{(y_0^{-2}- f_y \ I_\infty ) \prod_i (g_{i0}^2b_i)^{\frac{f_{g_i}}{b_i}} } ~~~~{\rm for}~t\to \infty \ ,
\eeq
and the Yukawa coupling decreases faster than the gauge couplings at large $t$. However, when $y_0^{-2} = f_y I_\infty$, the ratio between the Yukawa coupling and any gauge coupling $g_j$ is constant in the asymptotic region,
\beq
\frac{y^2}{g_j^2}= \frac{b_j}{f_y}\left[ \left( \sum_i \frac{f_{g_i}}{b_i} \right) -1\right] ~~~~{\rm for~}t \to \infty \ .
\label{solgg}
\eeq
Unlike the case of a single gauge coupling, in which the ratio $y^2/g^2$ was RG invariant on the
fixed-flow, \eq{solgg} is valid only in the asymptotic regime. It corresponds to an isolated RGE solution, which behaves as an attractor in the IR and is characterised by the low-energy value $y_0^2 = 1/f_y I_\infty$.

\subsubsection*{One quartic, with one Yukawa and multiple gauge couplings}\label{gggy}
In this case the RGE for the scalar quartic coupling $\lambda$ is 
\beq \frac{d}{dt}\lambda = \lambda \left(s_\lambda  \lambda + s_{\lambda y}  y^2 - \sum_i s_{\lambda_{g i}}  g_i^2\right) -s_y y^4 + \sum_{ij} s_{gij}  g_i^2 g_j^2,
\label{rgemult}
\eeq
where $s_\lambda, s_{\lambda y}, s_{\lambda g i}, s_y$ and $s_{gij}$ are non-negative constants.

Although we do not solve exactly \eq{rgemult}, we can easily obtain the asymptotic behaviour for $t\to \infty$.
When the Yukawa coupling does not satisfy the special initial condition $y_0^2 = 1/f_y I_\infty$, it can be neglected with respect to the gauge couplings in the deep UV. Then, the two possible asymptotic behaviours for the scalar quartic are 
\beq
\frac{g_k^2}{\lambda} = D_k \pm \sqrt{E_k} ~~~~~{\rm for}~t\to \infty \, ,
\eeq
\beq
C_k \equiv b_k \sum_{i,j} \frac{s_{g_{ij}}}{b_i b_j} ,~~~
D_k \equiv \frac{1}{2C_k}\left( \sum_i \frac{s_{\lambda g_i}}{b_i} -1 \right), ~~~
E_k \equiv D_k^2 -\frac{s_\lambda}{b_k C_k} .
\eeq
The parameters $C_k,D_k,E_k$ are the generalisation to multiple-gauge couplings of the parameters $C,D,E$ previously defined in \eq{defpar}. Indeed, the discussion of the solutions is completely analogous to the case of a single gauge coupling, once we translate the parameters $C,D,E$ into $C_k,D_k,E_k$. In particular $E_k >0$ is a necessary condition for {TAF}.

When the Yukawa coupling satisfies $y_0^2 = 1/f_y I_\infty$, it has the same asymptotic behaviour as the gauge couplings in the UV. In this case, the asymptotic solutions for the scalar quartic are
\beq
\frac{g_k^2}{\lambda} = {\hat D}_k \pm \sqrt{{\hat E}_k} ~~~~~{\rm for}~t\to \infty \, ,
\eeq
\bea
&{\hat C}_k \equiv b_k \left[ \sum_{i,j} \frac{s_{g_{ij}}}{b_i b_j}-\left( \sum_i \frac{f_{g_i}}{b_i}-1\right)^2\frac{s_y}{f_y^2}\right] \, ,& \nonumber \\
&{\hat D}_k \equiv \frac{1}{2{\hat C}_k}\left[ \sum_i \frac{s_{\lambda g_i}}{b_i}  -\left( \sum_i \frac{f_{g_i}}{b_i}-1\right)\frac{s_{\lambda y}}{f_y}-1 \right], ~~~
{\hat E}_k \equiv {\hat D}_k^2 -\frac{s_\lambda}{b_k {\hat C}_k} & .
\eea
Again, the discussion of the solutions is analogous to the case of a single gauge coupling, with the translation 
of the parameters $\hat{C},\hat{D},\hat{E}$ defined in \eq{defparhat} into the multiple-gauge parameters $\hat{C}_k,\hat{D}_k,\hat{E}_k$.

\subsection{Supersymmetric case}

In the context of a generic QFT, having the Yukawa coupling sitting exactly on the fixed-flow solution of the kind of \eq{soly0}, with the isolated asymptotic behaviour $y^2 \propto 1/t$, may seem a very special situation corresponding to an extreme fine-tuning of the initial condition for the $y/g$ ratio. It is interesting to remark that, in the context of supersymmetric theories, such a special situation could be dictated by symmetry properties. The Yukawa coupling $y_g$ of the fermion--sfermion--gaugino interaction automatically satisfies the condition to sit on the fixed-flow. This is because supersymmetry ensures that $y_g$ is proportional to the gauge coupling $g$ at all scales. On the contrary, as seen in section~\ref{gauge}, when the Yukawa coupling of a generic QFT is on its fixed-flow, the proportionality between $y$ and $g$ holds only in the asymptotic region $t\to\infty$, while a more complicated behaviour appears at finite energy.

The other interesting aspect about supersymmetry is that scalar quartic couplings are induced by $D$-terms and therefore are proportional to $g^2$. Gauge asymptotic freedom automatically ensures that such quartic couplings are free from Landau poles and satisfy the {TAF} conditions. Nevertheless, low-energy supersymmetry does not offer much practical advantage in the construction of TAF extensions of the SM. The reason is that, in low-energy supersymmetry, both $\U(1)_Y$ and $\SU(2)_L$ are non-asymptotically free and a TAF extension requires rather big gauge groups at the weak scale.



\mysection{Conditions for {TAF}: general case} \label{multi}
In section~\ref{simple} we analysed the RGE in some simple cases, finding
analytic solutions. However, as soon as we increase the number of couplings, the problem of solving exactly the RGE quickly becomes analytically cumbersome and intractable. 
Armed with the experience acquired from the simple cases, we can now present a  systematic procedure to identify all {TAF} conditions in a generic QFT with multiple couplings. 
In this section we will illustrate this method.
 
 We  consider a generic QFT with multiple gauge couplings $g_i$, Yukawa couplings $y_a$ and scalar quartic couplings $\lambda_m$. Our starting point is to define  new rescaled couplings $x_I=\{\tilde{g}_i,\tilde{y}_a,\tilde{\lambda}_m\}$ by factoring out the leading asymptotic behaviour $1/t$:
\beq g_i^2(t) = \frac{\tilde{g}_i^2(t)}{t}, \qquad
y_a^2(t) = \frac{\tilde{y}_a^2(t)}{t},\qquad
\lambda_m(t) = \frac{\tilde\lambda_m(t)}{t}\, , \eeq
where $t= \ln(\mu^2/\mu_0^2)/{(4\pi)^2}$.
The one-loop RGE for the couplings $x_I$ are
\beq \frac{d \tilde g_i}{d\ln t} = \frac{\tilde g_i}{2} +\beta_{g_i}(\tilde g),\qquad
\frac{d\tilde y_a}{d\ln t} = \frac{\tilde y_a}{2} + \beta_{y_a} (\tilde g, \tilde y),\qquad
 \frac{d\tilde\lambda_m}{d\ln t} = \tilde\lambda_m + \beta_{\lambda_m}  (\tilde g, \tilde y, \tilde\lambda) \, .
\label{1loor}
\eeq
The right-hand sides of the RGE in \eq{1loor} do not depend explicitly on $t$, but only through the functional dependence of the couplings $x_I$. This is true in the one-loop approximation, because $\beta_{g_i}$ is cubic in $g$; $\beta_{y_a}$ is cubic in $g$ and $y$; $\beta_{\lambda_m}$ is quadratic in $g^2$, $y^2$, $\lambda$. As discussed at the beginning of section~\ref{simple}, the one-loop approximation is adequate for our purposes. Thus, the RGE in \eq{1loor} take the form of a vector flow in the space of the rescaled $x$ couplings,
\beq \label{eq:flow}
\frac{d x_I}{d\ln t} = V_I(x)\, , ~~~~~~x_I=\{\tilde{g}_i,\tilde{y}_a,\tilde{\lambda}_m\}.
\eeq

The next step of the procedure is to identify the possible asymptotic behaviours by solving the system of {\em algebraic} equations 
\beq
V_I(x_{\infty})=0 ~~~\Rightarrow ~~~\left\{\begin{array}{l}{\tilde g}_{i\infty} =-2\beta_{g_i}({\tilde g}_\infty)\\
{\tilde y}_{a\infty} =-2\beta_{y_a}({\tilde g}_\infty , {\tilde y}_\infty )\\
{\tilde \lambda}_{m\infty} =-\beta_{\lambda_m}({\tilde g}_\infty , {\tilde y}_\infty ,{\tilde \lambda}_\infty )\end{array} \right.
\label{vzeri}.
\eeq
 The constants $x_\infty=\{{\tilde g}_{i\infty}, {\tilde y}_{a\infty},{\tilde \lambda}_{m\infty}\}$ are fixed points of the RG flow for the rescaled couplings $x$. We will call them {\em fixed-flows}, extending the terminology introduced in section~\ref{simple}, since  they describe special RG trajectories in which individual couplings run, but their ratio is fixed.
The constants $x_{\infty}$ correspond to RG solutions for the running couplings with infinite boundary conditions at the IR scale $t=0$. These solutions are especially useful to track the asymptotic behaviour of the RG running at $t\to \infty$.  When one of  the $x_\infty$ constants vanishes, it can either mean that the corresponding 
 running coupling vanishes (if the fixed-flow is UV-repulsive)
 or that there are running coupling solutions with subleading asymptotic behaviour $t^{-\alpha}$ with $\alpha >1$ (if the fixed-flow is UV-attractive).
 
\medskip 
 
The main qualitative  characterisation of each fixed-flow is its
UV-attractive or repulsive behaviour.
The nature of each fixed-flow can be understood by linearising \eq{eq:flow} in the neighbourhood of $x=x_\infty$, where the vector flow is approximated by 
\beq\label{Mmat}
V_I(x) \simeq \sum_J M_{IJ} (x_{J} - x_{J\infty}),\qquad
\hbox{where}\qquad M_{IJ} = \left.\frac{\partial V_I}{\partial x_{J}}\right|_{x=x_\infty} 
\eeq 
is a numerical matrix.
The RGE for the small deformation around the fixed-flow solution, $\Delta_I \equiv x_{I} - x_{I\infty}$, is 
\beq
\frac{d \Delta_I }{d\ln t} = \sum_J M_{IJ} \Delta_J  \, .
\label{deltarun}
\eeq
From \eq{deltarun} we obtain the following result on the nature of the solutions. {\em A fixed-flow $x_\infty$ is fully UV-attractive (IR-repulsive) if all the eigenvalues of the matrix $M(x_\infty)$ are negative, and is
fully UV-repulsive (IR-attractive) if all eigenvalues are positive}. 

A UV-repulsive fixed-flow corresponds to an isolated asymptotic behaviour: a small deviation from the fixed-flow will bring the solution further away as we move towards the UV, and therefore will lead to a different asymptotic behaviour. This means that the request of sitting on a UV-repulsive fixed-flow implies a precise determination of one combination of couplings in the IR. 

In general, some of the eigenvalues of the matrix $M(x_\infty)$ are positive, and others are negative.
The number of positive eigenvalues is the number of combinations of couplings that are univocally predicted
by demanding that the theory can reach infinite energy at that
fixed-flow.   The number of negative eigenvalues is the residual number of free parameters.\footnote{
Zero eigenvalues are also possible, and correspond to accidental global flavour symmetries of the theory
(an example is discussed in section~\ref{nonab}).
Imaginary eigenvalues are absent in all the examples that we computed, 
which means that the asymptotic RGE flow never performs cycles around a fixed point.} 

The matrix $M(x_\infty)$ has constant entries that depend only on $x_\infty$ and is given by
\beq
M(x_\infty) = \begin{pmatrix} \delta_{ij}(\frac12-\frac32 b_i{\tilde g}_i^2) & 0 & 0 \\ \frac{\partial \beta_{y_a}(x)}{\partial  {\tilde g}_j} & \frac{\delta_{ab}}{2}+\frac{\partial \beta_{y_a}(x)}{\partial  {\tilde y}_b} & 0 \\
\frac{\partial \beta_{\lambda_m}(x)}{\partial  {\tilde g}_j}  & \frac{\partial \beta_{\lambda_m}(x)}{\partial  {\tilde y}_b} &
\delta_{mn}+ \frac{\partial \beta_{\lambda_m}(x)}{\partial  {\tilde \lambda}_n} \end{pmatrix}_{x=x_\infty} \, ,
\label{matt}
\eeq 
where $b_i$ are the gauge $\beta$-function coefficients. The matrix is block triangular. This means that the nature of the gauge fixed-flows are not influenced by Yukawa and quartic couplings, and the nature of the Yukawa fixed-flows are not influenced by quartic couplings. This follows from the consideration that the eigenvalues of any triangular matrix are equal to its diagonal elements (as can be easily proved by induction). The special structure of the matrix $M(x_\infty)$ suggests that we can proceed in steps, solving in succession the cases of gauge, Yukawa, and scalar quartic couplings. In each case, {\em we first solve the system of equations (\ref{vzeri}), $V_I (x_\infty) =0$, to determine the fixed-flows and then, for each solution, we compute the eigenvalues of the matrix $M(x_\infty)$ in \eq{matt} to determine their UV-attractive or repulsive nature}.

\subsection{Gauge couplings}
For each gauge coupling, we find at most two fixed-flows that solve \eq{vzeri}:
  \beq 
  {\rm for}~b_i>0~~~~~ {\tilde g}_{i\infty}^2 = \left\{\begin{array}{ll}
  1/{b_i}  & \hbox{UV-attractive}\\
0 & \hbox{UV-repulsive}
\end{array}\right. 
\label{gaggio}
\eeq
\beq
{\rm for}~b_i<0~~~~~ {\tilde g}_{i \infty}^2 = 0 ~~~~~~~~~~~~\hbox{UV-repulsive}~~\, 
\label{gaggie}
\eeq
 The solution with ${\tilde g}_{i\infty}\neq0$, being  UV-attractive, does not imply any prediction for the value of the gauge coupling in the IR,
 and exists only when the gauge group is asymptotically free, $b_i>0$. 
If this is not the case, a vanishing gauge coupling is the only way to achieve {TAF}.
The solution ${\tilde g}_{i\infty}=0$ is always present and, being UV-repulsive, implies an IR prediction, which is $g_{i0}=0$.
Of course, these results are trivial and agree with what can be easily derived from the explicit solutions of \eq{eqg0multi}.

\subsection{Yukawa couplings}\label{ygen}
Next, let us consider the Yukawa couplings $y_a$, whose one-loop RGE are
\beq\label{RGEyj}
\frac{d }{dt} y_a= \beta_{y_a}(g,y)= \frac 12 \left(
f^y_{abcd}\, y_{b}\, y_{c}\, y_{d} - f^g_{ia}\, g_i^2\, y_a \right) \, ,
\eeq
where $f^y$ and $f^g$ are numerical coefficients. In the general case, \eq{RGEyj} cannot be analytically solved and our method becomes essential to analyse the problem.
For the Yukawa couplings \eq{vzeri}, which determines the fixed-flows, becomes
\beq
 {\tilde y}_{a\infty}= -2\, \beta_{y_a}({b}^{-1/2},{\tilde y}_\infty)=
  -f^y_{abcd}\, {\tilde y}_{b\infty}\, {\tilde y}_{c\infty}\, {\tilde y}_{d\infty} + \frac{f^g_{ia}} {b_i}\, {\tilde y}_{a\infty} \, .
\label{rgegeny}
\eeq
 
 For illustration, we can solve \eq{rgegeny}
 in the case of a single Yukawa and gauge coupling:
  \beq 
  {\rm for}~f_g>b~~~~~ {\tilde y}_\infty^2 = \left\{\begin{array}{ll}
  (f_g-b)/(bf_y)  & \hbox{UV-repulsive}\\
0 & \hbox{UV-attractive}
\end{array}\right. 
\label{gaggiu}
\eeq
\beq
{\rm for}~f_g<b~~~~~ {\tilde y}_\infty^2 = 0 ~~~~~~~~~~~~~~~~\hbox{UV-repulsive}\, .~~~~~~~~~~~~
\eeq
Comparing \eq{gaggiu} with \eq{gaggio} for $f_g>b>0$, we see that the nature of the non-vanishing (${\tilde y}_\infty \ne 0$) and vanishing (${\tilde y}_\infty = 0$) fixed-flows of the Yukawa coupling is reversed with respect to the case of the gauge coupling. So, in the neighbourhood of ${\tilde y}_\infty = 0$, we expect a family of solutions with an asymptotic behaviour subleading with respect to $1/t$. On the other hand, the non-vanishing fixed-flow corresponds to an isolated asymptotic behaviour $y^2={\tilde y}_\infty^2/t$ and leads to a prediction for a combination of couplings in the IR. As derived in section~\ref{simple}, the IR prediction is $y_0^2/g_0^2 = (f_g-b)/f_y$ in the case of a single gauge coupling and $y_0^{-2}=f_yI_\infty$ in the case of multiple couplings, where $I_\infty$ is defined after \eq{deffy}.

\medskip

Going back to the general case, the task is to find the set of real solutions of the system of cubic equations (\ref{rgegeny}). The problem 
is often simplified by the following observation:
$\beta_{y_a}$ vanishes for $y_a=0$ whenever the Lagrangian acquires a chiral symmetry in the limit $y_a=0$.
In such a case,  one of the fixed-flows is simply ${\tilde y}_{a\infty}=0$ (and is UV-attractive for $\sum_i f_{ia}^g/b_i >1$, as can be seen by computing the matrix $M$). The other fixed-flow is found
 by solving a linear equation in ${\tilde y}_{a\infty}^2$ and choosing the positive value of the Yukawa coupling. 
  If a chiral symmetry holds for each one of the $N_y$ Yukawa couplings,
 the full cubic system is reduced to a linear system that
 admits up to $2^{N_y}$ solutions.
This simplification holds in various theories of interest, when the number $N_y$ of Yukawa couplings
 is less than the number of fermionic fields (with associated chiral symmetries).
 In theories with multiple generations one has a continuum of solutions, 
trivially  obtained by acting on one representative solution with the flavour symmetry of the theory.


%

\subsection{Scalar quartic couplings}
The one-loop RGE for the scalar quartic couplings $\lambda_m$ is
\beq
\frac{d }{dt} \lambda_m= \beta_{\lambda_m} (g, y, \lambda )
= s^\lambda_{mnp} \lambda_{n} \lambda_{p}+ \lambda_m\left( s^{\lambda y}_{m ab} 
  y_a y_{b} 
  - s^{\lambda g}_{mi}  g^2_i\right) -s^y_{mabcd} y_a y_{b} y_{c} y_{d} + s^g_{mij}  g^2_i g^2_{j}
 \ .
\label{rgegen}
\eeq
The fixed-flows are obtained by 
solving \eq{vzeri}, 
\beq
{\tilde \lambda}_{m\infty} =-\beta_{\lambda_m}(b^{-1/2} , {\tilde y}_\infty ,{\tilde \lambda}_\infty ) \, ,
\label{pippo}
\eeq 
where ${\tilde y}_\infty$ are the solutions of \eq{rgegeny}. 

For illustration, we can solve \eq{pippo} in the case of a single quartic, Yukawa, and gauge couplings. We find that the fixed-flows are
\beq
{\tilde \lambda}_\infty ={\tilde \lambda}_\pm \equiv \frac{s_{\lambda g} -b-s_{\lambda y}b {\tilde y}_\infty^2 \pm \sqrt{( s_{\lambda g} -b-s_{\lambda y}b {\tilde y}_\infty^2)^2-4s_\lambda(s_g-s_yb^2{\tilde y}_\infty^4)}}{2s_\lambda b} \, ,
\label{paperino}
\eeq
where the two possible values of ${\tilde y}_\infty$ are given in \eq{gaggiu}. The solutions ${\tilde \lambda}_\pm$ exist only under the condition that the term inside the square root in \eq{paperino} is not negative. Inspection of the matrix $M(x_\infty)$ in \eq{matt} shows that ${\tilde \lambda}_+$ is UV-repulsive, while ${\tilde \lambda}_-$ is UV-attractive. Since both fixed-flows are non-vanishing, they correspond to running couplings with the same asymptotic behaviour $\lambda \sim 1/t$. The request of sitting exactly on the UV-repulsive fixed-flow  ${\tilde \lambda}_+$ implies an IR prediction, which is $\lambda_0/g_0^2=b{\tilde \lambda}_+$ in one-loop approximation.  These conclusions are in agreement with the full analytic study of the RGE presented in section~\ref{secqua}.

In general, \eq{pippo} is a system of quadratic equations in ${\tilde \lambda}_{m \infty}$, for any given 
${\tilde y}_{a \infty}$ and ${\tilde g}_{i \infty}$. The {TAF} conditions are obtained by requiring that at least one such system 
admits a solution where all coefficients ${\tilde \lambda}_{m \infty}$ are real. Usually  ${\tilde \lambda}_{m \infty}=0$ is not a solution, 
because of the additive renormalisation of quartic couplings due to gauge and Yukawa couplings.
In some models, a quartic coupling  can break an accidental global symmetry: only in such a case
${\tilde \lambda}_{m \infty}=0$ is a solution, providing an easy way to find the other solution.

\subsection{Basins of attraction}
The study of the fixed-flows determines the asymptotic behaviour of the RG trajectories and the necessary conditions on the field content of the theory to satisfy {TAF}. The next step of our procedure consists in determining the initial conditions of the coupling constants in the IR that insure that all couplings flow towards a vanishing fixed-point in the far UV, without being attracted towards a Landau pole at finite $t$. This is done by studying the basins of attraction of the fixed-flows in the space of the rescaled couplings  $x_I=\{\tilde{g}_i,\tilde{y}_a,\tilde{\lambda}_m\}$, defined
as the parameter range covered by stream lines that flow into such point.
Every positive eigenvalue of $M$ implies a reduced dimensionality of such parameter space,
and thereby one prediction for a combination of couplings.

As an example, we show in
fig.\fig{flows} at page~\pageref{fig:flows}
the flow $V_I(x)$ and the fixed points $x=x_\infty$ for the SM with vanishing hypercharge gauge coupling, a case which will be discussed in detail in section~\ref{SM}.
The basin of attraction of the fully UV-attractive fixed-flow (in blue) is the shaded (two-dimensional) region.
The basins of attraction of the mixed IR/UV fixed-flows (in magenta) are the magenta curves: one parameter is predicted and the region is one-dimensional.
The basin of the fully UV-repulsive fixed-flow (in red) is the point itself: two couplings are predicted and the region is zero-dimensional.

The RG flow of the scalar quartic couplings can cross the boundary that separates a stable from an unstable potential.
When a UV-repulsive fixed-flow corresponds to a stable potential 
and a UV-attractive fixed-flow corresponds to an unstable potential, 
the low-energy vacuum is meta-stable.
An example of such a situation is shown in fig.\fig{flows}b, in which there is a single quartic $\lambda$, and the stability condition is $\lambda>0$.
In the opposite situation (which is possible with multiple quartics)
a stable potential can become unstable at low energy,
signalling that the phenomenon of spontaneous symmetry breaking is taking place,
according to the Coleman-Weinberg mechanism~\cite{Coleman:1973jx}.
 
 \subsection{Mass parameters}\label{HH}
So far we have focused our discussion on the RG running of dimensionless parameters, {\it i.e.} the coupling constants of the theory. Mass parameters (scalar and fermion masses and cubic scalar couplings) become dynamically irrelevant at very high energy and so are not crucial for our considerations. Nevertheless, for completeness, we discuss now their RG evolution.

Let us first consider the simplest case of a theory with a single mass parameter $m$ for a scalar field. The one-loop RGE can be easily solved
\beq
\frac{d m^2}{dt} =\gamma  \,  m^2, \qquad \Rightarrow \qquad
m^2(t)  = m^2(t_0) \exp\int_{t_0}^{t} \gamma(t)~  d t  .
\eeq
Here $\gamma$ is the anomalous dimension which, in the case of the SM Higgs, is given by  
\beq
\gamma = 3 y_t^2+6\lambda-\frac94 \left(g^2_2+\frac {g_1^2}{5}\right) .
\eeq

In a TAF theory, the leading asymptotic behaviour is $\gamma(t)= {\tilde \gamma}/t$, where ${\tilde \gamma}$ is a constant given by the fixed-flows of the coupling constants (for instance, as discussed in sect.~\ref{SM}, in the TAF version of the SM with $g_1=0$, we find ${\tilde \gamma}\approx -0.75$). Therefore, the asymptotic RG behaviour of $m^2$ is  
\beq m^2(t) = m^2(t_0)\times (t/t_0)^{{\tilde \gamma}}\, .
\label{eqruf}
\eeq
For negative ${\tilde \gamma}$, the mass parameter $m^2$ flows to zero in the UV. For positive ${\tilde \gamma}$, there is an infinite 
multiplicative renormalisation as $t\to\infty$ (although $m^2(t)$ remains always negligible with respect to the renormalisation scale $\mu^2$ in the UV).

In ref.~\cite{Skiba}, it is claimed that this infinite renormalisation introduces a hierarchy problem, even for asymptotically-free theories.
We disagree with this conclusion. In an asymptotically-free theory with a single scalar mass, there are no physical mass scales larger than $m$. Therefore, no hierarchies can arise and the theory is natural.

The issue of naturalness comes up in theories with multiple mass scales. 
Let us consider a TAF model with two very different scalar mass parameters $m_1^2$ and $m_2^2$. The RG trajectory of the smallest of the two masses will start in the IR with a logarithmic running, until it meets the second mass scale, where it receives a threshold correction proportional to the large mass. At that scale, the RG trajectory has a sudden jump, and then follows a logarithmic running proportional to the heavy mass. This special RG trajectory is very sensitive to initial conditions and a large separation between $m_1^2$ and $m_2^2$ requires a careful tuning of parameters. The instability of the RG trajectory for very different $m_1^2$ and $m_2^2$ is a reincarnation of the naturalness problem. On the other hand, the RG trajectory in \eq{eqruf} does not exhibit any special sensitivity on initial conditions, confirming our conclusion that there is no naturalness problem in a theory with a single mass scale.

It is interesting to consider the RG flow towards the UV of the TAF theory with two mass scales. Given the RG evolution of the mass parameters $d m_i^2/dt = \gamma_{ij} m_j^2$, the ratio $r = m_1^2/m_2^2$ obeys the equation
\beq \frac{dr}{dt} =\gamma_{12} +( \gamma_{11}-\gamma_{22}) r  - \gamma_{21} r^2\, .  \label{RGEr}\eeq
In the asymptotic region, we can write $\gamma_{ij} (t)={\tilde \gamma}_{ij}/t$ with constant ${\tilde \gamma}_{ij}$ determined by the fixed-flows of the coupling constants.
In the special case  $\gamma_{12}=\gamma_{21}=0$, the two scalar particles belong to two different sectors with no common interactions and $r=r_0\, (t/t_0)^{{\tilde \gamma}_{11}-{\tilde \gamma}_{22}}$. Each mass parameter evolves independently and, for $t\to \infty$, $r$ flows to zero (when ${\tilde \gamma}_{22}>{\tilde \gamma}_{11}$) or to infinity (when ${\tilde \gamma}_{11}>{\tilde \gamma}_{22}$). 

On the other hand, when there are common interactions between the two scalars ($\gamma_{12},\gamma_{21}\ne 0$), the ratio $r$ can have a more complicated evolution and, in particular, cross $r=0$ or $1/r=0$ at finite $t$, leading to a dynamical generation of spontaneous symmetry breaking. When $\gamma_{12}$ and $\gamma_{21}$ are positive (which is always true for scalars with mutual quartic and trilinear interactions, but does not hold when there are other sources in the RGE from large masses of vector-like fermions), the asymptotic solution of \eq{RGEr} is\footnote{Note that \eq{RGEr} is formally identical to the RGE for the ratio between gauge and quartic couplings, see \eq{eqy1uffa}. An important difference is that, while \eq{eqy1uffa} is valid only for perturbative values of the coupling constants, \eq{RGEr} is valid for any value of $r$.}
\beq
r(t) = \frac{r_+(r_0-r_-)(t/t_0)^\Delta -r_-(r_0-r_+)}{(r_0-r_-)(t/t_0)^\Delta -(r_0-r_+)} \, ,
\label{soso}
\eeq
where
\beq
r_{\pm} \equiv \frac{{\tilde \gamma}_{11}-{\tilde \gamma}_{22}\pm \Delta}{2{\tilde \gamma}_{21}}\, ,\quad
\Delta\equiv \sqrt{({\tilde \gamma}_{11}-{\tilde \gamma}_{22})^2+4{\tilde \gamma}_{12}{\tilde \gamma}_{21}} \, .
\label{defdelta}
\eeq
A special behaviour of \eq{soso} is the UV-isolated constant solution $r(t) = r_-$, which acts as an IR-attractor. More generally,  solutions are attracted in the UV towards the point $r(t)=r_+$. In any case, for $t\to \infty$, the ratio $r$ is always finite and thus all masses become typically comparable, despite experiencing a common overall infinite rescaling. Mass hierarchies do not arise dynamically in this context.

As an additional remark, note the behaviour of \eq{soso}. For $r_0>0$, the ratio $r$ is always non-vanishing, finite, and positive at all scales: there is no dynamical mechanism of spontaneous symmetry breaking. For $r_-<r_0<0$ (recall that $r_-$ is always negative and $r_+$ positive), the solution crosses $r=0$, where $m_1^2$ changes sign, and asymptotically flows towards $r_+$. For $r_0<r_-$, the solution crosses $1/r=0$, where $m_2^2$ changes sign, before flowing to $r_+$.

Let us consider now the separate running of the mass parameters $m_{1,2}^2$ according to their RG evolution $d m_i^2/dt = \gamma_{ij} m_j^2$. In the asymptotic region where $\gamma_{ij} (t)={\tilde \gamma}_{ij}/t$ holds, we can express the low-energy value of $m_1^2$ as
\beq
m_1^2 (t_0) = \frac{m_1^2(t)}{a_+(t/t_0)^{c_+}+a_-(t/t_0)^{c_-}}+\frac{b\left[ (t/t_0)^{\Delta}-1\right]}{
a_++a_-(t/t_0)^{\Delta}}~m_2^2 (t_0)
\label{superpin}
\eeq
\beq
a_\pm \equiv \frac 12 \pm \frac{{\tilde \gamma}_{11}-{\tilde \gamma}_{22}}{2\Delta} \, ,~~~~~
b\equiv \frac{{\tilde \gamma}_{12}}{\Delta} \, ,~~~~~
c_\pm \equiv \frac{{\tilde \gamma}_{11}+{\tilde \gamma}_{22}\pm \Delta}{2} \, , 
\eeq
where $c_\pm$ are the two eigenvalues of the matrix ${\tilde \gamma}$ 
and $\Delta = c_+ -c_-$ is given in \eq{defdelta}.

The first term in the right-hand side of \eq{superpin} represents the boundary condition for $m_1^2$ in the far UV, once we take the limit $t\to \infty$. The interesting term is the second one, which measures how $m_2^2(t_0)$ affects $m_1^2(t_0)$ at the quantum level. In the case of a finite logarithmic running from the scale $\mu$ to some cut-off scale $\Lambda$, expanding \eq{superpin} we find
\beq
\delta m_1^2(\mu) \approx m_2^2(\mu)~ \frac{\tilde\gamma_{12}}{(4\pi)^2}~\ln \frac{\Lambda^2}{\mu^2} \, .
\eeq
This corresponds to the familiar result that the hierarchy between two scalar masses $m_1^2\ll m_2^2$ is destabilised by a one-loop correction $\delta m_1^2$ proportional to $m_2^2$ and to a logarithm, which becomes large whenever the theory can be extrapolated up to scales $\Lambda^2 \gg m_2^2$. For instance, this is the source of the naturalness problem in supersymmetry, where the stop mass feeds into the Higgs mass at one-loop with a logarithm of the ratio between the weak and the GUT scale.

Naively, one could expect that, since in a TAF theory we effectively take $\Lambda \to \infty$, the coefficient measuring the contribution of $m_2^2$ to $\delta m_1^2$ must blow up to infinity, leading to a situation in which mass hierarchies are completely out of control. Equation~(\ref{superpin}) shows that this is not the case. As we send $t\to \infty$, the coefficient in front of $m_2^2(t_0)$ (which measures $\delta m_1^2/m_2^2$ at $t_0$) remains finite, no matter what the sign of $\Delta$ is. Moreover, note that this coefficient is given by a ratio between $b$ and $a_\pm$: it is of order unity and it is no longer suppressed by the loop factor $(4\pi)^2$, independently of the size of the coupling constants. The quantum correction $\delta m_1^2$ is parametrically equal to $m_2^2$, with no coupling constant or loop suppression, no matter how small is the coupling involved.

In conclusion, we have found that the infinite renormalisation of the mass parameters always leads to  a finite result for the coefficient measuring how one scalar mass feeds into a smaller scalar mass. The infinite resummation eliminates the loop factor, insuring that all scalar masses at low energy be equal, within factors of order one. This has important consequences for the implementation of the naturalness criterion: in a TAF theory without special protection mechanisms
(such as dynamical generation of masses at low scale~\cite{follie}), any significant separation between mass scales entails a  naturalness problem.


\mysection{Asymptotic behaviour of the Standard Model}\label{SM} 
As an illustrative example, we apply the method described in section~\ref{multi} to the Standard Model, ignoring gravity.
The coupling constants are the gauge couplings $g_1,g_2,g_3$, the Yukawa couplings $y_t,y_b,y_\tau,y_\nu$ (for simplicity we set to zero the Yukawa couplings of the first two generations), and the Higgs quartic coupling $\lambda$. Since we want to use the SM to elucidate our method, we will follow the analysis of section~\ref{multi} step by step.

The approximation of ignoring interactions from the gravitational sector, or any other super-weak interactions from additional sectors, is justified by the assumptions of softened gravity. The feebleness of these interactions makes sure that they will never be able to cure the Landau poles of SM couplings. Of course, gravitational or super-weak interactions could affect the RG trajectories of SM couplings in the far UV, where the SM couplings asymptotically vanish, but cannot turn a non-TAF into a TAF theory. So our results, which ignore the effect of possible super-weak interactions, can be conservatively viewed as describing only necessary conditions for TAF.

\subsection{SM gauge couplings}
As is well known, the one-loop RGE for the SM gauge couplings are
\beq\frac{dg_1^2}{dt} = \frac{41}{10} g_1^4, \qquad
\frac{dg_2^2}{dt} = -\frac{19}{6} g_2^4,\qquad
\frac{dg_3^2}{dt}=-7 g_3^4.\eeq
According to eqs.~(\ref{gaggio})--(\ref{gaggie}), we have 4 possible fixed-flows
\beq
\begin{array}{c|cccc}
\rowcolor[cmyk]{0,0,0,0.05}
 & {\tilde g}_{1\infty}^2 & {\tilde g}_{2\infty}^2 & {\tilde g}_{3\infty}^2  &  M\hbox{-eigenvalues} \\ 
  \hline
 \hbox{Solution 1} & 0 & 6/19 & 1/7 &{+}{-}{-}\cr
\hbox{Solution 2} & 0 & 6/19 & 0 &{+}{-}{+}  \cr
\hbox{Solution 3} & 0 & 0 & 1/7 &{+}{+}{-} \cr
\hbox{Solution 4} & 0 & 0 & 0 &{+}{+}{+} \cr
\end{array}
\label{tableg}
\eeq
The last column in \eq{tableg} shows the signs of the respective eigenvalues of the matrix $M(x_\infty)$ defined in \eq{matt}. We recall that a negative eigenvalue corresponds to a UV-attractor (IR-repulsor), while a positive eigenvalue corresponds to a UV-repulsor (IR-attractor) and thus to an IR prediction.

Of course hypercharge is not asymptotically free, 
so all solutions are unphysical, since they require $g_1=0$.
Still, it is interesting to pursue the study of the SM because it presents a non-trivial structure of possible Landau poles for the Yukawas and the Higgs quartic, providing a good illustration of our method,  and
also because $g_1\approx 0$ might be viewed as a rough approximation for the SM at low energy or for extensions of the SM where hypercharge is embedded in a non-abelian gauge group.

We proceed by focusing on the phenomenologically most relevant case of the fixed-flow corresponding to solution 1 in in \eq{tableg}, with ${\tilde g}_{1\infty}=0$, which is IR-attractive (UV-repulsive) giving one IR prediction, and ${\tilde g}_{1\infty},{\tilde g}_{2\infty}\ne0$, which are UV-attractive giving no extra predictions.


\begin{figure}[t]
\begin{center}
$$\includegraphics[width=0.95\textwidth]{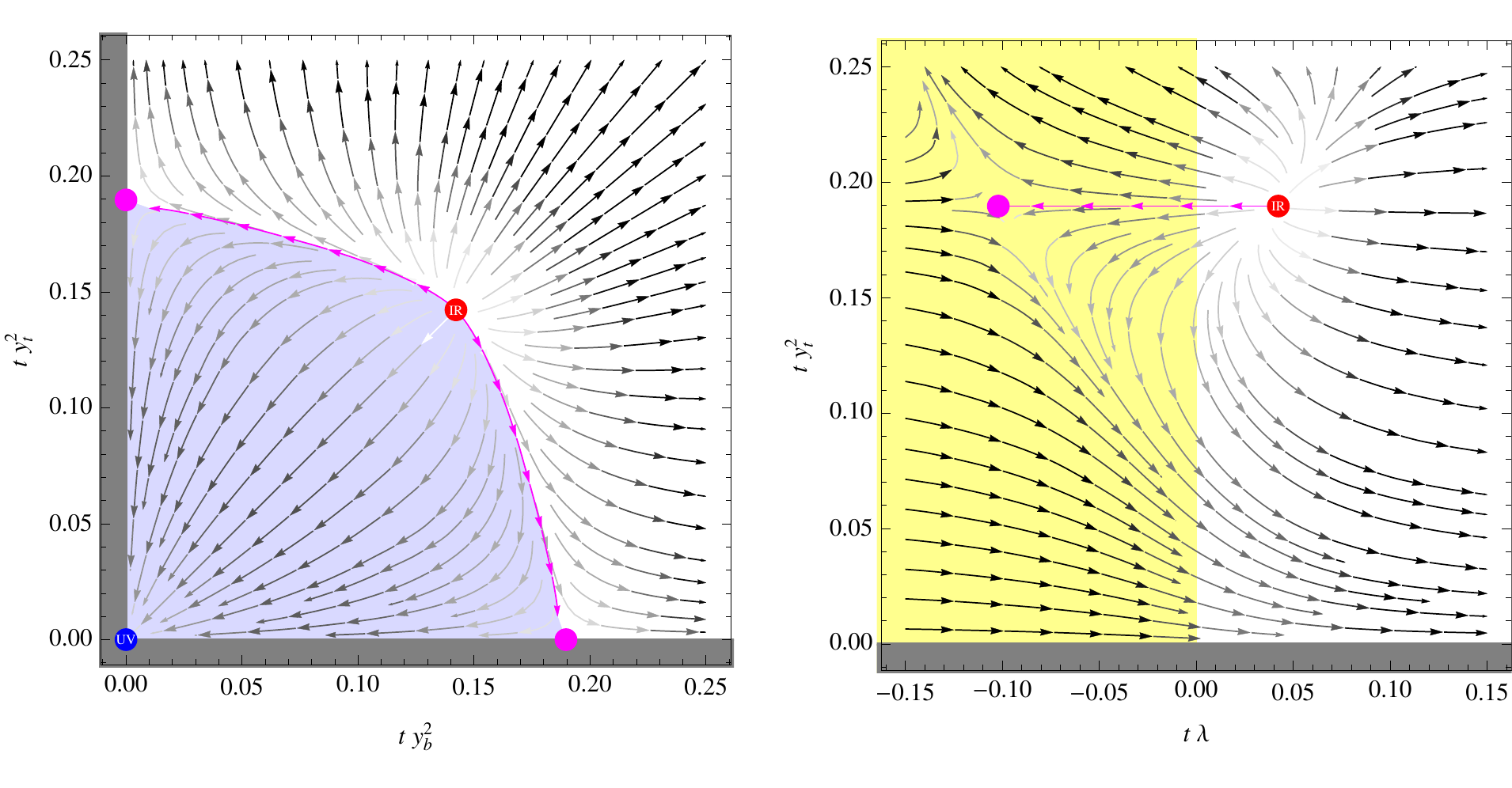}$$
\caption{\em RG flows for the SM with $g_1=0$ in the plane of rescaled couplings $\tilde{y}^2=ty^2$ and $\tilde{\lambda}=t\lambda$.  
The dots are the fixed-flows of the rescaled couplings,
with 
fully IR-attractive fixed points coloured red (2 combinations of couplings predicted),
fully UV-attractors  in blue,
and hybrid in magenta (1 combination of couplings predicted).
In the left plot we consider the top and bottom Yukawa couplings, setting to zero all other Yukawas. 
The basin of attraction  of the UV fixed-flow is the shaded region;
the basins of the hybrid fixed points are the magenta curves.
In the right plot we consider the top Yukawa coupling and the Higgs quartic.
No fully UV-attractive fixed-flow is present. The basic of attraction of the hybrid fixed-flow
 is the magenta curve.
\label{fig:flows}}
\label{default3}
\end{center}
\end{figure}

\subsection{SM Yukawa couplings}
The one-loop RGE for the Yukawa couplings of the top quark
($y_t$), the bottom ($y_b$), the tau lepton ($y_\tau$) and, if an interaction with a right-handed neutrino is present, of the neutrino ($y_\nu$) are
\begin{eqnsystem}{sys:RGESMy}
\frac{dy^2_t}{dt} &=& y_t^2 \left(-\frac{17 }{20}g_1^2-\frac{9 }{4}g_2^2-8 g_3^2+ \frac92 y_t^2 +\frac32 y_b^2 +y_{\tau }^2+y_\nu^2\right), 
\label{rrr1}\\
\frac{dy^2_b}{dt} &=& y_b^2 \left(-\frac{ 1}{4}g_1^2-\frac{9 }{4}g_2^2-8 g_3^2+ \frac32 y_t^2 +\frac92 y_b^2 +y_{\tau }^2+y_\nu^2\right), 
\label{rrr2}\\
\frac{dy^2_\tau}{dt} &=& y_\tau^2 \left(-\frac{9 }{4}g_1^2-\frac{9 }{4}g_2^2+3 y_t^2 + 3 y_b^2+ \frac52 y_{\tau }^2-\frac12 y_\nu^2\right), \\
\frac{dy^2_\nu}{dt} &=& y_\nu^2 \left(-\frac{9 }{20}g_1^2-\frac{9 }{4}g_2^2+3 y_t^2 + 3 y_b^2-\frac12 y_\tau^2+ \frac52 y_{\nu }^2\right) .
\end{eqnsystem}
For our choice of ${\tilde g}_\infty$, the system of equations that determines the Yukawa fixed-flows, given by eq~(\ref{rgegeny}), has 4 possible solutions
\beq
\begin{array}{c|ccccc}
\rowcolor[cmyk]{0,0,0,0.05}
 & {\tilde y}_{t\infty}^2 & {\tilde y}_{b\infty}^2 & {\tilde y}_{\tau\infty}^2 & {\tilde y}_{\nu\infty}^2 &  M\hbox{-eigenvalues} \\  \hline
\hbox{Solution 1} & 227/1197 & 0 & 0 & 0 &{+}{-}{+}{+}  \cr
\hbox{Solution 2} & 0 & 227/1197 & 0 & 0&{-}{+}{+}{+} \cr
\hbox{Solution 3} & 227/1596 & 227/1596 & 0 & 0 &{+}{+}{+}{+}\cr
\hbox{Solution 4} & 0 & 0 & 0 & 0&{-}{-}{+}{+} \cr
\end{array}
\label{tabley}
\eeq

The four fixed-flows and the flow restricted to the $({\tilde y}_t , {\tilde y}_b)$ plane are plotted in fig.\fig{flows}a, where solutions 1 and 2 are plotted in magenta,
point 3 (fully IR-attractive) in red, point 4 (UV-attractive) in blue.
Solution 3 makes four predictions in the IR: $y_t (M_t)=y_b (M_t)=0.879$, $y_\tau (M_t)=y_\nu (M_t)=0$. This corresponds to the pole masses $M_t = M_b\approx 163$~GeV and $M_\tau =M_\nu =0$. Solution 2 makes three predictions in the IR, giving $M_b\simeq 186$~GeV and $M_\tau =M_\nu =0$. Neither case gives a reasonable result for the bottom-quark mass.

Solution 4 corresponds to vanishing $\tilde y_\infty$ for all Yukawa couplings. However, there is an important difference between quark and lepton Yukawa couplings. Since ${\tilde y}_{\tau , \nu\infty} =0$ are UV-repulsive,  Landau poles can be avoided only if the lepton Yukawa couplings are exactly zero in the IR. On the other hand, since ${\tilde y}_{t , b_\infty} =0$ are UV-attractive, the quark Yukawa couplings satisfy {TAF} for a range of IR boundary conditions. Indeed, near the fixed-flow of solution 4 and for small Yukawas, eqs.~(\ref{rrr1})--(\ref{rrr2}) become
\beq 
\frac{dy_{t,b}^2}{dt} 
\simeq -c \frac{y_{t,b}^2}{t},\qquad c = \frac{493}{266}. \eeq
The solutions are $y_{t,b}^2 \propto t^{-c}$, which  in the asymptotic region $t\to\infty$ become negligible with respect to $g^2_{2,3}\sim 1/t$, since $c>1$.  Although solution 4 seems phenomenologically plausible as far as Yukawas are concerned, it is not compatible with {TAF} for the Higgs quartic, as we will show in the next section. For this reason, in the following we will focus on solution 1.  

Solution 1 gives three predictions: zero $\tau$ and neutrino masses, and top mass at its IR fixed-flow. The bottom Yukawa is near a UV-attractive fixed-flow, hence the bottom-quark mass is not determined. This can be seen explicitly by writing \eq{rrr2} near the fixed-flow of solution 1, for small $y_{b,\tau,\nu} \ll 1$
\beq 
\frac{dy_b^2}{dt} 
\simeq -c \frac{y_b^2}{t},\qquad c = \frac{626}{399}. 
\eeq
The solution, $y_{b}^2 \propto t^{-c}$ goes to zero faster than the gauge couplings for $t\to\infty$, since $c>1$.

\subsection{SM Higgs quartic coupling}
The one-loop RGE for the Higgs quartic coupling
that parameterises the potential  $\lambda|H|^4$ is
\begin{eqnarray}
 \frac{d\lambda}{dt} &=&
12 \lambda^2  +
\lambda\left(6 y_t^2+6 y_b^2+2y_{\tau }^2+2y_{\nu }^2-\frac92 g_2^2-\frac{9}{10} g_1^2\right) \br
-3 y_t^4-3 y_b^4-y_{\tau }^4 - y_\nu^4
+\frac{9}{16} g_2^4+\frac{27}{400} g_1^4+\frac{9}{40} g_2^2 g_1^2 .
\end{eqnarray}
The fixed-flows are obtained by solving \eq{pippo}. As anticipated, we find no solutions for ${\tilde \lambda}_\infty$ when all Yukawa couplings are on their vanishing fixed-flow ${\tilde y}=0$, {\it i.e.} solution 4 in \eq{tabley}. On the other hand, we find two solutions for ${\tilde \lambda}_\infty$ in each of the other three cases, {\it i.e.} solutions 1--3 in \eq{tabley}. This is an example of how Yukawa couplings sitting on some non-vanishing fixed-flow can save the running of a scalar quartic coupling, otherwise doomed to suffer from Landau poles, and produce asymptotically-free solutions. Among the three possible cases, we concentrate on solution 1 in  \eq{tabley} for the Yukawas, which is the most propitious from a phenomenological point of view. Then, \eq{pippo} admits the following two solutions for ${\tilde \lambda}_\infty$:
\beq
\begin{array}{c|ccc}
\rowcolor[cmyk]{0,0,0,0.05}
 & {\tilde \lambda}_{\infty} &  \hbox{$M$-eigenvalue} & \hbox{potential} \\   \hline
\hbox{Solution 1} &\displaystyle  \frac{-143 + \sqrt{119402}}{4788}    \approx +0.0423 &{+}  & \hbox{stable} \cr
\hbox{Solution 2} &\displaystyle \frac{-143 - \sqrt{119402}}{4788}  \approx -0.1020&{-}  & \hbox{unstable} \cr
\end{array}
\label{tablel}
\eeq

Figure\fig{flows}b shows the RG flow in the plane ${\tilde \lambda}$--${\tilde y}_{t}$. We find three possible behaviours. {\it (i)} A generic point in the plane flows towards a Landau pole of either $\lambda$ or $y_t$. {\it (ii)} If we select ${\tilde y}_{t}={\tilde y}_{t\infty}$, then ${\tilde \lambda}$ flows towards the UV-attractive fixed-flow, where $\lambda$ is negative, making the EW vacuum potentially unstable. In this case, one parameter is predicted in the IR (the top-quark mass) and the dimensionality of the basin of attraction is reduced by one. {\it (iii)} The solution sits on the IR-attractor, for which the basin of attraction is reduced by two and both the top-quark and the Higgs masses are predicted. The prediction of the IR-attractive solution 1 in \eq{tablel} corresponds to $\lambda (M_t)= 0.217$, {\it i.e.} to a pole Higgs mass $M_h = 163\GeV$.

\medskip

The negative value of $\lambda$ at the UV-attractive solution 2 in \eq{tablel} means that the EW vacuum is unstable. 
However, this situation is not necessarily ruled out if the tunnelling rate is slower than the age of the universe.
Considering a field direction along which the quantum-corrected potential is $V \approx \frac14 \lambda(\mu\approx h) h^4$,
the EW vacuum is sufficiently long-lived  provided that $\lambda$ does not become too much negative at large energy.
The Fubini bounce solution to the classical field equation, $h(r) = \sqrt{-2/\lambda}\times 2R/(r^2+R^2)$,
has tree-level action $S = 8\pi^2/3|\lambda|$ where  $R$ is a free parameter~\cite{SMvacdecay}.
Thereby, imposing a negligible probability for the vacuum to have decayed during its past history 
\beq e^{-S} \ll (R H_0)^4 \, ,\eeq
where $H_0$ is the present Hubble constant,
we obtain
\beq \lambda(\mu \sim\frac{1}{R}) > \frac{2\pi^2}{3\ln H_0 R} \stackrel{t\to\infty} \simeq  -\frac{1}{12 t}.
\label{metas}
\eeq
The fixed-flow in solution 2 of \eq{tablel} corresponds to an asymptotic behaviour $\lambda \approx -1/(9.8~t)$, which slightly violates the metastability constraint in \eq{metas}.

In conclusion, {TAF} imposes strong constraints on the Higgs quartic coupling: either $M_h = 163\GeV$, or $M_h < 163\GeV$ and the lifetime of the EW vacuum is shorter than the age of the universe. Neither possibility is realistic.


\begin{figure}[t]
\begin{center}
$$\includegraphics[width=0.65\textwidth]{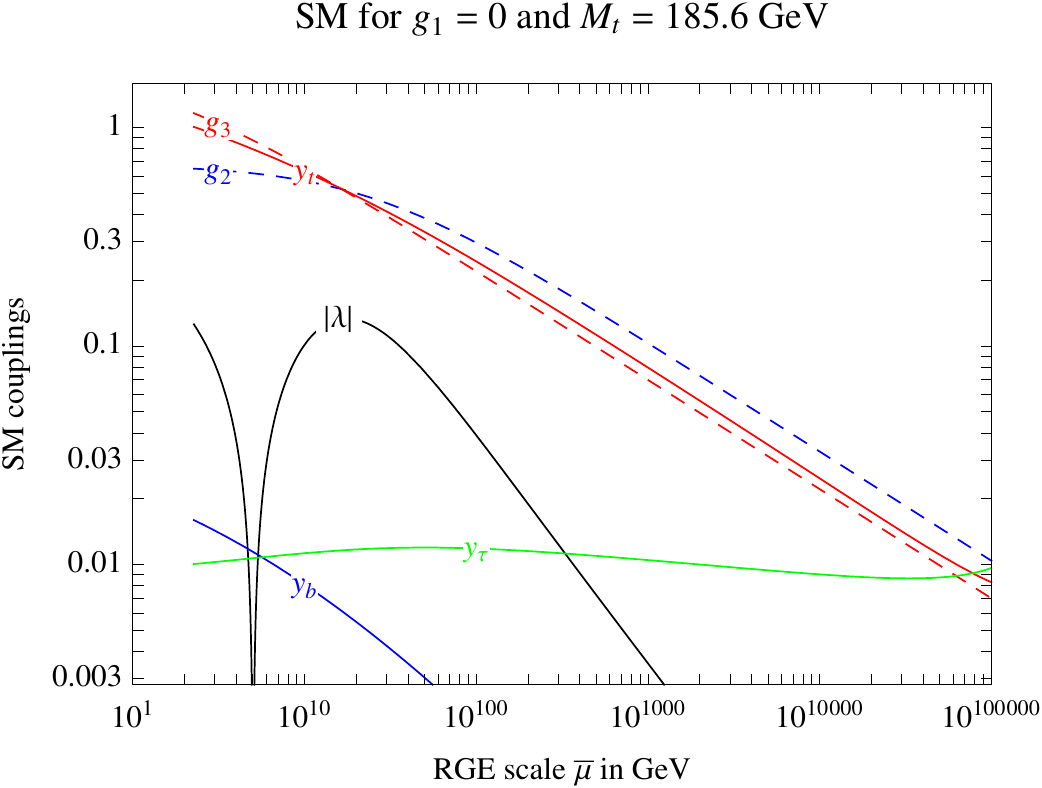}$$
\caption{\em RG running towards infinite energy (note the double-log scale for the $\overline{\rm MS}$ renormalisation scale $\bar{\mu}$)
in the SM for the measured values of $M_h,M_b,M_\tau,g_2,g_3$ and
for the values of $g_1$ and $M_t$ needed to achieve TAF:  $g_1=0$ and $M_t = 185.6\GeV$.
\label{fig:runSM}}
\label{default4}
\end{center}
\end{figure}

\subsection{RG flow of the SM couplings}
For physical values of its coupling constants, the SM is not an asymptotically free theory. We have found that the closest approximation to physical reality for the SM to be a {TAF} theory is that its coupling constants lie in the basin of attraction of the following fixed-flow 
\beq
\begin{array}{cccccccc}
\rowcolor[cmyk]{0,0,0,0.05}
 {\tilde g}_{1\infty}^2 & {\tilde g}_{2\infty}^2 & {\tilde g}_{3\infty}^2  & {\tilde y}_{t\infty}^2 & {\tilde y}_{b\infty}^2 & {\tilde y}_{\tau\infty}^2 & {\tilde y}_{\nu\infty}^2 & {\tilde \lambda}_{\infty}  \\   \hline
 0 & \frac{6}{19} & \frac17 & \frac{227}{1197} & 0 & 0 & 0 &\frac{-143 + \sqrt{119402}}{4788}  \\
 {+}&{-}&{-}&{+}&{-}&{+}&{+}&{+}  \cr
\end{array}
\label{tabletot}
\eeq
The matrix $M(x_\infty)$ has 5 positive eigenvalues; hence, within the 8-dimensional space of couplings $x=\{{\tilde g}_{1,2,3}, {\tilde y}_{t,b,\tau,\nu},{\tilde \lambda}\}$, the basin of attraction of the fixed-flow in \eq{tabletot} has dimensionality equal to $8-5=3$. As a result, 
{TAF} makes the following 5 predictions on physical parameters. The hypercharge gauge coupling must be zero; the tau lepton and neutrino must be massless; the top quark mass must be  $M_t=186$~GeV, which is $7\%$ higher than the observed value; the Higgs mass must be $M_h =163$~GeV, which is $30\%$ higher than the observed value (or $M_h <163$~GeV, but the EW vacuum is unstable). None of these predictions is correct, but they are not bad approximations of reality. It is conceivable that these wrong predictions can be cured in extensions of the SM where hypercharge is embedded in a non-abelian group and where we expect corrections at least of order $g_1^2/g_2^2$.

In fig.\fig{runSM} we show the RG flow of the SM coupling constants, taking $g_1$ and $y_t$ at their fixed-flows, while all other parameters are equal to their physical values in the IR. The figure is obtained by solving the SM RGE in the 3-loop approximation. For the physical value of the $\tau$ mass, we observe that $y_\tau$ starts small in the IR, but becomes the
largest coupling at $10^{10^5}\GeV$, where soon reaches a Landau pole.
For the physical value of the Higgs mass, the coupling $\lambda$ becomes negative at an intermediate scale, before flowing to zero in the deep UV, barring the effect of the Landau pole in the $\tau$ Yukawa. The coupling $\lambda$ can be asymptotically free and always positive only for the special IR condition $M_h =163$~GeV (with  $M_t=186$~GeV). This condition on $M_h$ and $M_t$ corresponds to the Pendleton-Ross point~\cite{Pendleton:1980as} or
to the tip of the SM phase diagram (shown in~\cite{SMphase}), although the numerical values of  $M_h$ and $M_t$ quoted here are somehow different, since in our calculation we set $g_1= 0$. 

\mysection{TAF extensions of the SM} \label{nonab}
\subsection{Grand unification}
We start our exploration of extensions of the SM by constructing asymptotically-free versions of grand-unified theories (GUT). These examples bear no relevance on the issue of naturalness in theories with softened gravity because they introduce in the observable sector a mass scale $M_{\rm GUT}$ much larger the weak scale, which feeds unnaturally large corrections to the Higgs mass $\delta M_h^2 \sim g_{\rm GUT}^2M_{\rm GUT}^2/(4\pi)^2$. Nevertheless, we find the study instructive for its own sake and we present it here. 

The simplest GUT is based on the gauge group SU(5), with 3 generations of chiral fermions in the 
${\bar 5}\oplus {10}$ representation of SU(5), and scalars $H$ and $\Sigma$ in the $5\oplus 24$ representation. This theory does not satisfy the TAF requirements. However, independently of TAF considerations, the theory cannot be considered realistic, since it gives some wrong predictions for quark and lepton masses. 

We then consider an extension of the minimal SU(5) setup that simultaneously satisfies the conditions of TAF and of an acceptable fermion mass spectrum. 
We add 3 generations of vector-like fermions 
$\psi_{5}$, $\psi_{\bar 5}$ and $\psi_{24}$ in the $5\oplus \bar 5 \oplus 24$ representation of SU(5) with mass terms $M_5  \psi_{5}\psi_{\bar 5} + M_{24} \psi_{24}^2/2$.
The Yukawa couplings, written for simplicity only for the third generation, are
\begin{eqnarray}
\Lag_Y &=&  - \frac{y_t}{8} 10~10~H + (y_b \bar 5 + y'_b \,\psi_{\bar 5})~10~H^* + (y_\nu \bar 5 + y'_\nu \,\psi_{\bar 5})\psi_{24} H \nonumber \\
&&+
y''_\nu \,\psi_5 \psi_{24} H^*+
(y_m \,\bar 5 +y_m' \,\psi_{\bar 5}   )  \Sigma \psi_5 +  y_\lambda \Tr (\Sigma\psi_{24}^2) \, .
 \eea
The coupling $y_t$ gives the top-quark Yukawa and $y_b$ gives the bottom-quark and $\tau$-lepton Yukawas.
The coupling $y_m$ induces an SU(5)-breaking mixing between $\bar 5$ and $\psi_{\bar 5}$, which modifies the phenomenologically wrong equality between
down-quark and the charged-lepton Yukawa couplings predicted by minimal SU(5).
The split multiplets contribute to threshold corrections to the SU(5) prediction for gauge coupling unification, which could make the result compatible with low-energy measurements.
The coupling $y_\nu$ generates a neutrino Yukawa, in which the right-handed neutrino is identified as the SM singlet in $\psi_{24}$. The simpler alternative of generating the neutrino Yukawa coupling with a right-handed neutrino as a singlet of SU(5) does not satisfy the TAF condition of \eq{TAFyuk}, unless the
$\beta$-function of $g_5$ is reduced by adding extra matter.
A two-generation SU(5) {TAF} model that ignores this issue was presented in~\cite{K5}.

\medskip

The most general quartic potential is
\beq V_4 = \lambda_H |H|^4 + \lambda_\Sigma \Tr(\Sigma^4)+\lambda'_\Sigma \Tr(\Sigma^2)^2  + \lambda_{H\Sigma}
H^\dagger \Sigma^2 H + \lambda'_{H\Sigma} |H|^2 \Tr(\Sigma^2).
 \eeq
The  purely quartic potential, restricted to $\Sigma$, is positive definite when\footnote{We used the identity
  $7/30 \le \Tr(\Sigma^4)/\Tr^2(\Sigma^2)\le 13/20$.  The stability conditions for the full potential $V_4$ are much more complicated~\cite{V>0}.}
\beq 
\lambda'_{\Sigma}>- \frac{7}{30} \lambda_\Sigma ~~~{\rm for}~\lambda_\Sigma>0
\qquad\hbox{and}\qquad
\lambda'_{\Sigma}>- \frac{13}{20} \lambda_\Sigma~~~{\rm for}~\lambda_\Sigma<0
\, .\eeq
The violation of the first condition leads to the symmetry breaking pattern $\SU(5)\to G_{\rm SM}$ \`a la Coleman-Weinberg;
the  violation of the second condition leads to $\SU(5)\to \SU(4)\otimes {\rm U}(1)$.

\medskip

The $\beta$-function coefficient for the unified gauge coupling is $b_5 = 4/3$. For the Yukawa couplings, we find various fixed-flow solutions and the phenomenologically most interesting case is 
\beq
\begin{array}{c|cccccccccc}
\rowcolor[cmyk]{0,0,0,0.05}
 & {\tilde g}_{5\infty}^2 & {\tilde y}_{t\infty}^2 & {\tilde y}_{b\infty}^2& {\tilde y}_{b\infty}^{\prime 2}  & {\tilde y}_{\nu\infty}^2 
 & {\tilde y}_{\nu\infty}^{\prime2} & {\tilde y}_{\nu\infty}^{\prime\prime2} & {\tilde y}_{m\infty}^2 & {\tilde y}_{m\infty}^{\prime 2}  & {\tilde y}_{\lambda\infty}^{ 2} \\
  \hline
  \hbox{Fixed-flow} &  3/4 & 38/15 & 0 &0& 0& 0 & 0 & 49/16 & 0 & 0\\
  \hbox{$M$-Eigenvalues} & - & + & -&- & - & - & - & + & 0 & -\\  
\end{array}
\label{SU5y}
\eeq
The lower row in \eq{SU5y} shows the signs of the eigenvalues of the $M$ matrix:
only two combinations of Yukawa couplings are predicted. 
The vanishing eigenvalue arises because the couplings are accidentally invariant
under global $\SU(2)$ rotations acting on $(\bar 5 ,\psi_{\bar 5})$.
All other Yukawa couplings with ${\tilde y}^2_\infty =0$ have negative eigenvalues and therefore their IR values are non-vanishing, but cannot be predicted by the TAF requirement.

One predicted combination of Yukawa couplings involves  the top coupling $y_t$.
Indeed its RGE is
\beq \frac{dy_t^2}{dt} = y_t^2 (-f_g g_5^2 + f_y y_t^2 - 2 y_b^2 + \frac{12}{5} y_\nu^2)\qquad\hbox{with}\qquad
f_g=\frac{108}{5},\qquad f_y=6.\eeq
Thereby, for $y_b,y_\nu\ll y_t$ one has the prediction $y_t^2 /g_5^2 \simeq (f_g - b_5)/f_y$.
However, there is no unique way of relating this prediction to the physical value of the top mass, because other Yukawa couplings can affect the IR value of $y_t$.

The solution for the Yukawa couplings in \eq{SU5y} allows for 4 fixed-flows for the quartics
\beq
\begin{array}{c|ccccccccc}
\rowcolor[cmyk]{0,0,0,0.05}
& \tilde\lambda_{H\infty}  & \tilde\lambda_{\Sigma\infty} & \tilde\lambda'_{\Sigma'\infty} & \tilde\lambda_{H\Sigma\infty} & \tilde\lambda'_{H\Sigma\infty}& \hbox{$M$-Eigenvalues} & \hbox{Potential} \\  \hline
\hbox{Solution 1} &-1.16 & -0.326 & 0.185 & 0.610 & -0.003 & ----- & \hbox{unstable}\\
\hbox{Solution 2} &-1.15 & -0.422 & 0.541 & 0.725 &  \phantom{-}0.116 & ----+& \hbox{unstable}  \\ 
\hbox{Solution 3} &0.831 & -0.315 & 0.215 & 0.989 & -0.562 & ---++ & \hbox{unstable} \\
\hbox{Solution 4} &0.821 & -0.334 & 0.500 & 1.617 & -0.597  & --+++ & \hbox{stable}
\end{array}
\label{SU5lambda}
\eeq
The potential is stable only in the case of solution 4, for which
  three combinations of quartic couplings are  predicted in the IR
by demanding that {TAF} is achieved.

In conclusion, it is not difficult to construct realistic TAF GUT models. The TAF conditions significantly constrain the field content of the theory, although any phenomenological prediction of low-energy parameters is highly model-dependent.


%

\subsection{TAF extensions of the SM at the weak scale}
\label{sect:Nonabel} 
%

As discussed in section~\ref{SM},
the SM does not satisfy TAF.  For theories with softened gravity to respect naturalness, the  SM must be modified at the weak scale and made compatible with TAF.  So we address now the question of how to construct such modifications.
The first problem is that hypercharge is not asymptotically free. This implies that, in any TAF extension of the SM, one must embed hypercharge in a non-abelian gauge group.
Such extensions have the additional advantage of explaining the observed quantisation of electric charge.

The hypercharges $Y$ of SM fermions satisfy the relation $Y=T_{3R} + (B-L)/2$, where
$T_{3R}$ is the third component of the right-handed isospin $\SU(2)_R$.
Thus, the most straightforward possibility is to promote $T_{3R}$ to the full non-abelian $\SU(2)_R$ gauge group, with SM field assignment as in table~\ref{eq:SMfields}.
Similarly, 
U(1)$_{B-L}$ is not asymptotically free, so one needs to embed it into a non-abelian group.
Given the known values of the $B-L$ charges, we find only two possibilities that do not lead to proton decay:
\begin{itemize}
\item Merging $B-L$ with $\SU(3)_c$ into the Pati-Salam $\SU(4)_{\rm PS}$, such that the full gauge group is
\beq G_{224}=\SU(2)_L\otimes\SU(2)_R\otimes\SU(4)_{\rm PS} .\eeq
 $B-L$ arises as the  diagonal $\diag(1,1,1,-3)/\sqrt{24}$ generator of $\SU(4)_{\rm PS}$.

\item Merging $B-L$ with $\SU(2)_L\otimes\SU(2)_R$ into $\SU(3)_L\otimes\SU(3)_R$, such that the full gauge group is
\beq G_{333}=\SU(3)_L\otimes\SU(3)_R\otimes\SU(3)_c  .\eeq
 $B-L$ arises as the combination of the diagonal $\diag(1,1,-2)/\sqrt{12}$ generators of $\SU(3)_L$ and of $\SU(3)_R$.
\end{itemize}
Before presenting specific models, 
in the rest of this section we assess some generic common features.
First, we discuss  in section~\ref{sect:HFCNC} the flavour-violations coming from the two-Higgs doublet structure implied by the left-right symmetry.
Then, in section~\ref{sect:WR} we study the phenomenological constraints on the existence
of the charged and neutral gauge bosons of $\SU(2)_R$.

\subsection{New heavy Higgs and flavour processes}\label{sect:HFCNC}
Both options we are considering for embedding the SM into a non-abelian group (Pati-Salam and trinification)  include $\SU(2)_R$, and thus the SM Higgs doublet must be extended at least into the structure
\beq \phi = \begin{pmatrix}H_U^0 & H_D^+ \\ H_U^- & H_D^0 \end{pmatrix}= 
\begin{pmatrix}H_U & H_D \end{pmatrix}
 \, ,
\eeq
which transforms as $(2_L,\bar2_R)$ under $\SU(2)_L\times \SU(2)_R$. The field $\phi$
contains two Higgs doublets, transforming under the SM $\SU(2)_L\times \U(1)_Y$ as $H_U\sim (2,-1/2)$ and $H_D\sim (2,1/2)$. The two-Higgs structure is often troublesome because it generates scalar-mediated flavour-changing neutral-current (FCNC) interactions.
We show here how the problem can be avoided with an appropriate  flavour structure of quark Yukawa couplings.

\smallskip

Denoting as $q_L = (u_L, d_L)$ and $q_R=(u_R,d_R)$ the SM quarks which transform as $(2_L,1_R)$ and $(1_L,2_R)$, the $\SU(2)_L\times \SU(2)_R$ invariant quark-Yukawa interactions are
\beq
-\Lag_{Y}^q= 
{\bar q}_L \left( Y\phi +Y_c \phi^c \right) q_R + {\rm h.c.} =
{\bar q}_L \left( Y H_U +Y_c H_D^c \right) u_R +
{\bar q}_L \left( Y H_D -Y_c H_U^c \right) d_R + {\rm h.c.} \,  \label{lagpu}
\eeq
where
\beq
\phi^c \equiv  \epsilon^T \phi^* \epsilon=\begin{pmatrix}H_D^{0*} & -H_U^+ \\ -H_D^- & H_U^{0*} \end{pmatrix} =\begin{pmatrix} H_D^c & -H_U^c \end{pmatrix} \, ,
\eeq
%
%
and $H_{U,D}^c =\epsilon H^*_{U,D}$, $\epsilon =i\sigma_2$, and $Y$ and $Y_c$ are two different Yukawa matrices in flavour space.
%
%
It is convenient to rewrite the two doublets $H_U$ and $H_D$ in terms of a SM-like doublet $h$ and of a heavy doublet $H$, 
defined such that $\langle H \rangle =(0,0)$ and
$\langle h \rangle =(0,v)$ with $v = (v_d^2+v_u^2)^{1/2}$:
\be
\left(\ba{c} h \\ H \ea \right) = \left( \ba{cc} \cos\beta & \sin\beta \\ -\sin\beta & \cos\beta \ea \right) \left(\ba{c} H_D \\ -H_U^c \ea \right),
\qquad  \sin\beta = \frac{v_u}{v}~, 
\qquad  \cos\beta = \frac{v_d}{v}~.
\ee
Defining the SM Yukawa couplings $Y_D$ and $Y_U$ as
\be
\left(\ba{c} Y_D\\ Y_U \ea \right) = \left( \ba{cc} \cos\beta & \sin\beta \\ \sin\beta & \cos\beta \ea \right) \left(\ba{c} Y \\  Y_c \ea \right) \, ,
\ee
the Lagrangian in \eq{lagpu} becomes
\bea
-\Lag_{Y}^q&=&
{\bar q}_L Y_D d_R\, h + {\bar q}_L 
\left( \frac{Y_U}{\cos 2\beta}-\tan 2\beta \, Y_D\right) d_R \, H+
\nonumber \\
&+&  {\bar q}_L Y_U u_R\, h^c +
{\bar q}_L 
\left( \frac{Y_D}{\cos 2\beta}-\tan 2\beta \, Y_U\right) u_R \, H^c \, . 
\label{lagpu2}
\eea
%
We see that down-type $H$-mediated FCNCs
are controlled by the off-diagonal entries of $Y_U$, in the basis where $Y_D$ is diagonal, and viceversa.
In the $\SU(2)_L\times \SU(2)_R$ invariant basis where $Y_{D}$ is diagonal we can write
\be
Y_U^{\rm d-base} = V^\dagger \lambda_u V_R~,  \qquad Y^{\rm d-base}_D= \lambda_d~,    
\label{eq:VRdef}
\ee
where $\lambda_{u,d}$ are diagonal matrices of the quark masses, $V$ is the usual CKM matrix, and $V_R$ is a new unitary matrix which controls all new flavour violations.

\medskip

The effects of $V_R$ are best understood by considering the accidental global flavour symmetry of the model: $\U(3)_L \times \U(3)_R$ explicitly
broken by $Y_{U}$ and $Y_{D}$ which both transform as $(3,\bar3)$.
From the SM point of view,
such flavour structure  is equivalent to 
 $\U(3)_{q_L} \times \U(3)_{u_R} \times \U(3)_{d_R}$,  with {\em three}
independent  spurions: $Y_U \sim (3,\bar 3, 1)$  and $Y_D \sim (3,1,\bar 3)$, as in Minimal Flavour Violation (MFV)~\cite{MFV}, plus $V_R \sim (1,  3, \bar 3)$~\cite{Buras:2010pz}.

In the quark mass eigenbasis, the interactions of the neutral Higgs $h^0$ and $H^0$, derived from \eq{lagpu2} are
\bea
-\Lag_{Y}^q&=&
{\bar d}_L \lambda_d d_R\, h^0 + {\bar d}_L 
\left( \frac{V^\dagger \lambda_u V_R}{\cos 2\beta}-\tan 2\beta \, \lambda_d\right) d_R \, H^0 
\nonumber \\
&+& {\bar u}_L \lambda_u u_R\, h^{0*} + 
{\bar u}_L 
\left( \frac{V\lambda_dV_R^\dagger}{\cos 2\beta}-\tan 2\beta \, \lambda_u\right) u_R \, H^{0*} \, . 
\label{lagpu3}
\eea
Again, we observe that the flavour violations mediated at tree level by the neutral scalar $H^0$ are proportional to $\lambda_u$ in the down sector and to $\lambda_d$ in the up sector, and are expressed in terms of the CKM matrix $V$ and the new rotation matrix $V_R$.

\subsubsection*{Right-handed flavour mixing can be naturally small}
On general grounds, $V_R$ can be parameterised as follows
\be
V_R = P_u \Vt_R P_d^\dagger~, \qquad P_q = {\rm diag}(e^{\phi^q_1}, e^{\phi^q_2}, e^{\phi^q_3})~,
\ee
where $ \Vt_R$ is a CKM-like matrix, with 3 rotational angles and one phase, and $P_u$ and $P_d$ are diagonal phase matrices.
One overall phase in $P_{u,d}$ is unphysical, while, in general, the other 5 are physical.\footnote{These phases could be moved into 
the eigenvalues of $\lambda_{u,d}$. 
However, to avoid confusion, we work in the usual basis where such eigenvalues are real and positive
and leave explicitly the phases in $P_{u,d}$.}

\medskip

Assuming a simple and natural form of Yukawa couplings ---
quasi-diagonal Yukawa matrices with small or negligible off-diagonal entries and no relations between them ---
the presence of left-handed CKM mixing implies a minimal amount of right-handed mixing given by
\beq\label{VRmin}\left\{\begin{array}{lll}
|(V_R)_{us}| \!\!\!& \approx   |V_{us}|  m_d/m_s \!\!\!\!\!&\approx 10^{-2},\\
|(V_R)_{cb}| \!\!\!&\approx  |V_{cb}| m_s/m_b \!\!\!\!\!& \approx 10^{-3},\\
|(V_R)_{ub}|\!\!\!&\approx  |V_{ub}| m_d/m_b \!\!\!!\!\!&\approx 10^{-5}.\end{array}
\right.\eeq
Furthermore, this small amount of right-handed mixing is radiatively stable and is generated by RG corrections
(more generally $\Vt_R \approx$ permutation matrix is radiatively stable too).
%
This can be seen by observing that $\Vt_R=\identity$ implies $[ Y_U^\dagger Y_U , Y_D^\dagger  Y_D]=0$.
As a result, entires proportional to $Y_U^\dagger Y_U$ and $Y_D^\dagger Y_D$ in the RG evolution 
of both $Y_U^\dagger Y_U$ and $Y_D^\dagger Y_D$ do not generate off-diagonal entries,
if starting from an initial condition with $\Vt_R=\identity$. Off-diagonal terms in the right-handed sector 
are generated only from the contribution of mixed terms, $Y_D^\dagger Y_U$ and $Y_U^\dagger Y_D$,
in the RG evolution of $Y_U^\dagger Y_U$ and $Y_D^\dagger Y_D$.
The latter give rises to small deviations from  $\Vt_R=\identity$, again suppressed by {\em both} 
off-diagonal CKM entires and small quark mass ratios:  $|(\Vt_R)_{i > j}| \lsim  (m_j/m_i)  |V_{i j}|  $. 
A similar argument holds for the CP-violating phases.

\medskip

\subsubsection*{Flavour bounds on heavy Higgs bosons}
In order to evaluate the strength of FCNCs, we assume that $\Vt_R$ is about equal to the identity matrix, 
\be
\Vt_R = 
\left(\begin{array}{ccc}
1 &s_{12} &s_{13} e^{-i\phi} \\ -s_{12} & 1& s_{23} \\ 
- s_{13} e^{i\phi} & -s_{23} & 1 
\end{array}\right)~+ ~{\cal O}( s_{ij}s_{kl} )~.
\label{eq:VRans}
\ee
In analogy with the CKM matrix, and motivated by the need to satisfy the strong constraints on FCNC (see below), we assume $|s_{13}| < |s_{23}| < |s_{12}| \ll 1$.

Integrating out the heavy Higgs $H^0$ (assuming a negligible $h$--$H$ mixing), the interactions in \eq{lagpu3} give the following $\Delta F=2$ dimension-six effective operators
\be
{\cal L}_{\Delta F=2}=\frac{X^{(d)}_{ij}}{M_H^2}    ~(\bar d_L^i  d^j_R) (\bar d_R^i  d^j_L),~~~~
X_{ij}^{(d)}= \frac{1}{\cos^2 2\beta } \Big( \sum_k \lambda_{u_k} V_{ki}^* V_{R_{kj}}\Big)  \Big(  \sum_\ell  \lambda_{u_\ell}  V_{\ell j} V^*_{R_{\ell i}}\Big) ,
\label{eq:Xdd}
\ee
\be
{\cal L}_{\Delta F=2}=\frac{X^{(u)}_{ij} }{M_H^2}   ~(\bar u_L^i  u^j_R) (\bar u_R^i  u^j_L),~~~~
X_{ij}^{(u)}= \frac{1}{\cos^2 2\beta }\Big( \sum_k   \lambda_{d_k}   V_{ik} V^*_{R_{jk}} \Big) \Big( \sum_\ell   \lambda_{d_\ell} V^*_{j\ell} V_{R_{i\ell}} \Big)  .
\label{eq:Xuu}
\ee
The $\tan\beta$ dependence gives an enhancement of the coefficients $X^{(u,d)}$ for $\tan\beta\approx 1$, but rapidly saturates for large values of $\tan\beta$. We will focus on the most conservative case $\tan\beta\gg 1$, but our results can be simply scaled by replacing $M_H$ with $M_H|\cos 2\beta |$.

%
\smallskip

\begin{table}[t]
\begin{center}
\begin{tabular}{c|c|l}
\rowcolor[cmyk]{0,0,0,0.05}
system  & effective operator &  bound from $\Delta M_{\rm meson}$   \\ \hline
$B_d^0$--$\bar B_d^0$ &   $ (\bar b_L d_R)(\bar b_R d_L)$  &   $|3.6 \times 10^{-8} + s_{13}| < 1.7\times 10^{-4}   ~(M_H/3~{\rm TeV})^2  $  \\
$B_s^0$--$\bar B_s^0$ &   $ (\bar b_L s_R)(\bar b_R s_L) $  &  $|1.5 \times 10^{-4}  - s_{23}  | <  1.0\times 10^{-3} ~(M_H/3~{\rm TeV})^2  $  \\
$K^0$--$\bar K^0$  &   $ (\bar s_L d_R)(\bar s_R d_L) $   &   $|6.2\times 10^{-4} -s_{12} + 11~s_{13} | <  1.0 \times 10^{-2} ~(M_H/3~{\rm TeV})^2 $   \\
$D^0$--$\bar D^0$  &   $ (\bar c_L u_R)(\bar c_R u_L) $   &   $|  1.1 \times 10^{-2} - s_{12} - 2.0~s_{13} | < 9.5  ~(M_H/3~{\rm TeV})^2  $  \\
\end{tabular}
\caption{\label{tab:MHbounds} \em
Constraints on right-handed mixing angles $(s_{12},s_{13},s_{23})$ as functions of the heavy Higgs mass ($M_H$) from $\Delta F=2$ processes, assuming $\tan\beta \gg 1$.  For generic values of $\tan\beta$, the constraints are obtained with the replacement
$M_H \to M_H|\cos 2\beta | $. }
\end{center}
\end{table}

\medskip

We first consider the bounds from the meson-antimeson mass differences, which depend only on 
the absolute value of the $X^{(u,d)}$ coefficients.
The results are summarised in table~\ref{tab:MHbounds}.\footnote{We 
use the updated list of bounds on the coefficients of $\Delta F=2$ operators from~\cite{Isidori:2013ez}.}
The entries in the table can be read in a twofold manner. On the one hand, we can derive absolute lower bounds on $M_H$
in the limit $s_{ij} \to 0$  ($V_R \to \identity$). In such a limit, the $X^{(u,d)}$ coefficients are necessarily suppressed by at
least one light Yukawa eigenvalue. As a result of the smallness of the Yukawa couplings of light quarks, the resulting constraints are not very stringent.  
The tightest bound is the one following from $\Delta M_K$, which implies
$\label{MHbound}
M_H > 0.75~{\rm TeV}$.

\smallskip

On the other hand, comparing the numerical values (independent of $s_{ij}$) with the terms linear in $s_{ij}$ in the bounds of table~\ref{tab:MHbounds},  we deduce 
the size of the right-handed mixing angles for which the bound on $M_H$ becomes more stringent with respect to the one derived in the $V_R = \identity$ limit. 
In the kaon system this happens for $|s_{12}| \gsim 10^{-3}$, while in the $B_d$ system the right-handed mixing gives the leading effect
already for $|s_{13}| \gsim 10^{-7}$.

For right-handed mixing angles equal to the corresponding left-handed CKM mixings --- $|s_{12}| =  |V_{us}|$,   $|s_{23}| =  |V_{cb}|$, and $|s_{13}| =  |V_{ub}|$ --- we find comparable and quite stringent bounds on $M_H$ from $\Delta M_K$ ($M_H> 14$~TeV),  $\Delta M_{B_s}$ ($M_H> 19$~TeV),  and  $\Delta M_{B_d}$ ($M_H>  14$~TeV). 

However,
as previously discussed, small values of the right-handed mixing angles are almost radiatively stable, and
for the smallest natural mixing angles of $V_R$ estimated in \eq{VRmin} the bound becomes 
\beq M_H\circa{>}3\TeV.\eeq 

Considering now the CP-violating effects,  if the $|s_{ij}|$ are close to the values that
saturate the $\Delta M$  bounds, the flavour-dependent phases in $P_{u,d}$ and the phase $\phi$ in (\ref{eq:VRans}) 
are significantly constrained.  These additional constraints are not particularly tight in the $B_{d,s}$ systems, where the 
bounds on CP-conserving and CP-violating $\Delta F=2$ amplitudes are comparable in size, but are quite relevant in the 
kaon system. The experimental constraint on $\epsilon_K$ implies $|\phi^q_{1,2}|  < 4\times 10^{-3}$, 
if $|s_{12}|$ is close to saturate the $\Delta M_K$ bound (barring cancellations among different contributions). 
 
 We finally mention that constraints from $\Delta F=1$ processes of the type $q_i \to  q_j \ell^+\ell^-$ or $q_i \to  q_j \gamma$
 are not very stringent: in the first case ($q_i \to  q_j \ell^+\ell^-$)  the corresponding scalar operators are suppressed by light 
 lepton masses, while in the second case ($q_i \to  q_j \gamma$) the transition is generated only beyond the tree level.

\begin{figure}
\begin{center}
\includegraphics[width=0.5\textwidth]{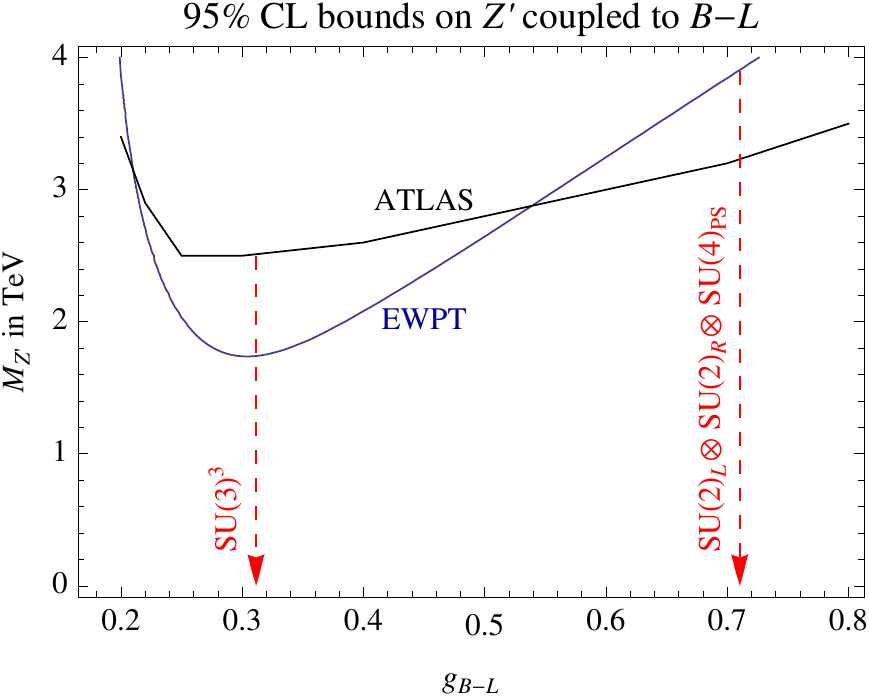}
\caption{\label{fig:Zprime}\em 
95\% CL bounds on the $Z'$ mass
from electroweak precision tests (EWPT, blue curve) and from LHC-Run1 data (ATLAS, black curve)
in extensions of the SM where ${\rm U}(1)_{B-L}\otimes \SU(2)_R\to {\rm U}(1)_Y$.
The red dashed vertical arrows indicate the predictions for $g_{B-L}$ coming from the non-abelian Pati-Salam and trinification models.
}
\label{default0}
\end{center}
\end{figure}

\subsection{New heavy SU(2)$_R$ vectors}\label{sect:WR}
In both $\SU(2)_L\otimes\SU(2)_R\otimes\SU(4)_{\rm PS}$ (Pati-Salam) and $\SU(3)_L\otimes\SU(3)_R\otimes\SU(3)_c$ (trinification)
the electroweak gauge group comes from  $\SU(2)_L\otimes \SU(2)_R\otimes \U(1)_{B-L}$,
with gauge couplings $g_L$, $g_R$ and $g_{B-L}$, respectively.
The SM electroweak gauge couplings are denotes as $g_2 = g_L$ and $g_Y = \sqrt{3/5} \, g_1$.
Both models predicts extra heavy $W_R^\pm$ and $Z'$ vectors.
In both non-abelian models, hypercharge is obtained as a combination of the $B-L$ and of the $T_{3R}$ vectors,
and the hypercharge gauge coupling is reproduced as
$1/g_Y^2 = 1/g_{R}^2 + 1/4g_{B-L}^2$, {\it i.e.}
\beq g_Y =  2g_{B-L}  \cos\theta_{B-L} = g_{R} \sin\theta_{B-L} \eeq
Each model implies a specific value of its gauge couplings and in particular of $g_{B-L}$:
\beq  \label{eq:gPS}
g_{B-L} = \sqrt{\frac38}g_3, \qquad g_R=\frac{g_Y}{\sqrt{1-2g_Y^2/3g_{3}^2}}  \approx 1.03~ g_Y\qquad \hbox{(Pati-Salam)}\eeq
and
\beq  \label{eq:gTri}
g_{B-L} = \frac{g_Y g_2 }{\sqrt{g_Y^2 + g_2^2}},\qquad g_R =\frac{2 g_Y g_2}{\sqrt{3g_2^2- g_Y^2}}\approx 1.22~ g_Y
\qquad \hbox{(Trinification)}.
\eeq

\subsubsection*{Bounds on heavy $Z'$ vectors}
The angle $\tan\theta_{B-L} = 2 g_{B-L}/g_R$ describes the mixing between the $B-L$ and $T_{3R}$ vectors,
in analogy to the weak mixing angle.
The heavy $Z'$  couples to SM fermions as
\beq  \frac{  g_Y[ (B-L) - 2 \cos^2\theta_{B-L} Y]  }{2\sin\theta_{B-L}\cos\theta_{B-L}} .  \eeq
LHC experiments directly searched for  $Z'$ vectors as di-lepton and di-jet resonances~\cite{Z'LHC}.
ATLAS reported the experimental bounds in terms of the relevant category of `minimal' $Z'$ vectors, 
which are combination of $B-L$ and $Y$~\cite{Zwirner},
such that we can extract from data the 95\% C.L.\ bounds: $M_{Z'}>3.2\TeV$ for the Pati-Salam $Z'$
($\tilde{g}_Y \approx -0.124$ and $\tilde{g}_{B-L}\approx 0.99$ in the notations of~\cite{Zwirner})
and $M_{Z'}>2.6\TeV$ for trinification
($\tilde{g}_Y \approx -0.33$ and $\tilde{g}_{B-L}\approx 0.51$).

\medskip

Furthermore, $Z'$ vectors are significantly constrained from precision electroweak data.
Using the results of~\cite{Caccia}, we performed a global fit of precision data 
finding the bound on the $Z'$ mass as function of $g_{B-L}$ plotted in fig.\fig{Zprime}.
The figure also shows the special values of $g_{B-L}$ predicted by the two non-abelian models.
The 95\% C.L.\ bound is $M_{Z'}\circa{>}4\TeV$ in Pati-Salam models and $M_{Z'}\circa{>}1.8\TeV$ in trinification models.

\subsubsection*{Bounds on heavy $W_R$ vectors}
The interactions of the $W^\pm_R$ bosons are described by 
\be
\Lag_{W_R} = \frac{g_R}{\sqrt{2}}  \left[ (V_R)_{ij} \bar u^i_R   \gamma_\mu d_R^j  +   \bar \nu^i_R   \gamma_\mu \ell^i_R
\right ]  W^\mu_R  ~+ ~{\rm h.c.}
\ee
where $V_R$ is the unitary matrix  introduced in section~\ref{sect:HFCNC}. 

LHC experiments searched for heavy $W_R^\pm$ gauge bosons.
For the predicted values of $g_R$, see
eqs.~(\ref{eq:gPS}) and (\ref{eq:gTri}),   CMS data~\cite{CMSWR} imply the bound\footnote{A
$3\sigma$ excess is present in the electron channel at the value of the mass corresponding to the bound.}
\be
M_{W_R}\circa{>}2.2\TeV.\label{WRbound}
\ee

Furthermore, we can identify two sets of indirect bounds on $M_{W_R}$: those from tree-level charged-current interactions, and those from one-loop FCNC processes. 
The latter are largely dominant if $V_R$ has a generic flavour structure, but evaporate when $V_R$ is sufficiently close to a permutation matrix. 
If we assume that $V_R$ is such that the bounds in Table~\ref{tab:MHbounds} are satisfied for $M_H \leq 3$~TeV, then the FCNC bounds on 
$M_{W_R}$ are automatically satisfied for $M_{W_R} \gsim 300$~GeV. This happens because $W_R$-mediated FCNCs appear
only at the one-loop level, and therefore the induced operators have an effective suppression scale $M_{\rm eff} \sim 4\pi  \times M_{W_R}$.

\smallskip

More specifically,
integrating out ${W_R}$ at tree level leads to the following charged-current  effective Lagrangian,
\be
\Lag^{\rm CC}_{\rm eff} = 
\frac{g^2_R}{ 2 M_{W_R}^2} \left[ ( \bar \nu_R \gamma_\mu \mu_R)(  \bar e_R \gamma^\mu \nu_R )+
(V_R)_{ud} (\bar u_R   \gamma_\mu d_R )( \bar e_R \gamma^\mu \nu_R) +\cdots \right]   + {\rm h.c.}
\ee
where we  wrote explicitly  the two most dangerous operators. The first term affects the determination of $G_{\rm F}$ from 
$\mu$ decays, while the second operator affects nuclear $\beta$ decays and LHC physics (see~\cite{Cirigliano:2012ab}).
The right-handed nature of these effective operators implies that they do not interfere with the left-handed SM contributions. As a result, the correction 
to both $G^{(\mu)}_{\rm F}$ and nuclear $\beta$ decays is  suppressed by $1/M^4_{W_R}$,
\be
G^{(\mu)}_{\rm F} \to G^{(\mu)}_{\rm F} \left[ 1 +  \frac{g_R^4 M_W^4}{ g_L^4 M_{W_R}^4  } \right]^{1/2} \approx G^{(\mu)}_{\rm F} \left[ 1 + 1.9\times 10^{-6} \left( \frac{1~{\rm TeV}}{M_{W_R}}
\right)^4 \right]~,
\ee
and does not leads to stringent bounds on $M_{W_R}$. Moreover, if $(V_R)_{ud}=1$ the correction is almost  universal in 
 $G^{(\mu)}_{\rm F}$ and nuclear $\beta$ decays, thereby not affecting the most stringent low-energy test of charged-current weak interactions, 
 namely the comparison between these two effective couplings.

%


\begin{table}
$$ \begin{array}{cccccc}
\hbox{Fields} & \hbox{spin}&{\rm U}(1)_{B-L}&\SU(2)_L&\SU(2)_R & \SU(3)_{\rm c}\cr \hline
( \nu_L, e_L) &1/2& -1& \bar 2 &1&1\cr
( \nu_R, e_R) &1/2&+1 & 1 &2&1  \cr
(u_L, d_L) &1/2&+ {1\over 3}  & \bar 2 & 1 & 3\cr 
( u_R,d_R ) &1/2& -{1 \over 3} & 1 &2&   \bar{3} \cr
\phi & 0 & 0 & 2 & \bar 2 &1
\end{array}
$$
\caption{\em Field content of extensions of the SM where ${\rm U}(1)_Y$ is embedded in $\SU(2)_R\times{\rm U}(1)_{B-L}$.
\label{eq:SMfields}}
\end{table}

\mysection{Towards a realistic weak-scale TAF theory}
\label{towards}

In this section we will attempt to construct realistic models of a TAF theory at the weak scale, based on Pati-Salam or trinification gauge groups. We describe all fermions as 2-component Weyl spinors and we use a short-hand notation in which the left-handed quark is denoted by $q_L$ and the conjugate right-handed quark by $q_R$ (with the symbol $c$ omitted). Moreover, for future convenience, we use a non-conventional assignment in which left-handed quarks transforms as anti-doublets of $\SU(2)_L$. A summary of the quantum number assignments of SM particles is given in table~\ref{eq:SMfields}.

\subsection{Pati-Salam SU(2)$_L\otimes\,$SU(2)$_R\otimes\,$SU(4)$_{\rm PS}$}\label{224}
\subsubsection*{Skeleton model}
The fermion and scalar field content of a `skeleton' Pati-Salam model, summarised in
the upper box  of table~\ref{tab:224}, is described by 
\begin{itemize}
\item The SM fermions are contained in the Weyl fermions $\psi_L\oplus \psi_R$.

\item The scalar $\phi_R$, in the same representation as $\psi_R$, can get a vev $v_R$ 
in its canonically-normalised  ${\rm Re}\, \tilde\nu_R$ component (we denote the entries as $\phi_R$
as those of $\psi_R$, adding a tilde symbol, like in supersymmetric models) breaking $G_{224} \to  G_{\rm SM}$.
Then, the scalars $\tilde e_R$, ${\rm Im}\,\tilde\nu_R$ and $\tilde{u}_R$ are respectively eaten by the vectors $W_{R\mu}^\pm$, $Z'_\mu$ and by the 6 leptoquarks $W'_\mu$
(coming from $\SU(4)_{\rm PS}/\SU(3)_c$), which acquire mass
\beq M_{W_R}^2 = \frac{g_R^2}{2} v_R^2,\qquad
M_{Z'}^2 = \frac{2 g_R^2 + 3 g_4^2}{4} v_R^2,\qquad
M_{W'}^2  = \frac{g_4^2}{2} v_R^2.
\eeq

\item
The SM Higgs is contained in the scalar $\phi$, in the  ($2_L,\bar 2_R$)  representation, which is real
given that $2_L$ and $\bar 2_R$ are pseudo-real, 
with  $ \phi^c \equiv \epsilon^T \phi^* \epsilon$ also transforming as ($2_L,\bar 2_R$).
One could impose the reality condition $\phi = \pm \phi^c$, and the representation $\phi$
would contain only one Higgs doublet with a single Higgs mass term. However, the left-right symmetry would constrain the Yukawa interactions to give identical masses for up and down quarks.
It is phenomenologically more interesting to treat $\phi$ as
a complex scalar field $\phi$, containing two Higgs doublets. In this case, there are two Higgs mass terms,
$\frac{m_1^2}{2} \Tr (\phi^\dagger \phi)+ \Re\frac{m_2^2}{4} \Tr(\phi^\dagger \phi^c)$ with mass eigenvalues $m_1^2 \pm m_2^2$.

\end{itemize}

\begin{table}
$$\begin{array}{|c|ccc|ccc|}\hline
 \rowcolor[cmyk]{0,0,0,0.05}
& {\hbox{Fields}}&\hbox{spin} &\hbox{generations}&
 \SU(2)_L&\SU(2)_R & \SU(4)_{\rm PS} \cr \hline
\parbox[t]{3mm}{\multirow{3}{*}{\rotatebox[origin=c]{90}{skeleton model  }}}
&{\psi_L} =\begin{pmatrix}\nu_L& e_L\\ u_L&d_L \end{pmatrix}& 1/2 &3& \bar 2 &1 &  4 \cr
&{\psi_R}=\begin{pmatrix}\nu_R& u_R\\   e_R& d_R \end{pmatrix} &1/2 &3& 1 &2  & \bar 4\cr 
& \phi_R & 0 &1& 1 & 2 & \bar 4\\
&  \phi = \begin{pmatrix}H_U^0 & H_D^+ \\ H_U^- & H_D^0 \end{pmatrix}& 0 &1& 2& \bar 2&1\\  
\hline  
\parbox[t]{3mm}{\multirow{3}{*}{\rotatebox[origin=c]{90}{extra fields~}}}
&  \psi & 1/2 &N_\psi  \le 3 &2 & \bar 2 & 1   \\
&{Q_{L}} & 1/2 & 2 & 1 & 1 & 10\\ 
&{ Q_R} & 1/2 & 2 & 1 & 1 & \overline{10}\\ 
& \Sigma & 0 &1& 1 &1 &15 \\ 
\hline
\end{array}$$
\caption{\em\label{tab:224} Field content of the skeleton Pati-Salam model (upper box) and the
 extra fields needed for a possible realistic TAF model (lower box).
}
\end{table}

After symmetry breaking, the Pati-Salam gauge couplings $g_L$, $g_R$, $g_4$ are related to the SM gauge couplings $g_3$, $g_2$, and $g_Y = \sqrt{3/5}\, g_1$ by
\beq g_L=g_2,\qquad g_R=\frac{g_Y}{\sqrt{1-2g_Y^2/3g_{3}^2}},\qquad g_4=g_3 .\eeq
The Yukawa couplings of the Pati-Salam skeleton model are
\beq  -\Lag_Y= Y \psi_R \psi_L \phi+Y_c\, \psi_R\psi_L \phi^c +{\rm h.c.} \eeq
As discussed in section~\ref{sect:HFCNC} the two-Higgs structure allows
for independent structures of the up-quark and down-quark Yukawa matrices $Y_U$ and $Y_D$.
However, in view of quark-lepton unification, the skeleton model predicts $Y_E = Y_D$, as in SU(5), and
 $Y_N = Y_U$, as in SO(10), where $Y_N$ is the Yukawa matrix of neutrinos.

\subsubsection*{Minimal extensions}
\label{MinPS}
There are at least two ways of avoiding the wrong quark-lepton mass predictions with the addition of new fields.
\begin{itemize}

\item Foot~\cite{Foot} proposed adding the vector-like fermions $\psi$ (in the same representation as the scalar $\phi$,
see table~\ref{tab:224}, containing a SM lepton doublet and an anti-lepton doublet)
and the scalar $\phi_L$ (in the same representation as the fermion $\psi_L$).
In this way the Yukawa interactions are extended to
\beq\label{eq:Yuk224}
 -\Lag_Y= 
Y \psi_R \psi_L \phi+Y_c\, \psi_R\psi_L \phi^c+Y_N\,\psi_L \psi \phi_R + Y_L\,\psi \psi_R \phi_L +{\rm h.c.}\eeq
The third term provides a mass term $Y_N\langle \phi_R\rangle$ pairing the lepton doublet in $\psi_L$ with the
anti-lepton doublet in $\psi$, while the lepton doublet in $\psi$ remains massless before EW breaking.
In this way, the SM quark doublet is embedded in $\psi_L$, while the SM lepton doublet is embedded in $\psi$, and lepton-quark unification is evaded in the left-handed sector. This breaking of lepton-quark unification also relaxes the bounds (discussed in section~\ref{TAFPS}) on the 
vector leptoquark mass $M_{W'}$, which can safely be about a few TeV~\cite{Foot}.
After EW breaking, the SM leptons acquire a mass $Y_L\langle \phi_L\rangle$ through the last term of the Yukawa interactions in \eq{eq:Yuk224}. Note that this mechanism requires the absence of a mass term $\psi \psi$, which is allowed by the gauge symmetry.

\item Volkas~\cite{Volkas} proposed adding the fermions $Q_L\oplus Q_R$ 
in the $10\oplus\overline{10}$ of $\SU(4)_{\rm PS}$ (see table~\ref{tab:224}). These fields
contain a vector-like copy
of a right-handed lepton $e_R$, of a right-handed quark $d_R$, and of an exotic quark in a colour sextuplet.
The Yukawa couplings and fermion mass terms are\footnote{Volkas presented a slightly different model with a real $\phi$.}
\beq \label{YPSV}
Y \psi_L \psi_R \phi+Y_c\, \psi_L\psi_R \phi^c  + Y_Q \, \psi_R Q_L \phi_R +  M Q_L Q_R + {\rm h.c.}
\eeq
The mass term $Y_Q \langle\phi_R\rangle$ induces a mass mixing between light and heavy states in the $e_R$ and $d_R$ sectors, which
differ by a group-theoretical factor $\sqrt{2}$.
As a consequence, the unwanted relation $Y_D=Y_E$ is avoided.
Even with a single $Q_L\oplus Q_R$ pair, the quark and charged-lepton masses can be made to agree with data.
Similarly, by adding a fermion singlet~\cite{Volkas}, one can avoid the unwanted prediction in the neutrino sector, $Y_N=Y_U$.

\end{itemize}

%

In both versions of the Pati-Salam model, gauge and Yukawa interactions are invariant under an accidental ${\rm U}(1)_{B'}$ global symmetry, 
defined by the following charge assignments: $B'(\psi_L)=B'(\phi_L)=1$, $B'(\psi_R)=B'(\phi_R)=-1$, $B'(\psi)=B'(\phi)=0$, $B'(Q_L)=-B'(Q_R)=2$. 
The vevs of $\phi_R$ and $\phi_L$ break spontaneously both $\SU(4)_{\rm PS}$ and ${\rm U}(1)_{B'}$, leaving 
unbroken a new ${\rm U}(1)$ 
global symmetry given by 
 $(3/4)[B^\prime + (B-L)]$ which, for SM particles, corresponds to baryon number. This
 symmetry
protects proton stability, and prevents the appearance of a massless state related to the spontaneous breaking 
of ${\rm U}(1)_{B'}$. 


We analysed the two minimal realistic Pati-Salam
models finding that, while the gauge couplings and the Yukawa couplings have appropriate 
TAF solutions, this is never the case for the quartic couplings in the scalar potential.
Indeed the most general quartic scalar potential is
\beq V(\phi,\phi_R) = V_{\phi_R}  + V_\phi  + V_{\phi\phi_R} \eeq
or, adding the scalar $\phi_L$ proposed by Foot:
\beq V(\phi,\phi_R,\phi_L) = V_{\phi_R} + V_{\phi_L}  + V_\phi + V_{\phi\phi_L} + V_{\phi\phi_R} + V_{\phi_L\phi_R}+ V_{\phi_L\phi_R}^B\eeq
which contains 18 real couplings and 6 complex couplings.
The various potential terms and the corresponding quartic couplings
are defined in eq.~(\ref{sys:VPS}).
The RGE for the quartics form a large system, listed in appendix~\ref{RGE224}.\footnote{It is interesting to observe that in this model where the scalar
$\phi_L$ is present, the coupling $\lambda_B$ in $V_{\phi_L\phi_R}^B$  is the only coupling that  violates baryon number. 
Therefore, it is only multiplicatively renormalised and its $\beta$-function must be proportional to the coupling itself (${d\lambda _{{B}}}/{d\ln\mu} \propto \lambda _ {{B}}$, as shown in appendix~\ref{RGE224}).
We found that the TAF conditions generally imply that $\lambda_B$ must  vanish, either asymptotically or identically, depending on the TAF solutions for the other couplings entering the $\beta$-function of $\lambda_B$. Similarly,
TAF often requires that CP-violating quartics must vanish.}
We do not find any TAF solution unless the $\beta$-function coefficient of $g_4$ is artificially reduced 
down to unacceptably small values.



\bigskip

\subsection{A TAF Pati-Salam model}
\label{TAFPS}

By considering non-minimal Pati-Salam models we found 
TAF models with 3 generations, which seems to be the maximum allowed in the present context.\footnote{
A Pati-Salam TAF model with 2 generations was build by Kalashnikov in 1977~\cite{K224}.
However, the model is incompatible with flavour data and cannot be extended to 3 generations because gauge couplings
would no longer be asymptotically free.
We confirm most of his results, although the author of~\cite{K224} missed the existence of the quartics
in our eq.~(\ref{eq:Kmiss}), which could have changed the result.
Its inclusion modifies the TAF conditions, and the model admits TAF solutions, including one with asymptotically positive potential.}
Table~\ref{tab:224} describes the matter content of the model, which employs
\begin{itemize} 
\item[-] the  matter content of the skeleton Pati-Salam model (upper box), 
\item[-]   $N_\psi\le 3$ generations of the $\psi$ fermions proposed by Foot~\cite{Foot}, 
\item[-]  $N_Q=2$ generations of the $Q_L\oplus Q_R$ fermion proposed by Volkas~\cite{Volkas},
\item[-]  
a scalar $\Sigma$ in the adjoint of $\SU(4)_{\rm PS}$.
\end{itemize}
Upper bounds $N_\psi \le 3$ and $N_Q\le 2$ arise from the request that gauge couplings are asymptotically free:
\beq \label{eq:bgPS}
b_L =3-\frac23 N_\psi,\qquad b_R =  \frac73-\frac23 N_\psi,\qquad
b_4 = \frac{29}{3}-4 N_Q.\eeq
The extra fields allow for a realistic fermion spectrum as described in section~\ref{MinPS} and, at the same time, modify the RG running of the scalar quartics in such a way that TAF solutions are found for $N_Q=2$.

The most general Yukawa interactions are\footnote{With the addition of three generations of massless fermion singlets $\psi_1 \sim (1,1,1)$, the Yukawa couplings $y_1\, \psi_1 \psi_R\phi_R^\dagger$
satisfy the TAF condition and allow us to obtain realistic neutrino masses avoiding the  $Y_N = Y_U$ relation.
An extra  $\psi_\Sigma \sim (1,1,15)$ could play a similar role or could be identified with the Dark Matter, if its Yukawa couplings vanish.}
\beq 
-\Lag_{\rm Y} = Y\phi \psi_L \psi_R+ Y_c \phi^c\psi_L \psi_R +  
Y_N \,  \psi_{L} \psi \phi_R+
Y_Q\,  \psi_R Q_L \phi_R+
Y_\Sigma\, Q_L Q_R\Sigma +{\rm h.c.}
\eeq
The Yukawa coupling $Y_Q$ gives mass to $Q_{L,R}$, once $\Sigma$ acquires a vev.
The most general scalar potential is
\beq V (\phi, \phi_R,\Sigma) = V_{\phi_R}  + V_\phi  + V_{\phi\phi_R} + V_\Sigma + V_{\phi \Sigma} + V_{\phi_R\Sigma}
\eeq
where the various terms are defined in eq.~(\ref{sys:VPS}).
The RGE are listed in appendix \ref{RGE224}.
The simplest TAF model corresponds to $N_\psi=0$  (no $\psi$ field and so $Y_N$ is absent).
TAF solutions for the quartics are found only if the gauge and Yukawa couplings are on the following fixed-flow
\beq
\begin{array}{c|ccccccc}
\rowcolor[cmyk]{0,0,0,0.05}
 & {\tilde g}_{L\infty}^2  & {\tilde g}_{R\infty}^2 & {\tilde g}_{4\infty}^2& {\tilde Y}_{\infty}^2 & {\tilde Y}_{c\infty}^2   & {\tilde Y}_{Q\infty}^{2} & {\tilde Y}_{\Sigma\infty}^2 \\
  \hline
  \hbox{Fixed-flow} &  3/5 & 1/3 & 3/7 & 0.432& 0 & 0.909 & 3.454\\
  \hbox{$M$-Eigenvalues} & - & - & - & + & + & + & + \\  
\end{array}
\label{SU5y2}
\eeq
There are 4 positive eigenvalues of the $M$ matrix, hence 4 Yukawa couplings are univocally predicted at low energy. 
The IR prediction $Y_c=0$ is incompatible with a realistic quark-mass spectrum, 
but can be evaded in the more complicated TAF model with $N_\psi>0$ and extra Yukawa couplings described in appendix~\ref{RGE224}.


The 19 quartics admit 25 different TAF solutions. If we set to zero the 4 CP-violating couplings, we find 15 TAF solutions for the 15 CP-conserving quartics, given by
$$
\tiny
\begin{array}{c|ccccccccccccccc} \rowcolor[cmyk]{0,0,0,0.05}
N_+ & \lambda _{\Sigma \phi _R1} & \lambda _{{\Sigma \phi 1}} & \lambda \
_{\Sigma \phi _R2} & \lambda _{{\Sigma \phi 2}} & \lambda _{{R1}} & \
\lambda _{{R2}} & \lambda _{{\Sigma 1}} & \lambda _{{\Sigma 2}} & \
\lambda _ 1 & \lambda _ 2 & \lambda _ 3 & \lambda _ 4 & \lambda \
_{\phi _R{\phi 1}} & \lambda _{\phi _R{\phi 2}} & \lambda _{\phi \
_R{\phi 3}} \\  \hline
 10 & -0.023 & -0.127 & -1.067 & 0 & -0.017 & 0.082 & 0.374 & -1.031 & 0 & -0.015 & -0.037 & -0.057 & 0.172 & 0 & -0.322 \\
8 & 0.210 & 0 & -1.174 & -0.024 & 0.036 & 0.042 & 0.661 & -1.621 & -0.033 & 0.018 & -0.034 & -0.013 & 0.210 & 0 & -0.344 \\
 10 & -0.040 & -0.018 & -1.021 & 0 & 0 & 0.056 & 0.313 & -0.973 & -0.028 & -0.031 & 0 & -0.095 & 0.195 & 0 & -0.381 \\
 8 & -0.036 & -0.053 & -1.036 & 0 & 0 & 0.068 & 0.341 & -1.009 & -0.051 & 0.019 & -0.031 & 0 & 0.197 & 0 & -0.357 \\
 6 & 0 & -0.184 & -1.110 & 0 & -0.028 & 0.095 & 0.404 & -1.033 & 0.022 & -0.022 & 0 & -0.101 & 0.152 & 0 & -0.277 \\
 8 & 0.212 & -0.129 & -1.197 & 0 & 0.033 & 0.047 & 0.660 & -1.585 & 0 & 0.027 & 0 & 0 & 0.192 & 0 & -0.327 \\
 7 & 0.206 & -0.181 & -1.233 & 0.036 & 0.021 & 0.059 & 0.656 & -1.513 & 0 & 0 & -0.024 & -0.054 & 0.194 & 0 & -0.286 \\
 9 & 0.215 & -0.077 & -1.179 & 0 & 0.038 & 0.040 & 0.662 & -1.616 & 0 & -0.029 & 0 & -0.106 & 0.197 & 0 & -0.347 \\
 8 & -0.027 & -0.095 & -1.059 & 0 & -0.021 & 0.083 & 0.374 & -1.045 & -0.085 & 0 & 0 & -0.032 & 0.211 & 0 & -0.324 \\
 9 & 0.185 & 0.054 & -1.165 & -0.089 & 0.037 & 0.039 & 0.621 & -1.516 & -0.052 & 0.013 & -0.017 & -0.023 & 0.219 & 0 & -0.354 \\
 9 & 0.139 & 0.164 & -1.151 & 0 & 0.016 & 0.062 & 0.599 & -1.475 & 0.024 & 0 & 0 & 0 & 0.218 & 0 & -0.312 \\
 10 & 0.185 & 0.054 & -1.165 & 0.089 & 0.037 & 0.039 & 0.621 & -1.516 & -0.052 & 0.013 & 0.017 & -0.023 & 0.219 & 0 & -0.354 \\
 9 & 0.194 & 0.046 & -1.174 & 0.030 & 0.029 & 0.051 & 0.653 & -1.599 & 0.010 & -0.016 & 0.037 & -0.059 & 0.206 & 0 & -0.326 \\
 9 & 0.167 & 0.149 & -1.156 & 0.060 & 0.032 & 0.044 & 0.607 & -1.488 & -0.015 & -0.019 & 0.026 & -0.062 & 0.222 & 0 & -0.348 \\
 9 & 0.167 & 0.149 & -1.156 & -0.060 & 0.032 & 0.044 & 0.607 & -1.488 & -0.015 & -0.019 & -0.026 & -0.062 & 0.222 & 0 & -0.348 \\
\end{array}
$$
The first column shows the number of positive eigenvalues of the $M$ matrix, {\it i.e.} the number of quartics predicted at low energy.
All these solutions give gauge, Yukawa, and quartic couplings that can be extrapolated up to infinite energy without hitting any Landau pole. However, all these solutions correspond to theories with metastable vacua, because the scalar potential has always some negative directions in the asymptotic region of large fields.

\subsubsection*{Flavour bounds from vector leptoquarks}
The experimental bounds on heavy Higgs bosons and on $W_R$ and $Z^\prime$ gauge bosons have been discussed in sections~\ref{sect:HFCNC} and~\ref{sect:WR}.
 The Pati-Salam gauge group also contains massive  vector leptoquarks  $W_{\mu}^{\prime}$, of charge $\pm 2/3$,
corresponding to the broken generators in  $\SU(4)_{\rm PS}/\SU(3)_{c}$. The gauge bosons $W_{\mu}^{\prime}$ are coupled to a right-handed current involving
 $d_R$ and  $e_R$ (since these two fields are unified in $\psi_R$,
see table~\ref{tab:224}), and to a left-handed current involving $q_L$ and $\ell_L$ doublets 
(only in versions of the model in which $q_L$ and $\ell_L$ are unified in the field $\psi_L$).
 
When both left-handed and right-handed interactions are present, data on $\pi\to e \nu$ give the strong constraint~\cite{Val}
\be
M_{W^\prime} > 250\TeV\qquad 
\hbox{(Pati-Salam with $q_L$-$\ell_L$ and $d_R$-$e_R$ unification).} 
\label{eq:LQboundL}
\ee
This bound can be avoided in the models discussed in section~\ref{MinPS} where the SM left-handed leptons $\ell_L$ are contained in the field $\psi$, while the quark doublet $q_L$ is in $\psi_L$. In this case, the gauge bosons $W_{\mu}^{\prime}$ are coupled only to right-handed currents of SM fermions. 
As discussed in appendix~\ref{sec:LQbounds}, the bounds from right-handed interactions can be significantly relaxed with an appropriate flavour structure and, under the most favourable assumptions, they give 
\be
M_{W^\prime} > 8.8~{\rm TeV}\qquad 
\hbox{(Pati-Salam with $d_R$-$e_R$ unification).} 
\label{eq:LQboundR}
\ee
Taking into account the relation $M_{W_R} \approx g_Y M_{W'}/g_3$, 
the bound in eq.~(\ref{eq:LQboundR})  implies $M_{W_R}> 2.7$~TeV, which is 
comparable to those from direct $W_R$ searches (see section~\ref{sect:WR}).

\bigskip

However,
the Pati-Salam TAF models that we discovered do not contain the scalar $\phi_L$ and 
thus lead to $q_L$-$\ell_L$ unification. This implies that these models
suffer from the strong bound of eq.~(\ref{eq:LQboundL})
and, consequently, from an unnaturally high degree of fine-tuning.
We have not found Pati-Salam TAF models containing the scalar $\phi_L$.

\begin{table}
$$
\begin{array}{|rccccc|}\hline
\rowcolor[cmyk]{0,0,0,0.05}
\multicolumn{2}{|c}{\hbox{Matter fields}}&\hbox{spin} & \SU(3)_L&\SU(3)_R & \SU(3)_{\rm c}\cr \hline
Q_R\,=\!\!&{\small \begin{pmatrix}u^1_R & u_R^2 &  u^3_R  \\ d_R^1 & d_R^2 &  d^3_R  \\    d^{\prime 1}_R & d^{\prime 2}_R & d^{\prime 3}_R\end{pmatrix}}&1/2& 1 & 3  & \bar 3\cr
Q_L\,=\!\!&{\small \begin{pmatrix}u^1_L & d^1_L &  \bar d^{\prime 1}_R  \\  u_L^2& d_L^2 &  \bar d^{\prime 1}_R   \\  u_L^3& d_L^3 & \bar d^{\prime 3}_R \end{pmatrix}} &1/2& \bar 3 &1 &   3 \cr
L\,=\!\!&{\small \begin{pmatrix}\bar\nu'_L & e'_L &  e_L \\ \bar e'_L &  \nu'_L & \nu_L  \\  e_R &\nu_R &\nu' \end{pmatrix}} & 1/2& 3 & \bar 3  & 1 \cr
\multicolumn{2}{|c}{H_1,H_2} &0&  3 & \bar 3  & 1 \cr \hline
\end{array}$$
\caption{\em\label{tab:333} Field content of the minimal trinification model.   Primed fermions correspond to new states beyond the SM.}
\end{table}

\subsection{Trinification SU(3)$_L\otimes\,$SU(3)$_R\otimes\,$SU(3)$_c$}\label{333}
Trinificaton is often considered as a unification model, after imposing a permutation symmetry among the three $\SU(3)$ factors that forces the
 gauge couplings $g_L$, $g_R$, $g_c$ to be equal.  While the permutation symmetry is respected by the SM fermions (see table~\ref{tab:333}),
it requires the addition of extra Higgs bosons with interactions that break baryon number.
We do not impose any permutation symmetry in order to insure that trinification interactions at the weak scale conserve baryon number. The relation between the trinification gauge coupling constants ($g_L$, $g_R$, $g_c$) and those of the SM ($g_3$, $g_2$, $g_Y = \sqrt{3/5}\, g_1$) is
\beq g_L = g_2, \qquad  g_R = \frac{2 g_2 g_Y}{\sqrt{3 g_2^2 - g_Y^2}} , \qquad g_c = g_3\, .\eeq

We consider the minimal model with the matter content summarised in table~\ref{tab:333}.
Since quarks are not unified with leptons, trinification is safer than Pati-Salam from the point of view of flavour.
Each generation of $Q_R\oplus Q_L \oplus L$ contains 27 chiral fermions  that decompose under the SM gauge group as the usual 15 SM fermions, plus a vector-like lepton doublet,
a vector-like right-handed down quark, and two singlets (denoted as primed fermions in table~\ref{tab:333}). These states correspond to the irreducible representation 27 of E(6).

\smallskip

A single Higgs $H_1$ in the $(3_L,\bar 3_R)$ representation contains 3 Higgs doublets.
The vev $\langle H_1\rangle_{33}=V_1/\sqrt{2}$
breaks $\SU(3)^3$ to $\SU(2)_L\otimes\SU(2)_R \otimes{\rm U}(1)_{B-L}\otimes \SU(3)_c$.
Two Higgs doublets and one singlet are eaten by the $9$ components of the vector bosons that acquire mass
\beq M^2_{H_L} = \frac{g_L^2}{2} V_1^2,\qquad 
 M^2_{H_R} = \frac{g_R^2}{2} V_1^2,\qquad
 M^2_{Z'} = \frac23 (g_L^2+g_R^2) V_1^2.
\label{eq:331br1} \eeq
The massive $Z'$ corresponds to the combination of gauge bosons $g_L A^8_{L\mu} - g_R A^8_{R\mu}$. 
The bound on its mass is in the range 2-6$\TeV$, depending on the $Z'$ charge of the SM Higgs.
The gauge boson of $B-L$ corresponds to $g_R A^8_{L\mu} + g_L A^8_{R\mu}$ with $g_{B-L} = (\sqrt{3}/2) g_Rg_L/\sqrt{g_R^2+g_L^2}$.

A second scalar is needed to break the left-right symmetry.  This is accomplished by a Higgs $H_2$ in the $(3_L,\bar 3_R)$ representation
with vev $\langle H_2\rangle_{32}=V_2/\sqrt{2}$.  The 12 components of the massive vectors 
can be grouped into the complex doublet $H_L$, which transforms as $(2_L,1/2_Y)\oplus ({\bar 2}_L, -1/2_Y)$ under the SM $\SU(2)_L\times \U(1)_Y$, two electrically-charged and one neutral complex gauge bosons $H_R^\pm$, $W_R$ $\sim (1_L,1_Y)\oplus (1_L, -1_Y)$ and 
$H_R^0$ $\sim (1_L,0_Y)\oplus (1_L, 0_Y)$, and two kinds of $Z^\prime$ gauge bosons. 
Their masses are given by
 \beq M^2_{H_L} = \frac{g_L^2}{2} (V_1^2+V_2^2)\qquad  
  M^2_{H_R^\pm} = \frac{g_R^2}{2} (V_1^2+V_2^2),\qquad
 M^2_{H_R^0} = \frac{ g_R^2}{2} V_1^2,\qquad M_{W_R}^2 = \frac{g_R^2}{2} V_2^2.\eeq
 The two $Z'$ have a mass matrix which, in the limit $V_2\ll V_1$, leads to a heavier state with mass as in \eq{eq:331br1}, and a lighter $B-L$ gauge boson with mass 
\beq M^2_{B-L} \simeq  \frac{(g_R^2-2 g_L^2)^2}{6(g_L^2+g_R^2)} V_2^2\, .\eeq
As shown in fig.\fig{Zprime}, the $B-L$ gauge boson is subjected 
 to the  bound $M_{B-L}\circa{>}2.6\TeV$ from ATLAS.
 The bounds from flavour processes mediated by the new gauge bosons are much milder than in the Pati-Salam models because no dangerous leptoquark interactions are present.
 
The SM Yukawa couplings are obtained from the $\SU(3)^3$-invariant interactions
\beq- \Lag_{Y} = \sum_{i=1}^2 (y_{Qi} ~Q_L Q_R  H_i + \frac{y_{Li}}{2} LL H^{*}_i) +{\rm h.c.}
\label{eqqqY}
\eeq
Similarly to the case discussed in section~\ref{sect:HFCNC}, bounds from flavour processes can be kept under control because there are only two independent Yukawa matrices and therefore small quark masses suppress the new flavour interactions. However, this raises the problem of generating sufficiently large masses for the new fermions. The new (primed) fermions get mass from \eq{eqqqY}, once the vev $V_1$ is generated, while mixing mass terms between new fermions and SM quarks/leptons are induced by the smaller vev $V_2$. Since all these terms are proportional to SM Yukawa couplings, the new fermions turn out to be too light, unless $V_1$ is unnaturally large. Additional fields and interactions are needed to construct realistic models.
\bigskip

Let us turn to the issue of the TAF properties of trinification models. We start by considering the simple, albeit unrealistic, case of a Higgs sector made of a single $H=H_1$ in the $(3_L,{\bar 3}_R)$ with the most general quartic potential 
\beq \label{eq:V1111}
V=\lambda_a \hbox{Tr}(H^\dagger H)^2+\lambda_b \hbox{Tr}(H ^\dagger H H^\dagger H).\eeq
The one-loop RGE for gauge, Yukawa, and quartic couplings are given in eq.~(\ref{sys:RGE3331})
and admit two different TAF solutions:
\beq
\begin{array}{c|ccccccc}
\rowcolor[cmyk]{0,0,0,0.05}
 & {\tilde g}_{L\infty}^2 & {\tilde g}_{R\infty}^2 & {\tilde g}_{c\infty}^2 & {\tilde y}_{Q\infty}^2 & {\tilde y}_{L\infty}^2 & {\tilde \lambda}_{a\infty}  & {\tilde \lambda}_{b \infty} \\
  \hline
  \hbox{Fixed-flow} &  2/9 & 2/9 & 1/5 &0& 23/54& \phantom{-}0.1628&-0.1732 \\
  \hbox{$M$-Eigenvalue} & - & - & -&- & + & - & + \\   \hline
    \hbox{Fixed-flow} &  2/9 & 2/9 & 1/5 &0& 23/54&  -0.0026& -0.0087 \\
  \hbox{$M$-Eigenvalue} & - & - & -&- & + & - & - \\  
\end{array}
\label{SU333y}
\eeq
The potential $V$ is definite positive for $\lambda_a \ge - \lambda_b$ (for $\lambda_b <0$) and $3\lambda_a\ge -\lambda_b$ (for $\lambda_b>0$).
Both fixed-flows in \eq{SU333y} violate the stability condition for the potential.
The RGE flow can produce 
spontaneous symmetry breaking at low energy \`a la Coleman-Weinberg from a  potential with no dimensional parameters.

\bigskip

Next, we consider the model with two Higgs multiplets $H_1$ and $H_2$ in the $(3_L,{\bar 3}_R)$. The scalar quartic potential is given in \eq{eq:VH1H2} and
contains 14 real couplings and 6 phases.    
The relevant one-loop RGE are reported in appendix~\ref{RGE333}.
The gauge couplings and the Yukawa admit TAF solutions, but
we do not find a TAF solution for the quartics.\footnote{TAF solutions appear if the $\SU(3)_c$ $\beta$ function coefficient is reduced to $b_3=1/6$.  However it is impossible to obtain such value by adding
fermionic multiples.}

The next step is to complicate the Higgs sector, with the twofold aim of recovering TAF and of generating sufficiently large masses for the new fermions.
One possibility is to consider one $H$ and one $\Sigma_R$, adjoint under $\SU(3)_R$.
A vev in its 23 component breaks $\SU(3)^3$ to the SM group.
We also add one vector-like fermion $Q'_R\oplus \bar Q'_R$ which, in the model with a single $H$ is needed to
obtain different masses for top and bottom quarks via right-handed quark mixing induced by  $Q^{(\prime)}_R \bar Q'_R \Sigma$ Yukawa couplings.
A TAF solution exists only for a single generation of $Q'_R\oplus \bar Q'_R$, which is not sufficient to produce a realistic fermionic mass spectrum. Once we add 
two or more generations of $Q'_R\oplus \bar Q'_R$, the gauge group is no longer asymptotically free. In conclusion, we were not able to identify a TAF trinification model with a realistic flavour structure.

\mysection{Conclusions}\label{concl}

The main point of this paper is to single out the implications of two severe difficulties of theories trying to solve the Higgs naturalness problem by having the weak scale as the only effective source of breaking of scale invariance at the classical level. 

The first well-known problem is gravity. The hope that transplanckian dynamics can miraculously cure the Higgs sensitivity to $\mpl$ seems to us unrealistic. In the absence of special symmetries like supersymmetry, any short-distance modification of general relativity is expected to affect the quantum corrections to the Higgs mass. Delaying such modifications to energy scales as high as $\mpl$ makes the problem too acute to leave us with any reasonable hope that new dynamics can rescue the situation: the problem must be tackled at energies lower than $\mpl$. 
From this line of reasoning, we have derived our first conclusion: {\it in the context of theories with no dynamical protection of the Higgs mass, naturalness requires a premature modification of gravity, at scales no larger than $\Lambda_G \sim 4\pi (M_h \mpl )^{1/2} \sim 10^{11}\GeV$}. In this paper we have called {\em softened gravity} a theory in which the communication between gravity and the SM sector remains weak at any scale larger than $\Lambda_G$.

\smallskip

The second problem, less explored in the literature, is that the quantum theory necessarily breaks scale invariance. In the SM, the problem manifests itself in the form of the Landau pole for the hypercharge gauge coupling. In softened versions of gravity, this dynamically-generated scale brings back the naturalness problem. To cure this problem one needs to extend the SM, below a few TeV, into a TAF theory. 
We have shown that the construction of such extensions is possible, but only at a high price:
many new particles around the TeV scale are needed, as well as elaborate model-building to avoid phenomenological constraints, especially related to the flavour sector. 

So our second (surprising) conclusion is that {\it theories intended to deal with naturalness without new dynamics in the TeV range actually need a large number of new particles around the TeV scale}.
This reopens the usual can of worms with the phenomenological difficulties in satisfying constraints from collider searches and flavour processes that plague traditional approaches to naturalness, like supersymmetry or composite Higgs. 
Indeed, we find that the degree of tuning in the models we constructed is typically worse than in traditional approaches.

Our two conclusions are based on common intuition derived from effective field theory and dimensional analysis. Only in special setups that defy such usual intuition, could our conclusions be proved wrong. The only example of such a setup we are aware of is the use of anthropic arguments in the multiverse. We cannot exclude the existence of other theoretical setups that evade our conclusions.

The positive side of our result is that the class of theories we have considered is not at all elusive, but can be tested at high-energy colliders. While the next probe comes from Run-2 LHC, the existing constraints on new-particle masses from rare processes suggest that a 100-TeV future collider is better positioned to explore the full parameter space. The other interesting aspect is that the experimental signals from softened-gravity models are distinct from those coming from traditional schemes with dynamical explanations of naturalness, but also differ from anthropic solutions, which are likely to have no light scalar particles other than the Higgs.

In a more technical vein, another result of our paper is the development of a systematic procedure to derive the asymptotic  behaviour of coupling constants in a perturbative QFT. The method is based on calculating the fixed-flows of the theory, which are special RG trajectories 
where couplings flow to zero in the UV with the slowest possible rate allowed by RG evolution.  The fixed-flows are determined by solving an algebraic system of equations, with no need to tackle the full differential equations describing the RG. This allows for a simple implementation of the method, even in models with many coupling constants. The second step of the procedure is the computation of the eigenvalues of the matrix $M$ in \eq{Mmat}, which determine whether the fixed-flow is UV attractive or repulsive.

UV-repulsive fixed-flows correspond to RG trajectories with isolated asymptotic behaviours, therefore singling out special values of some combination of coupling constants in the IR. For this reason, one could regard the requirement of sitting on these special RG trajectories either (pessimistically) as an extreme fine-tuning of parameters or (optimistically) as a novel way of predicting physical quantities. Our point of view is that this requirement provides a genuine prediction of the theory. The UV-repulsive fixed-flow is disconnected from the other RG trajectories in the asymptotic region and any small deformation leads to an ill-defined theory in the UV. As an analogy, take the case of a Yang-Mills theory. One cannot regard the relation between the cubic and quartic gluon couplings as a fine tuning, because such relation follows from a consistency condition of the theory in the UV. 

The existence of UV-repulsive fixed-flows is essential for TAF. This is because, in practice, TAF conditions for quartic couplings can often be satisfied only if some Yukawa couplings lie on UV-repulsive fixed-flows. Moreover, stability of the scalar potential often favours quartic couplings $\lambda$ on UV-repulsive fixed-flows rather than on UV-attractive ones, because $\lambda$ is generally larger on the former than on the latter. As a result, requiring the theory to be TAF usually leads to some precise predictions of certain coupling constants in the IR. These predictions are robust against deformations in the UV, since they correspond to IR-attractive solutions. In particular, gravitational or super-weak interactions could modify the exact locations of the fixed-flows, but would not change the numerical values of the IR predictions.

\medskip

 We  exemplified our technique in the case of the SM,  
studying under which conditions it satisfies TAF.  We  found that this happens for $g_1 = 0$,  $M_t = 186\GeV$, $M_\tau = 0$, $M_h = 163\GeV$ (or $M_h < 163\GeV$ if unstable vacua are permitted). Since these conditions are unphysical (although not too far from reality), we  searched for TAF extensions of the SM at the weak scale. 
The simplest candidates are based on the Pati-Salam group $\SU(2)_L\otimes \SU(2)_R\otimes\SU(4)_{\rm PS}$, for which we  found some explicit examples, and on the trinification group $\SU(3)_L\otimes \SU(3)_R\otimes\SU(3)_c$, for which we have not found a fully realistic model.
Our technique has been proven useful to perform systematic searches for TAF theories and determine their IR predictions.

\small

\subsubsection*{Acknowledgments}
This project started thanks to discussions with A.\ Arvanitaki, S.\ Dimopoulos, S.\ Dubovski, and G.\ Villadoro.
We thank K.\ Kannike, R.\ Rattazzi, W.\ Skiba and M.\ Strassler for useful discussions. This work was supported by the ESF grant MTT8.
The work of Alberto Salvio has been also supported by the Spanish Ministry of Economy and Competitiveness under grant FPA2012-32828, Consolider-CPAN (CSD2007-00042), the grant  SEV-2012-0249 of the ``Centro de Excelencia Severo Ochoa'' Programme and the grant  HEPHACOS-S2009/ESP1473 from the C.A. de Madrid.

\appendix

    \mysection{RGE for  Pati-Salam models}\label{RGE224}
 We give here the one-loop RGE for a wide class of Pati-Salam models containing the scalars $\phi_L,\phi_R,\phi,\Sigma$
 and the fermions $\psi,\psi_L,\psi_R,\psi_1,\psi_\Sigma, Q_L, Q_R$ defined in section~\ref{224}.
While this model does not allow for TAF solutions,
the RGE for Yukawa and quartic couplings of the TAF models proposed in section~\ref{TAFPS} are
found by dropping  all interactions involving the scalar $\phi_L$.    
The RGE for the gauge couplings depend on the chosen field content, see {\it e.g.} \eq{eq:bgPS}.

The Yukawa couplings for one generation of fermions are
\begin{eqnarray}
-\Lag_Y \!&\! = \!&\! y\, \psi_R \psi_L \phi+
y_c\, \psi_R\psi_L \phi^c+
y_N\,\psi_L \psi \phi_R + 
y_E\,\psi \psi_R \phi_L+
y_{RQ}\,  \psi_R Q_L \phi_R +
y_\psi \, \psi_{\Sigma}^2 \Sigma+~~~~\\
&&+y_{LQ}\,\psi_L Q_R\phi_L
+y_R \, \psi_{\Sigma}\psi_R \phi_R^*+
y_L \, \psi_{\Sigma}\psi_L \phi_L^*+
y_\Sigma\, Q_L Q_R\Sigma +
y_\nu \,\psi_1  \psi_R\phi_R^*+
{\rm h.c.} \nonumber
 \end{eqnarray}
The RGE for the Yukawa couplings are 
{\begin{eqnsystem}{sys:y224}
(4\pi)^2\frac{dy}{d\ln\mu} &=& y (20 y_{{c}}^2+2 y_{{E}}^2-\frac{9 g_{{L}}^2}{4}-\frac{9 g_{{R}}^2}{4}-\frac{45 g_4^2}{4}+\frac{15 y_{{L}}^2}{8}
\br
+\frac{5 y_{{LQ}}^2}{2}+2 y_{{N}}^2+\frac{15 y_{{R}}^2}{8}+\frac{5 y_{{RQ}}^2}{2}+y_{\nu }^2)+12 y^3 \\
(4\pi)^2\frac{dy_{{c}}}{d\ln\mu} &=& y_{{c}} (2 y_{{E}}^2-\frac{9 g_{{L}}^2}{4}-\frac{9 g_{{R}}^2}{4}-\frac{45 g_4^2}{4}+\frac{15 y_{{L}}^2}{8}
+\frac{5 y_{{LQ}}^2}{2}+2 y_{{N}}^2 \br +\frac{15 y_{{R}}^2}{8}+\frac{5 y_{{RQ}}^2}{2}+20 y^2+y_{\nu }^2)+12 y_{{c}}^3 \\
(4\pi)^2\frac{dy_{{R}}}{d\ln\mu} &=& y_{{R}} (2 y_{{c}}^2+2 y_{{E}}^2-\frac{9 g_{{R}}^2}{4}-\frac{141 g_4^2}{8}+y_{{L}}^2+4 y_{{N}}^2
+\frac{11 y_{{RQ}}^2}{2}+2 y^2+3 y_{\nu }^2+\frac{3 y_{\psi }^2}{8})\br +8 y_{{E}} y_{{L}} y_{{N}}+\frac{53 y_{{R}}^3}{8} \\
(4\pi)^2\frac{dy_{{L}}}{d\ln\mu} &=& y_{{L}} (2 y_{{c}}^2+4 y_{{E}}^2-\frac{9 g_{{L}}^2}{4}-\frac{141 g_4^2}{8}+\frac{11 y_{{LQ}}^2}{2}+2 y_{{N}}^2+y_{{R}}^2+2 y^2+\frac{3 y_{\psi }^2}{8})\br +8 y_{{E}} y_{{N}} y_{{R}}+\frac{53 y_{{L}}^3}{8} \\
(4\pi)^2\frac{dy_{{RQ}}}{d\ln\mu} &=& y_{{RQ}} (2 y_{{c}}^2+2 y_{{E}}^2-\frac{9 g_{{R}}^2}{4}-\frac{153 g_4^2}{8}+4 y_{{N}}^2
\br
+\frac{33 y_{{R}}^2}{8}+2 y^2-y_{\nu }^2+\frac{9 y_{\Sigma }^2}{16})+\frac{19 y_{{RQ}}^3}{2} \\
(4\pi)^2\frac{dy_{{LQ}}}{d\ln\mu} &=& y_{{LQ}} (2 y_{{c}}^2+4 y_{{E}}^2-\frac{9 g_{{L}}^2}{4}-\frac{153 g_4^2}{8}+\frac{33 y_{{L}}^2}{8}+2 y_{{N}}^2+2 y^2+\frac{9 y_{\Sigma }^2}{16})+\frac{19 y_{{LQ}}^3}{2} \\
(4\pi)^2\frac{dy_{{E}}}{d\ln\mu} &=& y_{{E}} (2 y_{{c}}^2-\frac{9 g_{{L}}^2}{4}-\frac{9 g_{{R}}^2}{2}-\frac{45 g_4^2}{8}+\frac{15 y_{{L}}^2}{4}+5 y_{{LQ}}^2+4 y_{{N}}^2+\frac{15 y_{{R}}^2}{8}
\br
+\frac{5 y_{{RQ}}^2}{2}+2 y^2+y_{\nu }^2)+10 y_{{E}}^3+\frac{15}{2} y_{{L}} y_{{N}} y_{{R}} \\
(4\pi)^2\frac{dy_{{N}}}{d\ln\mu} &=& y_{{N}} (2 y_{{c}}^2+4 y_{{E}}^2-\frac{9 g_{{L}}^2}{2}-\frac{9 g_{{R}}^2}{4}-\frac{45 g_4^2}{8}+\frac{15 y_{{L}}^2}{8}+\frac{5 y_{{LQ}}^2}{2}\br
+\frac{15 y_{{R}}^2}{4}+5 y_{{RQ}}^2+2 y^2+2 y_{\nu }^2)+\frac{15}{2} y_{{E}} y_{{L}} y_{{R}}+10 y_{{N}}^3 \\
(4\pi)^2\frac{dy_{\nu }}{d\ln\mu} &=& y_{\nu } (2 y_{{c}}^2+2 y_{{E}}^2-\frac{9 g_{{R}}^2}{4}-\frac{45 g_4^2}{8}+4 y_{{N}}^2+\frac{45 y_{{R}}^2}{8}-\frac{5 y_{{RQ}}^2}{2}+2 y^2)+11 y_{\nu }^3 \\
(4\pi)^2\frac{dy_{\psi }}{d\ln\mu} &=& y_{\psi } (-24 g_4^2+2 y_{{L}}^2+2 y_{{R}}^2+\frac{3 y_{\Sigma }^2}{2})+\frac{7 y_{\psi }^3}{4} \\
(4\pi)^2\frac{dy_{\Sigma }}{d\ln\mu} &=& y_{\Sigma } (-27 g_4^2+2 y_{{LQ}}^2+2 y_{{RQ}}^2+\frac{3 y_{\psi }^2}{4})+\frac{31 y_{\Sigma }^3}{8} 
\end{eqnsystem}}

The most general quartic potential is
$$
 V(\phi,\phi_R,\phi_L,\Sigma) = V_{\phi_R} + V_{\phi_L}  + V_\phi  + V_\Sigma+ V_{\phi\phi_L} + V_{\phi\phi_R} + V_{\phi_L\phi_R}+ V_{\phi_L\phi_R}^B
 + V_{\phi \Sigma} + V_{\phi_R\Sigma}+ V_{\phi_L\Sigma}+V_{X} 
$$
 where 
 \begin{eqnsystem}{sys:VPS}
 V_{\phi_R} &=&  \lambda_{R1} \Tr^2 (\phi_R \phi_R^\dagger) + \lambda_{R2}  \Tr( \phi_R \phi_R^\dagger\phi_R \phi_R^\dagger) ,\\ 
V_{\phi_L}&=& \lambda_{L1} \Tr^2 (\phi_L^\dagger \phi_L) + \lambda_{L2}  \Tr( \phi_L^\dagger \phi_L \phi_L^\dagger \phi_L),\\ 
V_\phi&=& \lambda_1 \Tr^2(\phi^\dagger \phi)
+\Re \lambda_2 \Tr^2(\phi^\dagger \phi^c) + \nonumber \\
&&+\Re\lambda_3  \Tr(\phi^\dagger\phi)\Tr(\phi^\dagger \phi^c) 
+(\lambda_4-2 \Re\lambda_2)|\Tr(\phi^\dagger \phi^c)|^2, ~~~~~~~\\ 
V_{\phi\phi_R}&=& \lambda_{R\phi1}\Tr(\phi_R \phi_R^\dagger)  \Tr (\phi^\dagger \phi) \!+ \!\Re  \lambda_{R\phi2}\Tr(\phi_R \phi_R^\dagger)  \Tr(\phi^\dagger \phi^c)+
\lambda_{R\phi3}\Tr(\phi_R \phi_R^\dagger \phi^\dagger\phi),
\\
V_{\phi\phi_L}&=& \lambda_{L\phi1}\Tr(\phi_L^\dagger \phi_L)  \Tr (\phi^\dagger \phi) + \Re \lambda_{L\phi2}\Tr(\phi_L^\dagger \phi_L)  \Tr(\phi^\dagger \phi^c)+
\lambda_{L\phi3} \Tr(\phi_L^\dagger \phi_L\phi \phi^\dagger ),\\
V_{\phi_L\phi_R}&=& \lambda_{LR1} \Tr(\phi_L^\dagger \phi_L) \Tr(\phi_R \phi_R^\dagger) + \lambda_{LR2} \Tr (\phi_L \phi_L^\dagger \phi_R^\dagger \phi_R)  +
\Re \lambda_{LR3} \Tr[(\phi_R\phi_L )(\phi_R\phi_L )^c],   \nonumber
\\ 
V_{\phi_L\phi_R}^B &=&\Re \lambda_B \epsilon_{IJKL}\epsilon_{ij}\epsilon_{k\ell} \phi_{LIi}\phi_{LJj} \phi_{RKk}^*\phi_{RL\ell}^*,   \\
V_{\Sigma} &=& \lambda_{\Sigma1}  \Tr^2(\Sigma^2 )+ \lambda_{\Sigma2}  \Tr(\Sigma^4 ) , \\
 V_{\phi\Sigma} &=& \lambda_{\Sigma \phi1} \Tr(\Sigma ^2) \Tr(\phi^\dagger \phi) + \Re \lambda_{\Sigma \phi2} \Tr(\Sigma ^2) \Tr(\phi^\dagger \phi^c) ,  \label{eq:Kmiss} \\
 V_{\phi_R\Sigma} &=&
\lambda_{\Sigma \phi_R1} \Tr(\Sigma^2) \Tr(\phi_R^\dagger \phi_R)+
\lambda_{\Sigma \phi_R2} \Tr ( \phi_R^\dagger \phi_R \Sigma^2) ,\\
 V_{\phi_L\Sigma} &=&
\lambda_{\Sigma \phi_L1} \Tr(\Sigma^2) \Tr(\phi_L^\dagger \phi_L)+
\lambda_{\Sigma \phi_L2} \Tr ( \phi_L \phi_L^\dagger \Sigma^2) ,\\
 V_{X}  &=& \Re \lambda_{X1} \Tr (\phi_L\phi_{LR} \phi_R\Sigma) + \Re \lambda_{X2} \Tr(\phi_L \phi_{LR}^c \phi_R\Sigma).
 \end{eqnsystem}
The symbol $\Re$ (real part) precedes quartic interactions that can support complex couplings $\lambda_i$. 
With an abuse of notation, in the RGE listed below we denote as $\lambda_i$ the real part and as $\theta_i$ the
imaginary part of  complex couplings $\lambda_i$. 
The RGE for the quartic couplings are 
{\tiny
\begin{eqnarray*}
(4\pi)^2\frac{d\lambda _{\Sigma \phi _L1}}{d\ln\mu} &=& \lambda _{\Sigma \phi _L1} (8 y_{{E}}^2-\frac{9 g_{{L}}^2}{2}-\frac{141 g_4^2}{4}+34 \lambda _{{\Sigma 1}}+\frac{29 \lambda _{{\Sigma 2}}}{2}+\frac{15 y_{{L}}^2}{2}+144 \lambda _{{L1}}+96 \lambda _{{L2}}+10 y_{{LQ}}^2+3 y_{\Sigma }^2+\frac{3 y_{\psi }^2}{2})
\br
+\frac{3 g_4^4}{2}-\frac{5}{8} y_{{L}}^2 y_{\psi }^2+32 \lambda _{{L1}} \lambda _{\Sigma \phi _L2}+16 \lambda _{{L2}} \lambda _{\Sigma \phi _L2}-\frac{1}{2} y_{{LQ}}^2 y_{\Sigma }^2+64 \lambda _{{LR1}} \lambda _{\Sigma \phi _R1}+16 \lambda _{{LR1}} \lambda _{\Sigma \phi _R2}+16 \lambda _{{LR2}} \lambda _{\Sigma \phi _R1}
\br
+8 \lambda _{\Sigma \phi _L1}^2+\frac{15 \lambda _{{\Sigma 1}} \lambda _{\Sigma \phi _L2}}{2}+\frac{19 \lambda _{{\Sigma 2}} \lambda _{\Sigma \phi _L2}}{8}+2 \lambda _{\Sigma \phi _L2}^2+32 \lambda _{{\Sigma \phi 1}} \lambda _{\phi _L{\phi 1}}+128 \theta _{{\Sigma \phi 2}} \theta _{\phi _L{\phi 2}}+128 \lambda _{{\Sigma \phi 2}} \lambda _{\phi _L{\phi 2}}+16 \lambda _{{\Sigma \phi 1}} \lambda _{\phi _L{\phi 3}} \\
(4\pi)^2\frac{d\lambda _{\Sigma \phi _R1}}{d\ln\mu} &=& \lambda _{\Sigma \phi _R1} (-\frac{9 g_{{R}}^2}{2}-\frac{141 g_4^2}{4}+34 \lambda _{{\Sigma 1}}+\frac{29 \lambda _{{\Sigma 2}}}{2}+8 y_{{N}}^2+\frac{15 y_{{R}}^2}{2}+144 \lambda _{{R1}}+96 \lambda _{{R2}}+10 y_{{RQ}}^2+4 y_{\nu }^2+3 y_{\Sigma }^2+\frac{3 y_{\psi }^2}{2})\br
+\frac{3 g_4^4}{2}+64 \lambda _{{LR1}} \lambda _{\Sigma \phi _L1}+16 \lambda _{{LR1}} \lambda _{\Sigma \phi _L2}+16 \lambda _{{LR2}} \lambda _{\Sigma \phi _L1}-\frac{5}{8} y_{{R}}^2 y_{\psi }^2+32 \lambda _{{R1}} \lambda _{\Sigma \phi _R2}+16 \lambda _{{R2}} \lambda _{\Sigma \phi _R2}-\frac{1}{2} y_{{RQ}}^2 y_{\Sigma }^2+8 \lambda _{\Sigma \phi _R1}^2\br
+\frac{15 \lambda _{{\Sigma 1}} \lambda _{\Sigma \phi _R2}}{2}+\frac{19 \lambda _{{\Sigma 2}} \lambda _{\Sigma \phi _R2}}{8}+2 \lambda _{\Sigma \phi _R2}^2+32 \lambda _{{\Sigma \phi 1}} \lambda _{\phi _R{\phi 1}}+128 \theta _{{\Sigma \phi 2}} \theta _{\phi _R{\phi 2}}+128 \lambda _{{\Sigma \phi 2}} \lambda _{\phi _R{\phi 2}}+16 \lambda _{{\Sigma \phi 1}} \lambda _{\phi _R{\phi 3}} \\
(4\pi)^2\frac{d\lambda _{{\Sigma \phi 1}}}{d\ln\mu} &=& \lambda _{{\Sigma \phi 1}} (16 y_{{c}}^2-\frac{9 g_{{L}}^2}{2}-\frac{9 g_{{R}}^2}{2}-24 g_4^2+34 \lambda _{{\Sigma 1}}+\frac{29 \lambda _{{\Sigma 2}}}{2}+80 \lambda _1-64 \lambda _2+32 \lambda _4+16 y^2+3 y_{\Sigma }^2+\frac{3 y_{\psi }^2}{2})+32 \theta _{{\Sigma \phi 2}}^2\br
+192 \theta _3 \theta _{{\Sigma \phi 2}}+8 \lambda _{{\Sigma \phi 1}}^2+32 \lambda _{{\Sigma \phi 2}}^2+192 \lambda _3 \lambda _{{\Sigma \phi 2}}+64 \lambda _{\Sigma \phi _L1} \lambda _{\phi _L{\phi 1}}+32 \lambda _{\Sigma \phi _L1} \lambda _{\phi _L{\phi 3}}+16 \lambda _{\Sigma \phi _L2} \lambda _{\phi _L{\phi 1}}+8 \lambda _{\Sigma \phi _L2} \lambda _{\phi _L{\phi 3}}\br
+64 \lambda _{\Sigma \phi _R1} \lambda _{\phi _R{\phi 1}}+32 \lambda _{\Sigma \phi _R1} \lambda _{\phi _R{\phi 3}}+16 \lambda _{\Sigma \phi _R2} \lambda _{\phi _R{\phi 1}}+8 \lambda _{\Sigma \phi _R2} \lambda _{\phi _R{\phi 3}}+8 \theta _{{X1}}^2+8 \lambda _{{X1}}^2+8 \theta _{{X2}}^2+8 \lambda _{{X2}}^2 \\
(4\pi)^2\frac{d\lambda _{\Sigma \phi _L2}}{d\ln\mu} &=& \lambda _{\Sigma \phi _L2} (8 y_{{E}}^2-\frac{9 g_{{L}}^2}{2}-\frac{141 g_4^2}{4}+4 \lambda _{{\Sigma 1}}+5 \lambda _{{\Sigma 2}}+\frac{15 y_{{L}}^2}{2}+16 \lambda _{{L1}}+32 \lambda _{{L2}}+10 y_{{LQ}}^2+16 \lambda _{\Sigma \phi _L1}+3 y_{\Sigma }^2+\frac{3 y_{\psi }^2}{2})\br
+6 g_4^4-\frac{1}{2} y_{{L}}^2 y_{\psi }^2-4 y_{{LQ}}^2 y_{\Sigma }^2+16 \lambda _{{LR2}} \lambda _{\Sigma \phi _R2}+6 \lambda _{\Sigma \phi _L2}^2+16 \theta _{{X1}}^2+16 \lambda _{{X1}}^2+16 \theta _{{X2}}^2+16 \lambda _{{X2}}^2 \\
(4\pi)^2\frac{d\lambda _{\Sigma \phi _R2}}{d\ln\mu} &=& \lambda _{\Sigma \phi _R2} (-\frac{9 g_{{R}}^2}{2}-\frac{141 g_4^2}{4}+4 \lambda _{{\Sigma 1}}+5 \lambda _{{\Sigma 2}}+8 y_{{N}}^2+\frac{15 y_{{R}}^2}{2}+16 \lambda _{{R1}}+32 \lambda _{{R2}}+10 y_{{RQ}}^2+16 \lambda _{\Sigma \phi _R1}+4 y_{\nu }^2+3 y_{\Sigma }^2+\frac{3 y_{\psi }^2}{2})\br
+6 g_4^4+16 \lambda _{{LR2}} \lambda _{\Sigma \phi _L2}-\frac{1}{2} y_{{R}}^2 y_{\psi }^2-4 y_{{RQ}}^2 y_{\Sigma }^2+6 \lambda _{\Sigma \phi _R2}^2+16 \theta _{{X1}}^2+16 \lambda _{{X1}}^2+16 \theta _{{X2}}^2+16 \lambda _{{X2}}^2 \\
(4\pi)^2\frac{d\lambda _{{\Sigma \phi 2}}}{d\ln\mu} &=& \lambda _{{\Sigma \phi 2}} (16 y_{{c}}^2-\frac{9 g_{{L}}^2}{2}-\frac{9 g_{{R}}^2}{2}-24 g_4^2+34 \lambda _{{\Sigma 1}}+\frac{29 \lambda _{{\Sigma 2}}}{2}+16 \lambda _{{\Sigma \phi 1}}+16 \lambda _1+64 \lambda _2+64 \lambda _4+16 y^2+3 y_{\Sigma }^2+\frac{3 y_{\psi }^2}{2})
\br
+192 \theta _2 \theta _{{\Sigma \phi 2}}+48 \lambda _3 \lambda _{{\Sigma \phi 1}}+64 \lambda _{\Sigma \phi _L1} \lambda _{\phi _L{\phi 2}}+16 \lambda _{\Sigma \phi _L2} \lambda _{\phi _L{\phi 2}}+64 \lambda _{\Sigma \phi _R1} \lambda _{\phi _R{\phi 2}}+16 \lambda _{\Sigma \phi _R2} \lambda _{\phi _R{\phi 2}}+8 \theta _{{X1}} \theta _{{X2}}+8 \lambda _{{X1}} \lambda _{{X2}} \\
(4\pi)^2\frac{d\theta _{{\Sigma \phi 2}}}{d\ln\mu} &=& \theta _{{\Sigma \phi 2}} (16 y_{{c}}^2-\frac{9 g_{{L}}^2}{2}-\frac{9 g_{{R}}^2}{2}-24 g_4^2+34 \lambda _{{\Sigma 1}}+\frac{29 \lambda _{{\Sigma 2}}}{2}+16 \lambda _{{\Sigma \phi 1}}+16 \lambda _1-320 \lambda _2+64 \lambda _4+16 y^2+3 y_{\Sigma }^2+\frac{3 y_{\psi }^2}{2})\br
+48 \theta _3 \lambda _{{\Sigma \phi 1}}+192 \theta _2 \lambda _{{\Sigma \phi 2}}+64 \lambda _{\Sigma \phi _L1} \theta _{\phi _L{\phi 2}}+16 \lambda _{\Sigma \phi _L2} \theta _{\phi _L{\phi 2}}+64 \lambda _{\Sigma \phi _R1} \theta _{\phi _R{\phi 2}}+16 \lambda _{\Sigma \phi _R2} \theta _{\phi _R{\phi 2}}+8 \lambda _{{X1}} \theta _{{X2}}-8 \theta _{{X1}} \lambda _{{X2}} \\
(4\pi)^2\frac{d\lambda _{{B}}}{d\ln\mu} &=& \lambda _{{B}} (8 y_{{E}}^2-\frac{9 g_{{L}}^2}{2}-\frac{9 g_{{R}}^2}{2}-\frac{45 g_4^2}{2}+\frac{15 y_{{L}}^2}{2}+16 \lambda _{{L1}}-16 \lambda _{{L2}}+10 y_{{LQ}}^2\br
+32 \lambda _{{LR1}}-32 \lambda _{{LR2}}+8 y_{{N}}^2+\frac{15 y_{{R}}^2}{2}+16 \lambda _{{R1}}-16 \lambda _{{R2}}+10 y_{{RQ}}^2+4 y_{\nu }^2) \\
(4\pi)^2\frac{d\theta _{{B}}}{d\ln\mu} &=& \theta _{{B}} (8 y_{{E}}^2-\frac{9 g_{{L}}^2}{2}-\frac{9 g_{{R}}^2}{2}-\frac{45 g_4^2}{2}+\frac{15 y_{{L}}^2}{2}+16 \lambda _{{L1}}-16 \lambda _{{L2}}+10 y_{{LQ}}^2\br
+32 \lambda _{{LR1}}-32 \lambda _{{LR2}}+8 y_{{N}}^2+\frac{15 y_{{R}}^2}{2}+16 \lambda _{{R1}}-16 \lambda _{{R2}}+10 y_{{RQ}}^2+4 y_{\nu }^2) \\
(4\pi)^2\frac{d\lambda _{{L1}}}{d\ln\mu} &=& 128 \theta _{{B}}^2+128 \lambda _{{B}}^2+\lambda _{{L1}} (16 y_{{E}}^2-9 g_{{L}}^2-\frac{45 g_4^2}{2}+15 y_{{L}}^2+192 \lambda _{{L2}}+20 y_{{LQ}}^2)+\frac{27}{32} g_4^2 g_{{L}}^2+\frac{9 g_{{L}}^4}{32}+\frac{27 g_4^4}{128}-5 y_{{L}}^2 y_{{LQ}}^2-\frac{y_{{L}}^4}{32}\br
+192 \lambda _{{L1}}^2+48 \lambda _{{L2}}^2-\frac{y_{{LQ}}^4}{2}+32 \lambda _{{LR1}}^2+16 \lambda _{{LR1}} \lambda _{{LR2}}+32 \theta _{{LR3}}^2+32 \lambda _{{LR3}}^2+\frac{15 \lambda _{\Sigma \phi _L1}^2}{2}+\frac{15 \lambda _{\Sigma \phi _L1} \lambda _{\Sigma \phi _L2}}{4}\br
+\frac{9 \lambda _{\Sigma \phi _L2}^2}{32}+16 \lambda _{\phi _L{\phi 1}}^2+16 \lambda _{\phi _L{\phi 1}} \lambda _{\phi _L{\phi 3}}+64 \theta _{\phi _L{\phi 2}}^2+64 \lambda _{\phi _L{\phi 2}}^2 \\
(4\pi)^2\frac{d\lambda _{{L2}}}{d\ln\mu} &=& -128 \theta _{{B}}^2-128 \lambda _{{B}}^2+\lambda _{{L2}} (16 y_{{E}}^2-9 g_{{L}}^2-\frac{45 g_4^2}{2}+15 y_{{L}}^2+96 \lambda _{{L1}}+20 y_{{LQ}}^2)-4 y_{{E}}^4-\frac{9}{16} g_4^2 g_{{L}}^2\br
+\frac{9 g_4^4}{16}+5 y_{{L}}^2 y_{{LQ}}^2-\frac{7 y_{{L}}^4}{4}+96 \lambda _{{L2}}^2-3 y_{{LQ}}^4+8 \lambda _{{LR2}}^2-32 \theta _{{LR3}}^2-32 \lambda _{{LR3}}^2+\frac{3 \lambda _{\Sigma \phi _L2}^2}{4}+8 \lambda _{\phi _L{\phi 3}}^2 \\
(4\pi)^2\frac{d\lambda _{{R1}}}{d\ln\mu} &=& 128 \theta _{{B}}^2+128 \lambda _{{B}}^2+\lambda _{{R1}} (-9 g_{{R}}^2-\frac{45 g_4^2}{2}+16 y_{{N}}^2+15 y_{{R}}^2+192 \lambda _{{R2}}+20 y_{{RQ}}^2+8 y_{\nu }^2)+\frac{27}{32} g_4^2 g_{{R}}^2+\frac{9 g_{{R}}^4}{32}+\frac{27 g_4^4}{128}+32 \lambda _{{LR1}}^2\br
+16 \lambda _{{LR1}} \lambda _{{LR2}}
+32 \theta _{{LR3}}^2+32 \lambda _{{LR3}}^2-5 y_{{R}}^2 y_{{RQ}}^2+\frac{1}{2} y_{{R}}^2 y_{\nu }^2-\frac{y_{{R}}^4}{32}+192 \lambda _{{R1}}^2+48 \lambda _{{R2}}^2-\frac{y_{{RQ}}^4}{2}\br
+\frac{15 \lambda _{\Sigma \phi _R1}^2}{2}+\frac{15 \lambda _{\Sigma \phi _R1} \lambda _{\Sigma \phi _R2}}{4}+\frac{9 \lambda _{\Sigma \phi _R2}^2}{32}+16 \lambda _{\phi _R{\phi 1}}^2+16 \lambda _{\phi _R{\phi 1}} \lambda _{\phi _R{\phi 3}}+64 \theta _{\phi _R{\phi 2}}^2+64 \lambda _{\phi _R{\phi 2}}^2-2 y_{\nu }^4 \\
(4\pi)^2\frac{d\lambda _{{R2}}}{d\ln\mu} &=& -128 \theta _{{B}}^2-128 \lambda _{{B}}^2+\lambda _{{R2}} (-9 g_{{R}}^2-\frac{45 g_4^2}{2}+16 y_{{N}}^2+15 y_{{R}}^2+96 \lambda _{{R1}}+20 y_{{RQ}}^2+8 y_{\nu }^2)-\frac{9}{16} g_4^2 g_{{R}}^2\br
+\frac{9 g_4^4}{16}+8 \lambda _{{LR2}}^2-32 \theta _{{LR3}}^2-32 \lambda _{{LR3}}^2-4 y_{{N}}^4+5 y_{{R}}^2 y_{{RQ}}^2-2 y_{{R}}^2 y_{\nu }^2-\frac{7 y_{{R}}^4}{4}+96 \lambda _{{R2}}^2-3 y_{{RQ}}^4+\frac{3 \lambda _{\Sigma \phi _R2}^2}{4}+8 \lambda _{\phi _R{\phi 3}}^2 \\
(4\pi)^2\frac{d\lambda _{{X1}}}{d\ln\mu} &=& \lambda _{{X1}} (8 y_{{c}}^2+4 y_{{E}}^2-\frac{9 g_{{L}}^2}{2}-\frac{9 g_{{R}}^2}{2}-\frac{93 g_4^2}{4}+4 \lambda _{{\Sigma \phi 1}}+\frac{15 y_{{L}}^2}{4}+5 y_{{LQ}}^2+8 \lambda _{{LR1}}+4 y_{{N}}^2+\frac{15 y_{{R}}^2}{4}+5 y_{{RQ}}^2+4 \lambda _{\Sigma \phi _L1}+7 \lambda _{\Sigma \phi _L2}\br
+4 \lambda _{\Sigma \phi _R1}+7 \lambda _{\Sigma \phi _R2}+8 \lambda _{\phi _L{\phi 1}}+16 \lambda _{\phi _L{\phi 3}}+8 \lambda _{\phi _R{\phi 1}}+16 \lambda _{\phi _R{\phi 3}}+8 y^2+2 y_{\nu }^2+\frac{3 y_{\Sigma }^2}{2}+\frac{3 y_{\psi }^2}{4})
-6 y_{{c}} y_{{LQ}} y_{{RQ}} y_{\Sigma }\br
-3 y y_{{L}} y_{{R}} y_{\psi }-16 \theta _{{LR3}} \theta _{{X2}}-16 \lambda _{{LR3}} \lambda _{{X2}}+16 \theta _{\phi _L{\phi 2}} \theta _{{X2}}+16 \lambda _{\phi _L{\phi 2}} \lambda _{{X2}}+16 \theta _{\phi _R{\phi 2}} \theta _{{X2}}+16 \lambda _{\phi _R{\phi 2}} \lambda _{{X2}}+8 \theta _{{\Sigma \phi 2}} \theta _{{X2}}+8 \lambda _{{\Sigma \phi 2}} \lambda _{{X2}} \\
(4\pi)^2\frac{d\theta _{{X1}}}{d\ln\mu} &=& \theta _{{X1}} (8 y_{{c}}^2+4 y_{{E}}^2-\frac{9 g_{{L}}^2}{2}-\frac{9 g_{{R}}^2}{2}-\frac{93 g_4^2}{4}+4 \lambda _{{\Sigma \phi 1}}+\frac{15 y_{{L}}^2}{4}+5 y_{{LQ}}^2+8 \lambda _{{LR1}}+4 y_{{N}}^2+\frac{15 y_{{R}}^2}{4}+5 y_{{RQ}}^2+4 \lambda _{\Sigma \phi _L1}+7 \lambda _{\Sigma \phi _L2}\br
+4 \lambda _{\Sigma \phi _R1}+7 \lambda _{\Sigma \phi _R2}+8 \lambda _{\phi _L{\phi 1}}+16 \lambda _{\phi _L{\phi 3}}+8 \lambda _{\phi _R{\phi 1}}+16 \lambda _{\phi _R{\phi 3}}+8 y^2+2 y_{\nu }^2+\frac{3 y_{\Sigma }^2}{2}+\frac{3 y_{\psi }^2}{4})+16 \lambda _{{LR3}} \theta _{{X2}}\br
-16 \theta _{{LR3}} \lambda _{{X2}}+16 \lambda _{\phi _L{\phi 2}} \theta _{{X2}}-16 \theta _{\phi _L{\phi 2}} \lambda _{{X2}}+16 \lambda _{\phi _R{\phi 2}} \theta _{{X2}}-16 \theta _{\phi _R{\phi 2}} \lambda _{{X2}}-8 \theta _{{\Sigma \phi 2}} \lambda _{{X2}}+8 \lambda _{{\Sigma \phi 2}} \theta _{{X2}} \\
(4\pi)^2\frac{d\lambda _{{X2}}}{d\ln\mu} &=& \lambda _{{X2}} (8 y_{{c}}^2+4 y_{{E}}^2-\frac{9 g_{{L}}^2}{2}-\frac{9 g_{{R}}^2}{2}-\frac{93 g_4^2}{4}+4 \lambda _{{\Sigma \phi 1}}+\frac{15 y_{{L}}^2}{4}+5 y_{{LQ}}^2+8 \lambda _{{LR1}}+4 y_{{N}}^2+\frac{15 y_{{R}}^2}{4}+5 y_{{RQ}}^2
+4 \lambda _{\Sigma \phi _L1}+7 \lambda _{\Sigma \phi _L2}\br
+4 \lambda _{\Sigma \phi _R1}+7 \lambda _{\Sigma \phi _R2}+8 \lambda _{\phi _L{\phi 1}}-8 \lambda _{\phi _L{\phi 3}}+8 \lambda _{\phi _R{\phi 1}}-8 \lambda _{\phi _R{\phi 3}}+8 y^2+2 y_{\nu }^2+\frac{3 y_{\Sigma }^2}{2}+\frac{3 y_{\psi }^2}{4})
-3 y_{{c}} y_{{L}} y_{{R}} y_{\psi }-6 y y_{{LQ}} y_{{RQ}} y_{\Sigma }\br -16 \theta _{{LR3}} \theta _{{X1}}
-16 \lambda _{{LR3}} \lambda _{{X1}}-16 \theta _{\phi _L{\phi 2}} \theta _{{X1}}+16 \lambda _{\phi _L{\phi 2}} \lambda _{{X1}}-16 \theta _{\phi _R{\phi 2}} \theta _{{X1}}+16 \lambda _{\phi _R{\phi 2}} \lambda _{{X1}}-8 \theta _{{\Sigma \phi 2}} \theta _{{X1}}+8 \lambda _{{\Sigma \phi 2}} \lambda _{{X1}} \\
(4\pi)^2\frac{d\theta _{{X2}}}{d\ln\mu} &=& \theta _{{X2}} (8 y_{{c}}^2+4 y_{{E}}^2-\frac{9 g_{{L}}^2}{2}-\frac{9 g_{{R}}^2}{2}-\frac{93 g_4^2}{4}+4 \lambda _{{\Sigma \phi 1}}+\frac{15 y_{{L}}^2}{4}+5 y_{{LQ}}^2+8 \lambda _{{LR1}}+4 y_{{N}}^2+\frac{15 y_{{R}}^2}{4}+5 y_{{RQ}}^2+4 \lambda _{\Sigma \phi _L1}+7 \lambda _{\Sigma \phi _L2}\br
+4 \lambda _{\Sigma \phi _R1}+7 \lambda _{\Sigma \phi _R2}+8 \lambda _{\phi _L{\phi 1}}-8 \lambda _{\phi _L{\phi 3}}+8 \lambda _{\phi _R{\phi 1}}-8 \lambda _{\phi _R{\phi 3}}+8 y^2+2 y_{\nu }^2+\frac{3 y_{\Sigma }^2}{2}+\frac{3 y_{\psi }^2}{4})
\br
+16 \lambda _{{LR3}} \theta _{{X1}}-16 \theta _{{LR3}} \lambda _{{X1}}+16 \lambda _{\phi _L{\phi 2}} \theta _{{X1}}+16 \theta _{\phi _L{\phi 2}} \lambda _{{X1}}+16 \lambda _{\phi _R{\phi 2}} \theta _{{X1}}+16 \theta _{\phi _R{\phi 2}} \lambda _{{X1}}+8 \theta _{{\Sigma \phi 2}} \lambda _{{X1}}+8 \lambda _{{\Sigma \phi 2}} \theta _{{X1}} \\
(4\pi)^2\frac{d\lambda _{{\Sigma 1}}}{d\ln\mu} &=& \lambda _{{\Sigma 1}} (-48 g_4^2+29 \lambda _{{\Sigma 2}}+6 y_{\Sigma }^2+3 y_{\psi }^2)+9 g_4^4+64 \theta _{{\Sigma \phi 2}}^2+46 \lambda _{{\Sigma 1}}^2+\frac{57 \lambda _{{\Sigma 2}}^2}{8}+16 \lambda _{{\Sigma \phi 1}}^2\br
+64 \lambda _{{\Sigma \phi 2}}^2+32 \lambda _{\Sigma \phi _L1}^2+16 \lambda _{\Sigma \phi _L1} \lambda _{\Sigma \phi _L2}+32 \lambda _{\Sigma \phi _R1}^2+16 \lambda _{\Sigma \phi _R1} \lambda _{\Sigma \phi _R2}-\frac{3 y_{\Sigma }^4}{8}-\frac{y_{\psi }^4}{2} \\
(4\pi)^2\frac{d\lambda _{{\Sigma 2}}}{d\ln\mu} &=& \lambda _{{\Sigma 2}} (-48 g_4^2+24 \lambda _{{\Sigma 1}}+6 y_{\Sigma }^2+3 y_{\psi }^2)+12 g_4^4+7 \lambda _{{\Sigma 2}}^2+8 \lambda _{\Sigma \phi _L2}^2+8 \lambda _{\Sigma \phi _R2}^2-\frac{3 y_{\Sigma }^4}{2}+\frac{y_{\psi }^4}{2} \\
(4\pi)^2\frac{d\lambda _1}{d\ln\mu} &=& \lambda _1 (32 y_{{c}}^2-9 g_{{L}}^2-9 g_{{R}}^2-128 \lambda _2+64 \lambda _4+32 y^2)-8 y_{{c}}^4+\frac{3}{16} g_{{L}}^2 g_{{R}}^2+\frac{9 g_{{L}}^4}{32}+\frac{9 g_{{R}}^4}{32}+256 \theta _2^2+192 \theta _3^2+\frac{15 \lambda _{{\Sigma \phi 1}}^2}{2}+128 \lambda _1^2\br
+512 \lambda _2^2+192 \lambda _3^2
+64 \lambda _4^2-256 \lambda _2 \lambda _4+32 \lambda _{\phi _L{\phi 1}}^2+32 \lambda _{\phi _L{\phi 1}} \lambda _{\phi _L{\phi 3}}+16 \lambda _{\phi _L{\phi 3}}^2+32 \lambda _{\phi _R{\phi 1}}^2+32 \lambda _{\phi _R{\phi 1}} \lambda _{\phi _R{\phi 3}}+16 \lambda _{\phi _R{\phi 3}}^2-8 y^4 \\
(4\pi)^2\frac{d\lambda _2}{d\ln\mu} &=& \lambda _2 (32 y_{{c}}^2-9 g_{{L}}^2-9 g_{{R}}^2+96 \lambda _1+192 \lambda _4+32 y^2)-4 y^2 y_{{c}}^2-\frac{15 \theta _{{\Sigma \phi 2}}^2}{2}-48 \theta _3^2\br
+\frac{15 \lambda _{{\Sigma \phi 2}}^2}{2}-384 \lambda _2^2+48 \lambda _3^2-32 \theta _{\phi _L{\phi 2}}^2+32 \lambda _{\phi _L{\phi 2}}^2-32 \theta _{\phi _R{\phi 2}}^2+32 \lambda _{\phi _R{\phi 2}}^2 \\
(4\pi)^2\frac{d\theta _2}{d\ln\mu} &=& \theta _2 (32 y_{{c}}^2-9 g_{{L}}^2-9 g_{{R}}^2+96 \lambda _1-384 \lambda _2+192 \lambda _4+32 y^2)+15 \theta _{{\Sigma \phi 2}} \lambda _{{\Sigma \phi 2}}+96 \theta _3 \lambda _3+64 \theta _{\phi _L{\phi 2}} \lambda _{\phi _L{\phi 2}}+64 \theta _{\phi _R{\phi 2}} \lambda _{\phi _R{\phi 2}} \\
(4\pi)^2\frac{d\lambda _3}{d\ln\mu} &=& \lambda _3 (32 y_{{c}}^2-9 g_{{L}}^2-9 g_{{R}}^2+192 \lambda _1+192 \lambda _4+32 y^2)-8 y^3 y_{{c}}-8 y y_{{c}}^3+384 \theta _2 \theta _3+15 \lambda _{{\Sigma \phi 1}} \lambda _{{\Sigma \phi 2}}\br
+64 \lambda _{\phi _L{\phi 1}} \lambda _{\phi _L{\phi 2}}+32 \lambda _{\phi _L{\phi 2}} \lambda _{\phi _L{\phi 3}}+64 \lambda _{\phi _R{\phi 1}} \lambda _{\phi _R{\phi 2}}+32 \lambda _{\phi _R{\phi 2}} \lambda _{\phi _R{\phi 3}} \\
(4\pi)^2\frac{d\theta _3}{d\ln\mu} &=& \theta _3 (32 y_{{c}}^2-9 g_{{L}}^2-9 g_{{R}}^2+192 \lambda _1-768 \lambda _2+192 \lambda _4+32 y^2)+15 \theta _{{\Sigma \phi 2}} \lambda _{{\Sigma \phi 1}}+384 \theta _2 \lambda _3+\br
64 \lambda _{\phi _L{\phi 1}} \theta _{\phi _L{\phi 2}}+32 \theta _{\phi _L{\phi 2}} \lambda _{\phi _L{\phi 3}}+64 \lambda _{\phi _R{\phi 1}} \theta _{\phi _R{\phi 2}}+32 \theta _{\phi _R{\phi 2}} \lambda _{\phi _R{\phi 3}} \\
(4\pi)^2\frac{d\lambda _4}{d\ln\mu} &=& \lambda _4 (32 y_{{c}}^2-9 g_{{L}}^2-9 g_{{R}}^2+96 \lambda _1+128 \lambda _2+32 y^2)-24 y^2 y_{{c}}^2+4 y_{{c}}^4+\frac{3}{8} g_{{L}}^2 g_{{R}}^2+512 \theta _2^2+\br
30 \lambda _{{\Sigma \phi 2}}^2+192 \lambda _3^2+64 \lambda _4^2+128 \lambda _{\phi _L{\phi 2}}^2-8 \lambda _{\phi _L{\phi 3}}^2+128 \lambda _{\phi _R{\phi 2}}^2-8 \lambda _{\phi _R{\phi 3}}^2+4 y^4 \\
(4\pi)^2\frac{d\lambda _{\phi _L{\phi 1}}}{d\ln\mu} &=& \lambda _{\phi _L{\phi 1}} (16 y_{{c}}^2+8 y_{{E}}^2-9 g_{{L}}^2-\frac{9 g_{{R}}^2}{2}-\frac{45 g_4^2}{4}+80 \lambda _1-64 \lambda _2+32 \lambda _4+\frac{15 y_{{L}}^2}{2}+144 \lambda _{{L1}}+96 \lambda _{{L2}}+10 y_{{LQ}}^2+16 y^2)\br
-4 y_{{c}}^2 y_{{E}}^2-\frac{15}{2} y_{{c}}^2 y_{{L}}^2-4 y^2 y_{{E}}^2+\frac{9 g_{{L}}^4}{16}+64 \lambda _{{L1}} \lambda _{\phi _L{\phi 3}}+16 \lambda _{{L2}} \lambda _{\phi _L{\phi 3}}-10 y^2 y_{{LQ}}^2+64 \lambda _{{LR1}} \lambda _{\phi _R{\phi 1}}+32 \lambda _{{LR1}} \lambda _{\phi _R{\phi 3}}
\br
+16 \lambda _{{LR2}} \lambda _{\phi _R{\phi 1}}+8 \lambda _{{LR2}} \lambda _{\phi _R{\phi 3}}+15 \lambda _{{\Sigma \phi 1}} \lambda _{\Sigma \phi _L1}+\frac{15 \lambda _{{\Sigma \phi 1}} \lambda _{\Sigma \phi _L2}}{4}+16 \lambda _{\phi _L{\phi 1}}^2+64 \theta _{\phi _L{\phi 2}}^2+192 \theta _3 \theta _{\phi _L{\phi 2}}\br
+64 \lambda _{\phi _L{\phi 2}}^2+192 \lambda _3 \lambda _{\phi _L{\phi 2}}+8 \lambda _{\phi _L{\phi 3}}^2+32 \lambda _1 \lambda _{\phi _L{\phi 3}}-64 \lambda _2 \lambda _{\phi _L{\phi 3}}+32 \lambda _4 \lambda _{\phi _L{\phi 3}}+\frac{15 \theta _{{X2}}^2}{2}+\frac{15 \lambda _{{X2}}^2}{2} \\
(4\pi)^2\frac{d\lambda _{\phi _L{\phi 2}}}{d\ln\mu} &=& \lambda _{\phi _L{\phi 2}} (16 y_{{c}}^2+8 y_{{E}}^2-9 g_{{L}}^2-\frac{9 g_{{R}}^2}{2}-\frac{45 g_4^2}{4}+16 \lambda _1+64 \lambda _2+64 \lambda _4+\frac{15 y_{{L}}^2}{2}+144 \lambda _{{L1}}+96 \lambda _{{L2}}+10 y_{{LQ}}^2
\br
+32 \lambda _{\phi _L{\phi 1}}+16 \lambda _{\phi _L{\phi 3}}+16 y^2)
-4 y y_{{c}} y_{{E}}^2-\frac{15}{4} y y_{{c}} y_{{L}}^2-5 y y_{{c}} y_{{LQ}}^2+64 \lambda _{{LR1}} \lambda _{\phi _R{\phi 2}}+16 \lambda _{{LR2}} \lambda _{\phi _R{\phi 2}}
+15 \lambda _{{\Sigma \phi 2}} \lambda _{\Sigma \phi _L1}\br
+\frac{15 \lambda _{{\Sigma \phi 2}} \lambda _{\Sigma \phi _L2}}{4}+48 \lambda _3 \lambda _{\phi _L{\phi 1}}+192 \theta _2 \theta _{\phi _L{\phi 2}}+24 \lambda _3 \lambda _{\phi _L{\phi 3}}+\frac{15 \theta _{{X1}} \theta _{{X2}}}{4}+\frac{15 \lambda _{{X1}} \lambda _{{X2}}}{4} \\
(4\pi)^2\frac{d\theta _{\phi _L{\phi 2}}}{d\ln\mu} &=& \theta _{\phi _L{\phi 2}} (16 y_{{c}}^2+8 y_{{E}}^2-9 g_{{L}}^2-\frac{9 g_{{R}}^2}{2}-\frac{45 g_4^2}{4}+16 \lambda _1-320 \lambda _2+64 \lambda _4+\frac{15 y_{{L}}^2}{2}+144 \lambda _{{L1}}+96 \lambda _{{L2}}+10 y_{{LQ}}^2+32 \lambda _{\phi _L{\phi 1}}+16 \lambda _{\phi _L{\phi 3}}+16 y^2)\br
+64 \lambda _{{LR1}} \theta _{\phi _R{\phi 2}}+16 \lambda _{{LR2}} \theta _{\phi _R{\phi 2}}+15 \theta _{{\Sigma \phi 2}} \lambda _{\Sigma \phi _L1}+\frac{15 \theta _{{\Sigma \phi 2}} \lambda _{\Sigma \phi _L2}}{4}\br
+48 \theta _3 \lambda _{\phi _L{\phi 1}}+192 \theta _2 \lambda _{\phi _L{\phi 2}}+24 \theta _3 \lambda _{\phi _L{\phi 3}}+\frac{15 \lambda _{{X1}} \theta _{{X2}}}{4}-\frac{15 \theta _{{X1}} \lambda _{{X2}}}{4} \\
(4\pi)^2\frac{d\lambda _{\phi _L{\phi 3}}}{d\ln\mu} &=& \lambda _{\phi _L{\phi 3}} (16 y_{{c}}^2+8 y_{{E}}^2-9 g_{{L}}^2-\frac{9 g_{{R}}^2}{2}-\frac{45 g_4^2}{4}+16 \lambda _1+64 \lambda _2-32 \lambda _4+\frac{15 y_{{L}}^2}{2}+16 \lambda _{{L1}}+64 \lambda _{{L2}}+10 y_{{LQ}}^2+32 \lambda _{\phi _L{\phi 1}}+16 y^2)\br
+\frac{15}{2} y_{{c}}^2 y_{{L}}^2-10 y_{{c}}^2 y_{{LQ}}^2-\frac{15}{2} y^2 y_{{L}}^2+10 y^2 y_{{LQ}}^2+16 \lambda _{\phi _L{\phi 3}}^2+\frac{15 \theta _{{X1}}^2}{2}+\frac{15 \lambda _{{X1}}^2}{2}-\frac{15 \theta _{{X2}}^2}{2}-\frac{15 \lambda _{{X2}}^2}{2} \\
(4\pi)^2\frac{d\lambda _{{LR1}}}{d\ln\mu} &=& \lambda _{{LR1}} (8 y_{{E}}^2-\frac{9 g_{{L}}^2}{2}-\frac{9 g_{{R}}^2}{2}-\frac{45 g_4^2}{2}+\frac{15 y_{{L}}^2}{2}+144 \lambda _{{L1}}+96 \lambda _{{L2}}+10 y_{{LQ}}^2+8 y_{{N}}^2+\frac{15 y_{{R}}^2}{2}+144 \lambda _{{R1}}+96 \lambda _{{R2}}+10 y_{{RQ}}^2+4 y_{\nu }^2)
\br
+256 \theta _{{B}}^2+256 \lambda _{{B}}^2-4 y_{{E}} y_{{L}} y_{{N}} y_{{R}}-4 y_{{E}}^2 y_{{N}}^2-2 y_{{E}}^2 y_{{R}}^2-2 y_{{E}}^2 y_{{RQ}}^2+\frac{27 g_4^4}{64}-2 y_{{L}}^2 y_{{N}}^2-\frac{17}{16} y_{{L}}^2 y_{{R}}^2+32 \lambda _{{L1}} \lambda _{{LR2}}
\br
+16 \lambda _{{L2}} \lambda _{{LR2}}-2 y_{{LQ}}^2 y_{{N}}^2+16 \lambda _{{LR1}}^2+8 \lambda _{{LR2}}^2+32 \lambda _{{LR2}} \lambda _{{R1}}+16 \lambda _{{LR2}} \lambda _{{R2}}+32 \theta _{{LR3}}^2+32 \lambda _{{LR3}}^2+15 \lambda _{\Sigma \phi _L1} \lambda _{\Sigma \phi _R1}\br
+\frac{15 \lambda _{\Sigma \phi _L1} \lambda _{\Sigma \phi _R2}}{4}+\frac{15 \lambda _{\Sigma \phi _L2} \lambda _{\Sigma \phi _R1}}{4}+\frac{9 \lambda _{\Sigma \phi _L2} \lambda _{\Sigma \phi _R2}}{16}+32 \lambda _{\phi _L{\phi 1}} \lambda _{\phi _R{\phi 1}}+16 \lambda _{\phi _L{\phi 1}} \lambda _{\phi _R{\phi 3}}
\br
+128 \theta _{\phi _L{\phi 2}} \theta _{\phi _R{\phi 2}}+128 \lambda _{\phi _L{\phi 2}} \lambda _{\phi _R{\phi 2}}+16 \lambda _{\phi _L{\phi 3}} \lambda _{\phi _R{\phi 1}}+8 \lambda _{\phi _L{\phi 3}} \lambda _{\phi _R{\phi 3}}+2 \theta _{{X1}}^2+2 \lambda _{{X1}}^2+2 \theta _{{X2}}^2+2 \lambda _{{X2}}^2 \\
(4\pi)^2\frac{d\lambda _{{LR2}}}{d\ln\mu} &=& \lambda _{{LR2}} (8 y_{{E}}^2-\frac{9 g_{{L}}^2}{2}-\frac{9 g_{{R}}^2}{2}-\frac{45 g_4^2}{2}+\frac{15 y_{{L}}^2}{2}+16 \lambda _{{L1}}+32 \lambda _{{L2}}+10 y_{{LQ}}^2+32 \lambda _{{LR1}}+8 y_{{N}}^2+\frac{15 y_{{R}}^2}{2}+16 \lambda _{{R1}}+32 \lambda _{{R2}}+10 y_{{RQ}}^2+4 y_{\nu }^2)\br
-256 \theta _{{B}}^2-256 \lambda _{{B}}^2+y_{{E}} y_{{L}} y_{{N}} y_{{R}}+\frac{1}{2} y_{{E}}^2 y_{{R}}^2-2 y_{{E}}^2 y_{{RQ}}^2-4 y_{{E}}^2 y_{\nu }^2+\frac{9 g_4^4}{8}+\frac{1}{2} y_{{L}}^2 y_{{N}}^2+\frac{1}{2} y_{{L}}^2 y_{{R}}^2-2 y_{{LQ}}^2 y_{{N}}^2+32 \lambda _{{LR2}}^2\br
+64 \theta _{{LR3}}^2+64 \lambda _{{LR3}}^2+\frac{3 \lambda _{\Sigma \phi _L2} \lambda _{\Sigma \phi _R2}}{2}-\frac{\theta _{{X1}}^2}{2}-\frac{\lambda _{{X1}}^2}{2}-\frac{\theta _{{X2}}^2}{2}-\frac{\lambda _{{X2}}^2}{2} \\
(4\pi)^2\frac{d\lambda _{{LR3}}}{d\ln\mu} &=& \lambda _{{LR3}} (8 y_{{E}}^2-\frac{9 g_{{L}}^2}{2}-\frac{9 g_{{R}}^2}{2}-\frac{45 g_4^2}{2}+\frac{15 y_{{L}}^2}{2}+16 \lambda _{{L1}}-16 \lambda _{{L2}}+10 y_{{LQ}}^2+32 \lambda _{{LR1}}\br
+48 \lambda _{{LR2}}+8 y_{{N}}^2+\frac{15 y_{{R}}^2}{2}+16 \lambda _{{R1}}-16 \lambda _{{R2}}+10 y_{{RQ}}^2+4 y_{\nu }^2)+\frac{5 \theta _{{X1}} \theta _{{X2}}}{2}-\frac{5 \lambda _{{X1}} \lambda _{{X2}}}{2} \\
(4\pi)^2\frac{d\theta _{{LR3}}}{d\ln\mu} &=& \theta _{{LR3}} (8 y_{{E}}^2-\frac{9 g_{{L}}^2}{2}-\frac{9 g_{{R}}^2}{2}-\frac{45 g_4^2}{2}+\frac{15 y_{{L}}^2}{2}+16 \lambda _{{L1}}-16 \lambda _{{L2}}+10 y_{{LQ}}^2+32 \lambda _{{LR1}}\br
+48 \lambda _{{LR2}}+8 y_{{N}}^2+\frac{15 y_{{R}}^2}{2}+16 \lambda _{{R1}}-16 \lambda _{{R2}}+10 y_{{RQ}}^2+4 y_{\nu }^2)-\frac{5 \lambda _{{X1}} \theta _{{X2}}}{2}-\frac{5 \theta _{{X1}} \lambda _{{X2}}}{2} \\
(4\pi)^2\frac{d\lambda _{\phi _R{\phi 1}}}{d\ln\mu} &=& \lambda _{\phi _R{\phi 1}} (16 y_{{c}}^2-\frac{9 g_{{L}}^2}{2}-9 g_{{R}}^2-\frac{45 g_4^2}{4}+80 \lambda _1-64 \lambda _2+32 \lambda _4+8 y_{{N}}^2+\frac{15 y_{{R}}^2}{2}+144 \lambda _{{R1}}+96 \lambda _{{R2}}+10 y_{{RQ}}^2+16 y^2+4 y_{\nu }^2)\br
-4 y_{{c}}^2 y_{{N}}^2-\frac{15}{2} y_{{c}}^2 y_{{R}}^2-4 y_{{c}}^2 y_{\nu }^2+\frac{9 g_{{R}}^4}{16}+64 \lambda _{{LR1}} \lambda _{\phi _L{\phi 1}}+32 \lambda _{{LR1}} \lambda _{\phi _L{\phi 3}}+16 \lambda _{{LR2}} \lambda _{\phi _L{\phi 1}}+8 \lambda _{{LR2}} \lambda _{\phi _L{\phi 3}}-4 y^2 y_{{N}}^2\br
+64 \lambda _{{R1}} \lambda _{\phi _R{\phi 3}}+16 \lambda _{{R2}} \lambda _{\phi _R{\phi 3}}-10 y^2 y_{{RQ}}^2+15 \lambda _{{\Sigma \phi 1}} \lambda _{\Sigma \phi _R1}+\frac{15 \lambda _{{\Sigma \phi 1}} \lambda _{\Sigma \phi _R2}}{4}+16 \lambda _{\phi _R{\phi 1}}^2+64 \theta _{\phi _R{\phi 2}}^2+192 \theta _3 \theta _{\phi _R{\phi 2}}\br
+64 \lambda _{\phi _R{\phi 2}}^2+192 \lambda _3 \lambda _{\phi _R{\phi 2}}+8 \lambda _{\phi _R{\phi 3}}^2+32 \lambda _1 \lambda _{\phi _R{\phi 3}}-64 \lambda _2 \lambda _{\phi _R{\phi 3}}+32 \lambda _4 \lambda _{\phi _R{\phi 3}}+\frac{15 \theta _{{X2}}^2}{2}+\frac{15 \lambda _{{X2}}^2}{2} \\
(4\pi)^2\frac{d\lambda _{\phi _R{\phi 2}}}{d\ln\mu} &=& \lambda _{\phi _R{\phi 2}} (16 y_{{c}}^2-\frac{9 g_{{L}}^2}{2}-9 g_{{R}}^2-\frac{45 g_4^2}{4}+16 \lambda _1+64 \lambda _2+64 \lambda _4+8 y_{{N}}^2+\frac{15 y_{{R}}^2}{2}+144 \lambda _{{R1}}+96 \lambda _{{R2}}+10 y_{{RQ}}^2+32 \lambda _{\phi _R{\phi 1}}\br
+16 \lambda _{\phi _R{\phi 3}}
+16 y^2+4 y_{\nu }^2)-4 y y_{{c}} y_{{N}}^2-\frac{15}{4} y y_{{c}} y_{{R}}^2-5 y y_{{c}} y_{{RQ}}^2-2 y y_{{c}} y_{\nu }^2+64 \lambda _{{LR1}} \lambda _{\phi _L{\phi 2}}+16 \lambda _{{LR2}} \lambda _{\phi _L{\phi 2}}\br +15 \lambda _{{\Sigma \phi 2}} \lambda _{\Sigma \phi _R1}
+\frac{15 \lambda _{{\Sigma \phi 2}} \lambda _{\Sigma \phi _R2}}{4}+48 \lambda _3 \lambda _{\phi _R{\phi 1}}+192 \theta _2 \theta _{\phi _R{\phi 2}}+24 \lambda _3 \lambda _{\phi _R{\phi 3}}+\frac{15 \theta _{{X1}} \theta _{{X2}}}{4}+\frac{15 \lambda _{{X1}} \lambda _{{X2}}}{4} \\
(4\pi)^2\frac{d\theta _{\phi _R{\phi 2}}}{d\ln\mu} &=& \theta _{\phi _R{\phi 2}} (16 y_{{c}}^2-\frac{9 g_{{L}}^2}{2}-9 g_{{R}}^2-\frac{45 g_4^2}{4}+16 \lambda _1-320 \lambda _2+64 \lambda _4+8 y_{{N}}^2+\frac{15 y_{{R}}^2}{2}+144 \lambda _{{R1}}+96 \lambda _{{R2}}+10 y_{{RQ}}^2+32 \lambda _{\phi _R{\phi 1}}\br
+16 \lambda _{\phi _R{\phi 3}}+16 y^2+4 y_{\nu }^2)+64 \lambda _{{LR1}} \theta _{\phi _L{\phi 2}}+16 \lambda _{{LR2}} \theta _{\phi _L{\phi 2}}+15 \theta _{{\Sigma \phi 2}} \lambda _{\Sigma \phi _R1}+\frac{15 \theta _{{\Sigma \phi 2}} \lambda _{\Sigma \phi _R2}}{4}+48 \theta _3 \lambda _{\phi _R{\phi 1}}\br
+192 \theta _2 \lambda _{\phi _R{\phi 2}}+24 \theta _3 \lambda _{\phi _R{\phi 3}}+\frac{15 \lambda _{{X1}} \theta _{{X2}}}{4}-\frac{15 \theta _{{X1}} \lambda _{{X2}}}{4} \\
(4\pi)^2\frac{d\lambda _{\phi _R{\phi 3}}}{d\ln\mu} &=& \lambda _{\phi _R{\phi 3}} (16 y_{{c}}^2-\frac{9 g_{{L}}^2}{2}-9 g_{{R}}^2-\frac{45 g_4^2}{4}+16 \lambda _1+64 \lambda _2-32 \lambda _4+8 y_{{N}}^2+\frac{15 y_{{R}}^2}{2}+16 \lambda _{{R1}}+64 \lambda _{{R2}}+10 y_{{RQ}}^2+32 \lambda _{\phi _R{\phi 1}}+16 y^2+4 y_{\nu }^2)\br
+\frac{15}{2} y_{{c}}^2 y_{{R}}^2-10 y_{{c}}^2 y_{{RQ}}^2+4 y_{{c}}^2 y_{\nu }^2-\frac{15}{2} y^2 y_{{R}}^2+10 y^2 y_{{RQ}}^2+16 \lambda _{\phi _R{\phi 3}}^2+\frac{15 \theta _{{X1}}^2}{2}+\frac{15 \lambda _{{X1}}^2}{2}-\frac{15 \theta _{{X2}}^2}{2}-\frac{15 \lambda _{{X2}}^2}{2}-4 y^2 y_{\nu }^2 
\end{eqnarray*}
}

\mysection{Flavour bounds on Pati-Salam vector leptoquarks}\label{W'}
\label{sec:LQbounds}
In this section we discuss the experimental bounds on the massive  vector leptoquarks  $W_{\mu}^{\prime}$, of electric charge $\pm 2/3$,
coming from   $\SU(4)_{\rm PS}/\SU(3)_{c}$.
These vector bosons are coupled to the left-handed current ${\bar q}_L \gamma^\mu \ell_L$ only in the Pati-Salam models of ref.~\cite{Volkas} , where $q_L$ and $\ell_L$  are unified in $\psi_L$ [see \eq{YPSV})], but not in the models of ref.~\cite{Foot} where $q_L$ is in $\psi_L$ and $\ell_L$ in $\psi$ [see \eq{eq:Yuk224}]. On the other hand, the right-handed current interaction is present in both frameworks and is given by
%
%
\be
\Lag_{W^\prime} = \frac{g_3}{\sqrt{2}}   ~W_{ia} ~\bar d^i_R   \gamma_\mu e_R^a  ~W^\prime_\mu  ~+ ~{\rm h.c.}\, ,
\ee
using Dirac notation for the fermions.
Here $i$ and $a$ are flavour indices and $W_{ia}$ is a new unitary matrix that describes the right-handed misalignment
between the effective down-quark and charged-lepton Yukawa couplings.
Integrating out the heavy $W^\prime$
fields leads to the following effective interaction
\be
\Lag^{d\ell}_{\rm eff} = -\frac{g^2_3  }{2 m^2_{W^\prime} }  ~W_{ia} W^\dagger_{bj} ~
(\bar d^i_R   \gamma_\mu e_R^a)(  \bar e^b_R   \gamma_\mu d_R^j  )   ~+ ~{\rm h.c.}
\ee
The bounds on these effective operators (expressed as bounds on $M_{W^\prime}$ always assuming the maximal possible mixing 
$W_{ia} W^\dagger_{bj}=1$) from various LFV and FCNC processes are reported in 
table~\ref{tab:LQbounds}.\footnote{For a previous version of this table see~\cite{Carpentier:2010ue}.}
In order to deduce from this table the lowest allowed value for $M_{W^\prime}$ we need to determine the structure 
of the right-handed quark-lepton mixing matrix $W_{ia}$. To this purpose, we distinguish two cases.

\begin{table}[t]
\begin{center}
\begin{tabular}{l|l|c}
\rowcolor[cmyk]{0,0,0,0.05}
 Flavour &
 Experimental constraint & Bound on $M_{W^\prime}$ in  TeV\\ 
\hline
$dd~e\mu $ &  
$ \sigma(\mu~\hbox{Ti}\rightarrow e~\hbox{Ti})/\sigma_0(\mu~\hbox{Ti}) < 4.3\times10^{-12}$   & 120
   \\
$ss~e\mu$ &  
$ \sigma(\mu~\hbox{Ti}\rightarrow e~\hbox{Ti})/\sigma_0(\mu~\hbox{Ti}) < 4.3\times10^{-12}$   & $12\times \sqrt{P_{s\bar s}/1\%}$   \\
$dd~e \tau$ & 
BR$(\tau \rightarrow \pi^{0} e)  < 8.0\times10^{-8}$  & \ 3.8  \\
$dd~\mu\tau$ &  
BR$(\tau \rightarrow \pi^{0}\mu) < 1.1\times 10^{-7}$  & \ 3.5 \\
\hline
$sd~\mu\mu$   &     
BR$(K_{L}\rightarrow \bar{\mu}\mu)_{\rm SD}  <   2.5  \times10^{-9}$  & 50\phantom{.0} \\
$sd~ee$   &  
BR$(K_{L}\rightarrow \bar{e}e) = (9.0 \pm 6.0)\times10^{-12}$ & 13.4  \\
$bd~\mu\mu$   &    
BR$(B_d \rightarrow \bar{\mu}\mu) = (3.6 \pm 1.6)\times 10^{-10}$  & 12.7 \\
$bs~\mu\mu$   &  
BR$(B_s \rightarrow \bar{\mu}\mu) = (2.9 \pm 0.7)\times 10^{-9}$  & 10.1 \\
\hline
$sd~e\mu$       &  
BR$(K_{L}\rightarrow \bar{e}\mu) <4.7\times10^{-12}$  & 200\phantom{.00}  \\
\hline
$sd~e\tau$       &
BR$( \tau \rightarrow K^* e)  < 3.2 \times 10^{-8}$  & 10.3  \\
$sd~\mu \tau$ &  
BR$( \tau \rightarrow K^* \mu )  < 5.9  \times10^{-8}$   & \ 8.8  \\
$bs~e\mu$ &     
BR$(B^{+}\rightarrow K^{+}\bar{e}\mu) < 9.1\times10^{-8}$ & \ 8.3  \\
$bd~e\mu$ &       
BR$(B^{+}\rightarrow \pi^{+}\bar{e}\mu) <  1.7\times10^{-7}$  & \ 7.1 \\
$bd~\mu\tau$ &
BR$(B_d \rightarrow \bar{\mu}\tau) <2.2\times10^{-5}$  & \ 3.0 \\
$bd~e\tau$ & 
BR$(B_d \rightarrow \bar{e}\tau) < 2.8\times10^{-5}$ &  \ 2.8  \\

\end{tabular}
\caption{\label{tab:LQbounds}\em
Bounds on the vector leptoquark mass $M_{W^\prime}$ assuming maximal mixing angles in each case
($W_{ia} W^\dagger_{bj} =1$)
from lepton-flavour-violating and flavour-changing-neutral-current processes. Here $P_{s\bar s}$  denotes the ratio of  
the strange-quark density in nuclei, normalised to that of down-type quarks $(P_{s\bar s}= \langle N | \bar s s | N \rangle/\langle N | \bar d  d | N \rangle)$,
for which we have assumed a reference value of $1\%$ (see e.g.~\cite{Bali:2011ks}).}
\end{center}
\end{table}

The first case corresponds to the Yukawa interaction in eq.~(\ref{eq:Yuk224}), where
we have two completely independent Yukawa couplings for down quarks and charged leptons.
As a result, $W_{ia}$ is arbitrary and we can tune it in order to minimize 
the  bounds on $M_{W^\prime}$ given in table~\ref{tab:LQbounds}.
The two most severe bounds (on the $dd e\mu$ and $sd e\mu$ flavour structures) 
force us to consider a mixing matrix of the type
\be
W = 
\left(\begin{array}{ccc}
0 & 0 & 1  \\ s_{s e } & c_{s e }  & 0 \\ 
-c_{s e }  & s_{se} & 0
\end{array}\right)
\ee
to forbid any $d \leftrightarrow e$ and  $d \leftrightarrow \mu$ mixing.
 All  bounds from processes with flavour conservation either in the 
quark (rows 1-4 in table~\ref{tab:LQbounds}) or in the lepton sector (rows 5-8 in table~\ref{tab:LQbounds}) 
disappear if we assume that $W_{ia}$ is a permutation matrix. 
As a result, the  constraint on $M_{W^\prime}$ is minimised for $s_{se}=0$ and it is set by 
$\hbox{BR}( \tau \rightarrow K^* \mu )$:
\be
M_{W^\prime} > 8.8~{\rm TeV}~. 
\label{eq:LQbound}
\ee

The second case corresponds to the Yukawa interaction in \eq{YPSV} with $Q_L\oplus Q_R$ pairs.
Now the structure of $W_{ia}$ is not arbitrary: the effective down-type and charged-lepton Yukawa
couplings are necessarily quasi aligned. The resulting $W_{ia}$ is close to a diagonal matrix, 
with a mixing angle in the $1-2$ sector of the order of the Cabibbo angle. 
In this case we cannot avoid the bound on $M_{W'}$ in the range of 100-200~TeV dictated by the 
$dd e\mu$ and $sd e\mu$ entries in Table~\ref{tab:LQbounds}.
The situation is even worse due to the simultaneous presence of both  
$d_R$--$e_R$--$W^\prime$ and $u_L$--$\nu_L$--$W^\prime$ interactions.
The combination of left-handed and right-handed currents leads to an effective interaction that breaks lepton chirality, without being proportional to the corresponding lepton mass. This gives rise to an effective breaking of lepton universality 
in $\Gamma(\pi\to e \nu)/\Gamma(\pi\to \mu \nu)$ that implies
$
M_{W^\prime} > 250\TeV
$~\cite{Val}.

    \mysection{RGE for trinification models}\label{RGE333}
We first list the one-loop RGE for the trinification models described in section~\ref{333}.
The gauge $\beta$-function coefficients are $b_L=b_R=5 - n_H/2$ and $b_3=5$,
where $n_H$ is the number of Higgs multiplets.
The RGE for the Yukawa couplings of \eq{eqqqY} are
\begin{eqnsystem}{sys:y333}
(4\pi)^2\frac{dy_{Q_1}}{d\ln\mu} &=& y_{Q_1} \left(-4 g_L^2-4 g_R^2-8 g_3^2+2 y_{L_1}^2+6 y_{Q_2}^2\right)+2 y_{L_1} y_{L_2} y_{Q_2}+6 y_{Q_1}^3 \\
(4\pi)^2\frac{dy_{Q_2}}{d\ln\mu} &=& y_{Q_2} \left(-4 g_L^2-4 g_R^2-8 g_3^2+2 y_{L_2}^2+6 y_{Q_1}^2\right)+2 y_{L_1} y_{L_2} y_{Q_1}+6 y_{Q_2}^3 \\
(4\pi)^2\frac{dy_{L_1}}{d\ln\mu} &=& y_{L_1} \left(-8 g_L^2-8 g_R^2+6 y_{L_2}^2+3 y_{Q_1}^2\right)+6 y_{L_1}^3+3 y_{L_2} y_{Q_1} y_{Q_2} \\
(4\pi)^2\frac{dy_{L_2}}{d\ln\mu} &=& y_{L_2} \left(-8 g_L^2-8 g_R^2+6 y_{L_1}^2+3 y_{Q_2}^2\right)+3 y_{L_1} y_{Q_1} y_{Q_2}+6 y_{L_2}^3 
\end{eqnsystem}
For the model with a single Higgs multiplet, $n_H=1$, one needs to set $y_{Q_1}=y_Q$, $y_{L_1}=y_L$ and $y_{Q_2}=y_{L_2}=0$.
The RGE for the two quartic couplings of the scalar potential in \eq{eq:V1111} are
    \begin{eqnsystem}{sys:RGE3331}
(4\pi)^2\frac{d\lambda _ a}{d\ln\mu} &=& \lambda _a \left(-16 g_L^2-16 g_R^2+48 \lambda _b+12 y_Q^2+8 y_ {L }^2\right)+\\ \nonumber
&&+\frac{10}{3} g_L^2 g_R^2+\frac{11 g_L^4}{12}+\frac{11 g_R^4}{12}+52 \lambda _a^2+12 \lambda _b^2-2 y_ {L }^4 \\
(4\pi)^2\frac{d\lambda _ b}{d\ln\mu} &=& \lambda_b  \left(-16 g_L^2-16 g_R^2+24 \lambda _a +12 y_Q^2+8 y_ {L }^2\right)+\\ \nonumber
&&-2 g_L^2 g_R^2+\frac{5 g_L^4}{4}+\frac{5 g_R^4}{4}+24 \lambda _b^2-6 y_U^4-2 y_ {L }^4 
\end{eqnsystem}
For the model with two Higgs fields $H_1$ and $H_2$ ($n_H=2$),
the most generic scalar potential becomes
\beq \label{eq:VH1H2}
V = V_{1111}+V_{2222}+V_{1122} + V_{1222}+ V_{1222}\eeq
where
\begin{eqnsystem}{sys:V333}
 V_{1111} &=& \lambda_{1a} \mbox{Tr}(H_1^\dagger H_1)^2 + \lambda_{1b}  \mbox{Tr}(H_1^\dagger H_1H_1^\dagger H_1), \\
   V_{2222} &=&  \lambda_{2a} \mbox{Tr}(H_2^\dagger H_2)^2 + \lambda_{2b}  \mbox{Tr}(H_2^\dagger H_2H_2^\dagger H_2) ,\\
   V_{1222} &=& \Re   \lambda_{3a}\Tr(H_1^\dagger H_2)\Tr(H_2^\dagger H_2)+  \Re \lambda_{3b}\Tr(H_1^\dagger H_2 H_2^\dagger H_2),  \\
   V_{1112}&=&  \Re   \lambda_{4a}\Tr(H_1^\dagger H_2)\Tr(H_1^\dagger H_1)+ \Re \lambda_{4b}\Tr(H_1^\dagger H_1 H_1^\dagger H_2) ,  \\
   V_{1122} &=&  \lambda_{a} \Tr(H_1^\dagger H_1)\Tr( H_2^\dagger H_2)+   \lambda_b |  \Tr(H_1^\dagger H_2)|^2 + \\
&&+  \lambda_{c} \Tr(H_1^\dagger H_1 H_2^\dagger H_2)+ \lambda_d \Tr(H_1 H_1^\dagger H_2 H_2^\dagger)\nonumber   \\ &&
+   \Re \lambda_{e} \Tr(H_1^\dagger H_2)^2 + \Re \lambda_{f} \Tr(H_1^\dagger H_2 H_1^\dagger H_2) \, .\nonumber
    \end{eqnsystem}
The RGE for the quartics, setting for simplicity to zero those that violate CP and/or baryon number, are:
{\tiny
\begin{eqnsystem}{sys:lambda333}
(4\pi)^2\frac{d\lambda _{{1a}}}{d\ln\mu} &=& \lambda _{{1a}} \left(48 \lambda _{{1b}}-16 g_L^2-16 g_R^2+8 y_{L_1}^2+12 y_{Q_1}^2\right)+52 \lambda _{{1a}}^2+12 \lambda _{{1b}}^2+56 \lambda _{{4b}}^2+2 \lambda _{{a}} \lambda _{{b}}+6 \lambda _{{a}} \lambda _{{c}}+
      \nonumber\\ &&
      6 \lambda _{{a}} \lambda _{{d}}+9 \lambda _{{a}}^2+\lambda _{{b}}^2+2 \lambda _{{c}} \lambda _{{d}}+4 \lambda _{{e}}^2+4 \lambda _{{f}}^2+\frac{10}{3} g_L^2 g_R^2+\frac{11 g_L^4}{12}+\frac{11 g_R^4}{12}-2 y_{L_1}^4 \\
(4\pi)^2\frac{d\lambda _{{1b}}}{d\ln\mu} &=& \lambda _{{1b}} \left(24 \lambda _{{1a}}-16 g_L^2-16 g_R^2+8 y_{L_1}^2+12 y_{Q_1}^2\right)+24 \lambda _{{1b}}^2+24 \lambda _{{4b}}^2+2 \lambda _{{b}} \lambda _{{c}}+2 \lambda _{{b}} \lambda _{{d}}+      \nonumber\\ &&
3 \lambda _{{c}}^2+3 \lambda _{{d}}^2+8 \lambda _{{e}} \lambda _{{f}}-2 g_L^2 g_R^2+\frac{5 g_L^4}{4}+\frac{5 g_R^4}{4}-2 y_{L_1}^4-6 y_{Q_1}^4 \\
(4\pi)^2\frac{d\lambda _{{2a}}}{d\ln\mu} &=& \lambda _{{2a}} \left(48 \lambda _{{2b}}-16 g_L^2-16 g_R^2+8 y_{L_2}^2+12 y_{Q_2}^2\right)+52 \lambda _{{2a}}^2+24 \lambda _{{2b}} \lambda _{{3a}}+18 \lambda _{{2b}}^2+26 \lambda _{{3a}}^2+2 \lambda _{{a}} \lambda _{{b}}+
      \nonumber\\ &&
      6 \lambda _{{a}} \lambda _{{c}}+6 \lambda _{{a}} \lambda _{{d}}+9 \lambda _{{a}}^2+\lambda _{{b}}^2+2 \lambda _{{c}} \lambda _{{d}}+4 \lambda _{{e}}^2+4 \lambda _{{f}}^2+\frac{10}{3} g_L^2 g_R^2+\frac{11 g_L^4}{12}+\frac{11 g_R^4}{12}-2 y_{L_2}^4 \\
(4\pi)^2\frac{d\lambda _{{2b}}}{d\ln\mu} &=& \lambda _{{2b}} \left(24 \lambda _{{2a}}-16 g_L^2-16 g_R^2+8 y_{L_2}^2+12 y_{Q_2}^2\right)+12 \lambda _{{2b}} \lambda _{{3a}}+36 \lambda _{{2b}}^2+2 \lambda _{{b}} \lambda _{{c}}+2 \lambda _{{b}} \lambda _{{d}}+
      \nonumber\\ &&
      3 \lambda _{{c}}^2+3 \lambda _{{d}}^2+8 \lambda _{{e}} \lambda _{{f}}-2 g_L^2 g_R^2+\frac{5 g_L^4}{4}+\frac{5 g_R^4}{4}-2 y_{L_2}^4-6 y_{Q_2}^4 \\
(4\pi)^2\frac{d\lambda _{{3a}}}{d\ln\mu} &=& \lambda _{{3a}} \left(52 \lambda _{{2a}}+24 \lambda _{{2b}}+6 \lambda _{{a}}+22 \lambda _{{b}}+6 \lambda _{{c}}+6 \lambda _{{d}}+48 \lambda _{{e}}+24 \lambda _{{f}}-16 g_L^2-16 g_R^2+2 y_{L_1}^2+6 y_{L_2}^2+3 y_{Q_1}^2+9 y_{Q_2}^2\right)+
      \nonumber\\ &&
      24 \lambda _{{2a}} \lambda _{{2b}}+12 \lambda _{{2b}} \lambda _{{b}}+4 \lambda _{{2b}} \lambda _{{c}}+4 \lambda _{{2b}} \lambda _{{d}}+24 \lambda _{{2b}} \lambda _{{e}}+8 \lambda _{{2b}} \lambda _{{f}}+12 \lambda _{{2b}}^2+32 \lambda _{{4b}} \lambda _{{a}}+
            \nonumber\\ &&
            4 \lambda _{{4b}} \lambda _{{b}}+8 \lambda _{{4b}} \lambda _{{c}}+8 \lambda _{{4b}} \lambda _{{d}}+4 \lambda _{{4b}} \lambda _{{e}}+4 \lambda _{{4b}} \lambda _{{f}}-4 y_{L_1} y_{L_2}^3 \\
(4\pi)^2\frac{d\lambda _{{2b}}}{d\ln\mu} &=& \lambda _{{2b}} \left(12 \lambda _{{2a}}+24 \lambda _{{2b}}+6 \lambda _{{a}}+2 \lambda _{{b}}+6 \lambda _{{c}}+6 \lambda _{{d}}+8 \lambda _{{e}}+24 \lambda _{{f}}-16 g_L^2-16 g_R^2+2 y_{L_1}^2+6 y_{L_2}^2+3 y_{Q_1}^2+9 y_{Q_2}^2\right)+
      \nonumber\\ &&
      12 \lambda _{{2b}} \lambda _{{3a}}+4 \lambda _{{3a}} \lambda _{{c}}+4 \lambda _{{3a}} \lambda _{{d}}+8 \lambda _{{3a}} \lambda _{{f}}+4 \lambda _{{4b}} \lambda _{{b}}+
      8 \lambda _{{4b}} \lambda _{{c}}+8 \lambda _{{4b}} \lambda _{{d}}+4 \lambda _{{4b}} \lambda _{{e}}+4 \lambda _{{4b}} \lambda _{{f}}-4 y_{L_1} y_{L_2}^3-12 y_{Q_1} y_{Q_2}^3 \\
(4\pi)^2\frac{d\lambda _{{4b}}}{d\ln\mu} &=& \lambda _{{4b}} \left(12 \lambda _{{1a}}+24 \lambda _{{1b}}+6 \lambda _{{a}}+2 \lambda _{{b}}+6 \lambda _{{c}}+6 \lambda _{{d}}+8 \lambda _{{e}}+24 \lambda _{{f}}-16 g_L^2-16 g_R^2+6 y_{L_1}^2+2 y_{L_2}^2+9 y_{Q_1}^2+3 y_{Q_2}^2\right)+
      \nonumber\\ &&
      12 \lambda _{{1b}} \lambda _{{4b}}+4 \lambda _{{2b}} \lambda _{{b}}+6 \lambda _{{2b}} \lambda _{{c}}+6 \lambda _{{2b}} \lambda _{{d}}+4 \lambda _{{2b}} \lambda _{{e}}+2 \lambda _{{3a}} \lambda _{{c}}+2 \lambda _{{3a}} \lambda _{{d}}+4 \lambda _{{3a}} \lambda _{{f}}+
            \nonumber\\ &&
            4 \lambda _{{4b}} \lambda _{{c}}+4 \lambda _{{4b}} \lambda _{{d}}+8 \lambda _{{4b}} \lambda _{{f}}-4 y_{L_2} y_{L_1}^3-12 y_{Q_1}^3 y_{Q_2} \\
(4\pi)^2\frac{d\lambda _{{4b}}}{d\ln\mu} &=& \lambda _{{4b}} \left(52 \lambda _{{1a}}+24 \lambda _{{1b}}+6 \lambda _{{a}}+22 \lambda _{{b}}+6 \lambda _{{c}}+6 \lambda _{{d}}+48 \lambda _{{e}}+24 \lambda _{{f}}-16 g_L^2-16 g_R^2+6 y_{L_1}^2+2 y_{L_2}^2+9 y_{Q_1}^2+3 y_{Q_2}^2\right)+
      \nonumber\\ &&
      24 \lambda _{{1a}} \lambda _{{4b}}+12 \lambda _{{1b}} \lambda _{{4b}}+12 \lambda _{{2b}} \lambda _{{a}}+2 \lambda _{{2b}} \lambda _{{c}}+2 \lambda _{{2b}} \lambda _{{d}}+4 \lambda _{{2b}} \lambda _{{f}}+20 \lambda _{{3a}} \lambda _{{a}}+
            \nonumber\\ &&
            4 \lambda _{{3a}} \lambda _{{b}}+6 \lambda _{{3a}} \lambda _{{c}}+6 \lambda _{{3a}} \lambda _{{d}}+4 \lambda _{{3a}} \lambda _{{e}}+12 \lambda _{{4b}} \lambda _{{b}}+4 \lambda _{{4b}} \lambda _{{c}}+4 \lambda _{{4b}} \lambda _{{d}}+24 \lambda _{{4b}} \lambda _{{e}}+8 \lambda _{{4b}} \lambda _{{f}}-4 y_{L_2} y_{L_1}^3 \\
(4\pi)^2\frac{d\lambda _{{a}}}{d\ln\mu} &=& \lambda _{{a}} \left(40 \lambda _{{1a}}+24 \lambda _{{1b}}+40 \lambda _{{2a}}+24 \lambda _{{2b}}-16 g_L^2-16 g_R^2+4 y_{L_1}^2+4 y_{L_2}^2+6 y_{Q_1}^2+6 y_{Q_2}^2\right)+4 \lambda _{{1a}} \lambda _{{b}}+12 \lambda _{{1a}} \lambda _{{c}}+
      \nonumber\\ &&
      12 \lambda _{{1a}} \lambda _{{d}}+4 \lambda _{{1b}} \lambda _{{c}}+4 \lambda _{{1b}} \lambda _{{d}}+4 \lambda _{{2a}} \lambda _{{b}}+12 \lambda _{{2a}} \lambda _{{c}}+12 \lambda _{{2a}} \lambda _{{d}}+32 \lambda _{{2b}} \lambda _{{4b}}+4 \lambda _{{2b}} \lambda _{{c}}+
            \nonumber\\ &&
            4 \lambda _{{2b}} \lambda _{{d}}+2 \lambda _{{2b}}^2+68 \lambda _{{3a}} \lambda _{{4b}}+4 \lambda _{{3a}}^2+6 \lambda _{{4b}}^2+4 \lambda _{{a}}^2+2 \lambda _{{b}}^2+2 \lambda _{{c}}^2+2 \lambda _{{d}}^2+8 \lambda _{{e}}^2+\frac{2}{3} g_L^2 g_R^2+\frac{11 g_L^4}{6}+\frac{11 g_R^4}{6}-4 y_{L_1}^2 y_{L_2}^2 \\
(4\pi)^2\frac{d\lambda _{{b}}}{d\ln\mu} &=& 2 \bigg(2 \lambda _{{1a}} \lambda _{{b}}+2 \lambda _{{2a}} \lambda _{{b}}+12 \lambda _{{2b}} \lambda _{{3a}}+3 \lambda _{{2b}}^2+2 \lambda _{{3a}} \lambda _{{4b}}+12 \lambda _{{3a}}^2+27 \lambda _{{4b}}^2+
4 \lambda _{{a}} \lambda _{{b}}+6 \lambda _{{b}} \lambda _{{c}}+
6 \lambda _{{b}} \lambda _{{d}}-8 \lambda _{{b}} g_L^2+
      \nonumber\\ &&
      -8 \lambda _{{b}} g_R^2+9 \lambda _{{b}}^2+2 \lambda _{{b}} y_{L_1}^2+2 \lambda _{{b}} y_{L_2}^2+3 \lambda _{{b}} y_{Q_1}^2+3 \lambda _{{b}} y_{Q_2}^2+4 \lambda _{{c}} \lambda _{{d}}+48 \lambda _{{e}} \lambda _{{f}}+44 \lambda _{{e}}^2+8 \lambda _{{f}}^2+3 g_L^2 g_R^2-2 y_{L_1}^2 y_{L_2}^2\bigg) \\
(4\pi)^2\frac{d\lambda _{{c}}}{d\ln\mu} &=& \lambda _{{c}} \left(4 \lambda _{{1a}}+12 \lambda _{{1b}}+4 \lambda _{{2a}}+12 \lambda _{{2b}}+8 \lambda _{{a}}-16 g_L^2-16 g_R^2+4 y_{L_1}^2+4 y_{L_2}^2+6 y_{Q_1}^2+6 y_{Q_2}^2\right)+4 \lambda _{{1b}} \lambda _{{b}}+8 \lambda _{{2b}} \lambda _{{3a}}+ 16 \lambda _{{2b}} \lambda _{{4b}}
      \nonumber\\ &&
     +4 \lambda _{{2b}} \lambda _{{b}}+6 \lambda _{{2b}}^2+4 \lambda _{{3a}} \lambda _{{4b}}+14 \lambda _{{4b}}^2+4 \lambda _{{b}} \lambda _{{d}}+6 \lambda _{{c}}^2+16 \lambda _{{e}} \lambda _{{f}}+24 \lambda _{{f}}^2-2 g_L^2 g_R^2+\frac{5 g_R^4}{2}-4 y_{L_1}^2 y_{L_2}^2-12 y_{Q_1}^2 y_{Q_2}^2 \\
(4\pi)^2\frac{d\lambda _{{d}}}{d\ln\mu} &=& \lambda _{{d}} \left(4 \lambda _{{1a}}+12 \lambda _{{1b}}+4 \lambda _{{2a}}+12 \lambda _{{2b}}+8 \lambda _{{a}}-16 g_L^2-16 g_R^2+4 y_{L_1}^2+4 y_{L_2}^2+6 y_{Q_1}^2+6 y_{Q_2}^2\right)+4 \lambda _{{1b}} \lambda _{{b}}+8 \lambda _{{2b}} \lambda _{{3a}}+      16 \lambda _{{2b}} \lambda _{{4b}}+
      \nonumber\\ &&
+4 \lambda _{{2b}} \lambda _{{b}}+6 \lambda _{{2b}}^2+4 \lambda _{{3a}} \lambda _{{4b}}+14 \lambda _{{4b}}^2+4 \lambda _{{b}} \lambda _{{c}}+6 \lambda _{{d}}^2+16 \lambda _{{e}} \lambda _{{f}}+24 \lambda _{{f}}^2-2 g_L^2 g_R^2+\frac{5 g_L^4}{2}-4 y_{L_1}^2 y_{L_2}^2-12 y_{Q_1}^2 y_{Q_2}^2 \\
(4\pi)^2\frac{d\lambda _{{e}}}{d\ln\mu} &=& 4 \lambda _{{1a}} \lambda _{{e}}+4 \lambda _{{1b}} \lambda _{{f}}+4 \lambda _{{2a}} \lambda _{{e}}+12 \lambda _{{2b}} \lambda _{{3a}}+2 \lambda _{{2b}} \lambda _{{4b}}+4 \lambda _{{2b}} \lambda _{{f}}+2 \lambda _{{2b}}^2+
      \nonumber\\ &&
      2 \lambda _{{3a}} \lambda _{{4b}}+12 \lambda _{{3a}}^2+26 \lambda _{{4b}}^2+8 \lambda _{{a}} \lambda _{{e}}+40 \lambda _{{b}} \lambda _{{e}}+24 \lambda _{{b}} \lambda _{{f}}+12 \lambda _{{c}} \lambda _{{e}}+4 \lambda _{{c}} \lambda _{{f}}+12 \lambda _{{d}} \lambda _{{e}}+
            \nonumber\\ &&
            4 \lambda _{{d}} \lambda _{{f}}-16 \lambda _{{e}} g_L^2-16 \lambda _{{e}} g_R^2+4 \lambda _{{e}} y_{L_1}^2+4 \lambda _{{e}} y_{L_2}^2+6 \lambda _{{e}} y_{Q_1}^2+6 \lambda _{{e}} y_{Q_2}^2-2 y_{L_1}^2 y_{L_2}^2 \\
(4\pi)^2\frac{d\lambda _{{f}}}{d\ln\mu} &=& \lambda _{{f}} \left(4 \lambda _{{1a}}+4 \lambda _{{2a}}+8 \lambda _{{a}}+12 \lambda _{{c}}+12 \lambda _{{d}}-16 g_L^2-16 g_R^2+4 y_{L_1}^2+4 y_{L_2}^2+6 y_{Q_1}^2+6 y_{Q_2}^2\right)+4 \lambda _{{1b}} \lambda _{{e}}+4 \lambda _{{2b}} \lambda _{{3a}}+
      \nonumber\\ &&
      2 \lambda _{{2b}} \lambda _{{4b}}+
4 \lambda _{{2b}} \lambda _{{e}}+6 \lambda _{{2b}}^2+2 \lambda _{{3a}} \lambda _{{4b}}+10 \lambda _{{4b}}^2+4 \lambda _{{c}} \lambda _{{e}}+4 \lambda _{{d}} \lambda _{{e}}-2 y_{L_1}^2 y_{L_2}^2-6 y_{Q_1}^2 y_{Q_2}^2 
\end{eqnsystem}}

\bigskip

\small

\begin{thebibliography}{nn}

\bibitem{pap}
   M.~Farina, D.~Pappadopulo and A.~Strumia,
  JHEP {1308} (2013) 022
  [\hhref{1303.7244}].

\bibitem{follie}
   J.~P.~Fatelo, J.~M.~Gerard, T.~Hambye and J.~Weyers,
  Phys.\ Rev.\ Lett.\  {74} (1995) 492.
  T.~Hambye,
  Phys.\ Lett.\ B {371} (1996) 87
  [\hhref{hep-ph/9510266}].
  R. Hempfling,
  Phys.\ Lett.\ B {379} (1996) 153
  [\hhref{hep-ph/9604278}].
   F.~Vissani,
  Phys.\ Rev.\ D {57} (1998) 7027
  [\hhref{hep-ph/9709409}].
    P.~H.~Frampton and C.~Vafa,
  \hhref{hep-th/9903226}.
    K.~A.~Meissner and H.~Nicolai,
  Phys.\ Lett.\ B {648} (2007) 312
  [\hhref{hep-th/0612165}].
  W.~-F.~Chang, J.~N.~Ng and J.~M.~S.~Wu,
  Phys.\ Rev.\ D {75} (2007) 115016
  [\hhref{hep-ph/0701254}].
  R.~Foot, A.~Kobakhidze and R.~R.~Volkas,
  Phys.\ Lett.\ B {655} (2007) 156
  [\hhref{0704.1165}].
R.~Foot, A.~Kobakhidze, K.~L.~McDonald, R.~R.~Volkas,
  Phys.\ Rev.\ D {77} (2008) 035006
  [\hhref{0709.2750}].
    S.~Iso, N.~Okada, Y.~Orikasa,
  Phys.\ Rev.\ D {80} (2009) 115007
  [\hhref{0909.0128}].
    M.~Shaposhnikov and C.~Wetterich,
  Phys.\ Lett.\ B {683} (2010) 196
  [\hhref{0912.0208}].
  S.~Iso and Y.~Orikasa,
  PTEP {2013} (2013) 023B08
  [\hhref{1210.2848}].
T.~Hur and P.~Ko,
  Phys.\ Rev.\ Lett.\  {106}, 141802 (2011)
  [\hhref{1103.2571}].  
    M.~Shaposhnikov,
  Theor.\ Math.\ Phys.\  {170} (2012) 229
   [Teor.\ Mat.\ Fiz.\  {170} (2012) 280].
  C.~Englert, J.~Jaeckel, V.~V.~Khoze and M.~Spannowsky,
  \hhref{1301.4224}.
   E.J.~Chun, S. Jung, H.M.~Lee,
  \hhref{1304.5815}.
    M.~Heikinheimo, A.~Racioppi, M.~Raidal, C.~Spethmann and K.~Tuominen,
  Mod.\ Phys.\ Lett.\ A {29} (2014) 1450077
  [\hhref{1304.7006}]. 
T.~Hambye and A.~Strumia,
  Phys.\ Rev.\ D {88} (2013) 055022
  [\hhref{1306.2329}].
  C.~D.~Carone and R.~Ramos,
  Phys.\ Rev.\ D {88} (2013) 055020
  [\hhref{1307.8428}].
     A.~Farzinnia, H.~J.~He and J.~Ren,
  Phys.\ Lett.\ B {727} (2013) 141
  [\hhref{1308.0295}].
   R.~Foot, A.~Kobakhidze, K.~L.~McDonald and R.~R.~Volkas,
  Phys.\ Rev.\ D {89} (2014) 115018
  [\hhref{1310.0223}].
    R.~Dermisek, T.~H.~Jung and H.~D.~Kim,
  Phys.\ Rev.\ Lett.\  {113} (2014) 051801
  [\hhref{1308.0891}].
    M.~Holthausen, J.~Kubo, K.~S.~Lim and M.~Lindner,
  JHEP {1312} (2013) 076
  [\hhref{1310.4423}].
  C.~T.~Hill,
  Phys.\ Rev.\ D {89} (2014) 073003
  [\hhref{1401.4185}].
  J.~Guo and Z.~Kang,
  \hhref{1401.5609}.
  S.~Benic and B.~Radovcic,
  Phys.\ Lett.\ B {732} (2014) 91
  [\hhref{1401.8183}].
    P.~H.~Chankowski, A.~Lewandowski, K.~A.~Meissner and H.~Nicolai,
  \hhref{1404.0548}.
   H.~Davoudiasl, I.M.~Lewis, 
  Phys.\ Rev.\ D {90}, 033003 (2014)   [\hhref{1404.6260}].
K.~Allison, C.~T.~Hill and G.~G.~Ross,   \hhref{1404.6268}.
   G.~M.~Pelaggi,
  \hhref{1406.4104}. 
  W.~Altmannshofer, W.~A.~Bardeen, M.~Bauer, M.~Carena and J.~D.~Lykken,
  \hhref{1408.3429}.
    R.~Foot, A.~Kobakhidze and A.~Spencer-Smith,
\hhref{1409.4915}.
    M.~B.~Einhorn and D.~R.~T.~Jones,
  \hhref{1410.8513}.
  O.~Antipin, M.~Redi and A.~Strumia,
  \hhref{1410.1817}.


\bibitem{low}
  I.~Antoniadis,
  Phys.\ Lett.\ B {246} (1990) 377.


\bibitem{fat}
  R.~Sundrum,
  Nucl.\ Phys.\ B {690} (2004) 302
  [\hhref{hep-th/0310251}].


\bibitem{agravity}  A.~Salvio and A.~Strumia, 
  JHEP {1406} (2014) 080
  [\hhref{1403.4226}]



\bibitem{Sannino}
D.~F.~Litim and F.~Sannino,
  \hhref{1406.2337}.

\bibitem{Suslov}
P.J. Redmond, J.L. Uretsky, Phys. Rev. Lett. 4 (1958) 147.
N.N. Bogolyubov, A.A. Logunov, D.V. Shirkov, Sov. Phys. JETP 37 (1960) 574.
I.~M.~Suslov,
  J.\ Exp.\ Theor.\ Phys.\  {107} (2008) 413
   [Zh.\ Eksp.\ Teor.\ Fiz.\  {134} (2008) 490]
  [\hhref{1010.4081}].
I.~M.~Suslov,
  J.\ Exp.\ Theor.\ Phys.\  {108} (2009) 980
  [\hhref{0804.2650}].
  
 


\bibitem{Seiberg}
N.~Seiberg, E.~Witten,
  Nucl.\ Phys.\ B {426} (1994) 19
   [Erratum-ibid.\ B {430} (1994) 485]
  [\hhref{hep-th/9407087}]. 


\bibitem{Skiba}
  G.~Marques Tavares, M.~Schmaltz and W.~Skiba,
  Phys.\ Rev.\ D {89} (2014) 015009
  [\hhref{1308.0025}].


\bibitem{Gross}
  D.~J.~Gross and F.~Wilczek,
  Phys.\ Rev.\ D {8} (1973) 3633;
  Phys.\ Rev.\ Lett.\  {30} (1973) 1343;
  Phys.\ Rev.\ D {9} (1974) 980.
   H.~D.~Politzer,
  Phys.\ Rev.\ Lett.\  {30} (1973) 1346.
  S.~R.~Coleman and D.~J.~Gross,
  Phys.\ Rev.\ Lett.\  {31} (1973) 851.
  
  
  
\bibitem{Cheng:1973nv}  T.~P.~Cheng, E.~Eichten and L.~-F.~Li,
  Phys.\ Rev.\ D {9} (1974) 2259.
    
  
  


\bibitem{K5}
  E.S.~Fradkin and O.K.~Kalashnikov,
  Phys.\ Lett.\ B {64} (1976) 177.
   O.K.~Kalashnikov,
  Lebedev-77-206.  
  E.S.~Fradkin, O.K.~Kalashnikov and S.E.~Konshtein,
 Lett.\ Nuovo Cim.\  {21} (1978) 5.


\bibitem{K224}
  O.~K.~Kalashnikov,
  Phys.\ Lett.\ B {72} (1977) 65.
    
  
\bibitem{Oehme:1984yy}
  R.~Oehme and W.~Zimmermann,
  Commun.\ Math.\ Phys.\  {97} (1985) 569.
  J.~Kubo, K.~Sibold and W.~Zimmermann,
  Nucl.\ Phys.\ B {259} (1985) 331.
  J.~Kubo, K.~Sibold and W.~Zimmermann,
  Phys.\ Lett.\ B {220} (1989) 185.

\bibitem{Zimmermann:2001pq}
  W.~Zimmermann,
  Commun.\ Math.\ Phys.\  {219} (2001) 221.
  
  
  
  
\bibitem{Litim:2015iea}
  D.~F.~Litim, M.~Mojaza and F.~Sannino,
  \hhref{1501.03061}.
   
 


\bibitem{Pendleton:1980as}
  B.~Pendleton and G.~G.~Ross,
  Phys.\ Lett.\ B {98} (1981) 291.
  
  

   
    


\bibitem{Coleman:1973jx}
  S.~R.~Coleman and E.~J.~Weinberg,
  Phys.\ Rev.\ D {7} (1973) 1888.
  
  

\bibitem{SMvacdecay}
C.G.~Callan and S.~Coleman, Phys. Rev. D16 (1977) 1762.
  G.~Isidori, G.~Ridolfi and A.~Strumia,
  Nucl.\ Phys.\ B {609} (2001) 387
  [\hhref{hep-ph/0104016}].
  J.~R.~Espinosa, G.~F.~Giudice and A.~Riotto,
  JCAP {0805} (2008) 002
  [\hhref{0710.2484}].
  J.~Elias-Miro, J.~R.~Espinosa, G.~F.~Giudice, G.~Isidori, A.~Riotto and A.~Strumia,
  Phys.\ Lett.\ B {709} (2012) 222
  [\hhref{1112.3022}].
  

\bibitem{SMphase}
   G.~Degrassi, S.~Di Vita, J.~Elias-Miro, J.~R.~Espinosa, G.~F.~Giudice, G.~Isidori and A.~Strumia,
  JHEP {1208} (2012) 098
  [\hhref{1205.6497}].
  D.~Buttazzo, G.~Degrassi, P.~P.~Giardino, G.~F.~Giudice, F.~Sala, A.~Salvio and A.~Strumia,
  JHEP {1312} (2013) 089
  [\hhref{1307.3536}].
  
%
  


\bibitem{V>0}
S.~C.~Frautschi and J.~Kim,
  Nucl.\ Phys.\ B {196} (1982) 301.
K. G. Klimenko,   Theor. and Mat. Phys. 62 (1985) 58.
 K.~Kannike,
  Eur.\ Phys.\ J.\ C {72} (2012) 2093
  [\hhref{1205.3781}].
  
  


\bibitem{MFV} G.~D'Ambrosio, G.~F.~Giudice, G.~Isidori and A.~Strumia,
  Nucl.\ Phys.\ B {645} (2002) 155
  [\hhref{hep-ph/0207036}].
  
  


\bibitem{Buras:2010pz}
  A.~J.~Buras, K.~Gemmler and G.~Isidori,
  Nucl.\ Phys.\ B {843} (2011) 107 [\hhref{1007.1993}].
  
  


\bibitem{Isidori:2013ez}
  G.~Isidori,
  \hhref{1302.0661}.
  
  


\bibitem{Z'LHC}   
 CMS Collaboration,
  Phys.\ Lett.\ B {720} (2013) 63
  [\hhref{1212.6175}].
  CMS Collaboration, Phys.\ Rev.\ D {87}  11 (2013)  114015
  [\hhref{1302.4794}].
  ATLAS Collaboration,
  \hhref{1405.412} and \hhref{1407.1376}.
 


\bibitem{Zwirner}
  E.~Salvioni, G.~Villadoro and F.~Zwirner,
  JHEP {0911} (2009) 068
  [\hhref{0909.1320}].
  E.~Salvioni, A.~Strumia, G.~Villadoro and F.~Zwirner,
  JHEP {1003} (2010) 010
  [\hhref{0911.1450}].
  
  


\bibitem{Caccia} 
  R.~Barbieri, A.~Pomarol, R.~Rattazzi and A.~Strumia,
  Nucl.\ Phys.\ B {703} (2004) 127
  [\hhref{hep-ph/0405040}].
  G.~Cacciapaglia, C.~Csaki, G.~Marandella and A.~Strumia,
  Phys.\ Rev.\ D {74} (2006) 033011
  [\hhref{hep-ph/0604111}].
  
  
 


\bibitem{CMSWR}
CMS collaboration, \hhref{1407.3683}
  
  


\bibitem{Cirigliano:2012ab}
  V.~Cirigliano, M.~Gonzalez-Alonso and M.~L.~Graesser,
  JHEP {1302} (2013) 046
  [\hhref{1210.4553}].

  
  


\bibitem{Foot}
   R.~Foot,
  Phys.\ Lett.\ B {420} (1998) 333
  [\hhref{hep-ph/9708205}].
  
  


\bibitem{Volkas}
  R.~R.~Volkas,
  Phys.\ Rev.\ D {53} (1996) 2681
  [\hhref{hep-ph/9507215}].
  
  


\bibitem{Val}
  G. Valencia, S. Willenbrock, Phys. Rev. D 50 (1994) 6843.
  
  


\bibitem{Carpentier:2010ue}
  M.~Carpentier and S.~Davidson,
  Eur.\ Phys.\ J.\ C {70} (2010) 1071 [\hhref{1008.0280}].
  
  


\bibitem{Bali:2011ks}
QCDSF Collaboration,
  Phys.\ Rev.\ D {85} (2012) 054502
  [\hhref{1111.1600}].
  
 


\end{thebibliography}
\end{document}